\documentclass[reprint,preprintnumbers,nofootinbib,superscriptaddress,amsmath,amssymb,aps,prd]{revtex4-2}
\pdfoutput=1
\usepackage{graphicx,hyperref,physics,orcidlink, algorithm2e, amssymb, amsmath, dsfont,subfigure,multirow}
\usepackage{lineno}
\usepackage{siunitx}
\usepackage{tikz}
\usepackage{bm}
\usetikzlibrary{quantikz2}
\newcommand{\us}{\unit{\micro\second}}
\graphicspath{{figs/}}
\hypersetup{
    unicode=true,
    pdftoolbar=true,
    pdfmenubar=true,
    pdffitwindow=false,
    pdfstartview={FitH},
    pdftitle={spin-boson},
    pdfauthor={J.Y.~Araz, M.~Grau, J.~Montgomery, F.~Ringer}, 
    pdfsubject={spin-boson},
    pdfcreator={J.Y.~Araz, M.~Grau, J.~Montgomery, F.~Ringer},
    pdfproducer={}, 
    pdfkeywords={},
    pdfnewwindow=true,
    colorlinks=true,
    linkcolor=blue,
    citecolor=olive,
    filecolor=magenta,
    urlcolor=cyan
}

\begin{document}


\title{Toward hybrid quantum simulations with qubits \& qumodes on trapped-ion platforms}

\author{Jack Y. Araz\orcidlink{0000-0001-8721-8042}}
\email{jack.araz@stonybrook.edu}
\affiliation{Thomas Jefferson National Accelerator Facility, Newport News, VA 23606, USA}
\affiliation{Department of Physics, Old Dominion University, Norfolk, VA 23529, USA}
\affiliation{Department of Physics and Astronomy,
Stony Brook University, New York 11794, USA}

\author{Matt Grau\orcidlink{0000-0002-2684-6923}}
\email{mgrau@odu.edu}
\affiliation{Department of Physics, Old Dominion University, Norfolk, VA 23529, USA}

\author{Jake Montgomery\orcidlink{0000-0003-1715-1107}}
\email{jake.montgomery@stonybrook.edu}
\affiliation{Department of Physics, Old Dominion University, Norfolk, VA 23529, USA}
\affiliation{Department of Physics and Astronomy,
Stony Brook University, New York 11794, USA}

\author{Felix Ringer\orcidlink{0000-0002-5939-3510}}
\email{felix.ringer@stonybrook.edu}
\affiliation{Thomas Jefferson National Accelerator Facility, Newport News, VA 23606, USA}
\affiliation{Department of Physics, Old Dominion University, Norfolk, VA 23529, USA}
\affiliation{Department of Physics and Astronomy,
Stony Brook University, New York 11794, USA}


\preprint{JLAB-THY-24-4200}

\begin{abstract}
We explore the feasibility of gate-based hybrid quantum computing using both discrete (qubit) and continuous (qumode) variables on trapped-ion platforms. Trapped-ion systems have demonstrated record one- and two-qubit gate fidelities and long qubit coherence times, while qumodes, which can be represented by the collective vibrational modes of the ion chain, have remained relatively unexplored for their use in computing. Using numerical simulations, we show that high-fidelity hybrid gates and measurement operations can be achieved for existing trapped-ion quantum platforms. As an exemplary application, we consider quantum simulations of the Jaynes-Cummings-Hubbard model, which is given by a one-dimensional chain of interacting spin and boson degrees of freedom. Using classical simulations, we study its real-time evolution and develop a suitable variational quantum algorithm for ground state preparation. Our results motivate further studies of hybrid quantum computing in this context, which may lead to direct applications in condensed matter and fundamental particle and nuclear physics.
\end{abstract}
\maketitle
\tableofcontents

\section{Introduction~\label{sec:intro}}

In recent years, significant progress has been made in developing different quantum hardware platforms. The most frequently studied quantum resources are qubits, which are two-level quantum systems. Unlike bits used in classical computations, a qubit can be in a superposition of two states. In addition, multiple qubits can be entangled, allowing for significant quantum speed-ups of certain calculations. For various computational problems \cite{bullock2005, rico2018}, it can be advantageous to have access to $d$-dimensional quantum systems, which are known as qudits. When taking the limit $d\to\infty$, the infinite-dimensional spectrum of a quantum mechanical harmonic oscillator is obtained. The corresponding bosonic quantum system is typically referred to as a qumode. Having access to an infinite-dimensional Hilbert space per computational unit is advantageous, for example, for quantum simulations of bosonic systems, and it can play a role in quantum error-correcting codes~\cite{Gottesman:2000di}. Due to the infinite-dimensional Hilbert space in the Fock basis, the states of a qumode can also be represented in terms of continuous variables analogous to the eigenvalues of the position and momentum operators of the harmonic oscillator~\cite{Lloyd:1998jk}. Although photonics are often the focus of continuous variable quantum computing~\cite{bartlett2002, menicucci2008, tasca2011}, qumodes can also be realized using other quantum platforms such as superconducting circuits~\cite{heeres2017, Stavenger:2022wzz, Wang:2020ghj}, trapped-ions~\cite{lau2012,fluhmann2019, sutherland2021a, neeve2022, matsos2024}, and cold atoms~\cite{bouchoule1999, kendell2024, shaw2024, bohnmann2024}.

In this work, we focus on hybrid qubit-qumode quantum computing with the aim of leveraging the advantages of each quantum system. Here, the combined Hilbert space can be written in terms of discrete (qubit) and continuous (qumode) variables ${\cal H}={\cal H}_{\rm qubit}\otimes {\cal H}_{\rm qumode}$. Potential applications include quantum error-correcting codes~\cite{Hu:2019zbe,Campagne-Ibarcq:2019nmy}, calculations of Franck-Condon spectra of molecules~\cite{Wang:2020ghj}, quantum simulations of condensed matter systems involving spin and boson degrees of freedom, as well as topological models~\cite{Cai_2021}. In addition, quantum field theories in fundamental particle and nuclear physics often involve fermion, scalar, and gauge field degrees of freedom. Within the Hamiltonian formulation of lattice field theories~\cite{Kogut:1974ag}, fermions can naturally be mapped to qubits using, for example, a Jordan-Wigner transformation~\cite{Jordan:1928wi}. Instead, scalar and gauge fields that are discretized on a spatial lattice require an infinite-dimensional Hilbert space per lattice site or link variable. As a result, they either require a truncation and subsequent mapping to a set of qubits or the relevant degrees of freedom can be mapped to the formally infinite-dimensional Hilbert space of qumodes~\cite{Chandrasekharan:1996ih,Jordan:2012xnu,Ercolessi:2017jbi,Banerjee:2012pg,Zohar:2015hwa,Hauke:2013jga,deJong:2021wsd,Jordan:2017lea,Zohar:2015hwa,Klco:2018kyo,Tong:2021rfv,Marshall:2015mna,Bauer:2021gek,Yeter-Aydeniz:2021mol,Barata:2020jtq,Spagnoli:2024mib,Honda:2021aum,Yeter-Aydeniz:2017ubh,Li:2021kcs,Cohen:2021imf,Jha:2023ecu,Florio:2023dke,Belyansky:2023rgh,Briceno:2023xcm,Davoudi:2022xmb,Funcke:2023lli,Araz:2022tbd,Araz:2023ngh,Asaduzzaman:2023wtd,Fromm:2023npm,Florio:2024aix,Grieninger:2024cdl,Lee:2024jnt,Briceno:2020rar,Barata:2024apg,Neill:2024klc,Farrell:2024fit,Gustafson:2024bww,Hardy:2024ric}.

We focus specifically on trapped-ion platforms where both qubits and qumodes can be realized. Qubits can be hosted by pairs of electronic states or hyperfine split states in an individual ion, and the collective motional degrees of freedom (phonons) of a chain of ions in a linear trap can be used to realize qumodes~\cite{porras2004, serafini2009, debnath2018, Chen_2021, Katz:2022gra, chen2023scalable}. In purely qubit-based quantum computing with trapped ions, the motional degrees of freedom are used, for instance, to realize two-qubit gates~\cite{cirac1995, molmer1999, sorensen1999}. Although it is possible to encode qumodes in the discrete spin degrees of freedom \cite{Macridin:2023xny}, here we focus on applications where simultaneous access to qubits and qumodes resources may give computational advantages. We will discuss qubit, qumode, and hybrid quantum gates that couple both degrees of freedom and suitable measurement schemes that can be implemented using existing trapped-ion platforms. Using an exemplary trapped-ion setup, we estimate gate times and the fidelities of different gate operations. 

This work complements the hybrid qubit-qumode framework that was recently developed by the Co-design Center for Quantum Advantage (C2QA)~\cite{Stavenger:2022wzz,Liu:2024mbr, Crane:2024tlj}, which focuses primarily on implementations using superconducting circuits. 
While many of the salient features of hybrid computation can be generalized to different physical implementations, there are several key differences between trapped-ion and superconducting cavity hardware that introduce practical trade-offs when designing circuits or choosing an optimal hybrid gate set.
Trapped-ion setups typically have chains of ions with a high degree of connectivity between qubits and modes. This connectivity generally comes at the expense of potential scalability, as it is currently only practical to work with highly connected chains of $\sim$20 ions or fewer. Additionally, the coherence times of trapped-ions hybrid systems exist in a completely different region of parameter space compared to superconducting systems, necessitating different considerations for control. Qumodes of superconducting niobium cavities have demonstrated coherence times of over a second~\cite{romanencko2020} and transmon qubit coherence times of several hundred microseconds~\cite{schreier2008}. Conversely, trapped-ions have qubit coherence times that are longer than the motional mode coherence times, e.g., with demonstrated qubit coherence times of 1 hour~\cite{wang2021} and motional mode coherence times of 10-100 milliseconds in linear Paul traps, with times reaching 1 second in Penning traps~\cite{Jarlaud_2021}. Instead of analog-digital hybrid simulations considered in previous work~\cite{Casanova:2011wh,Lv:2017tmn,Davoudi:2021ney}, here we focus on gate-based simulations using qubit-qumode hybrid platforms. See also Ref.~\cite{Liu:2024mbr}.

The experimental toolbox for manipulating trapped-ion qubits is sophisticated and precise, with high-fidelity state preparation, state readout, and single and multi-qubit operations.
In the Lamb-Dicke regime, the Jaynes-Cummings and anti-Jaynes-Cummings interactions can be realized natively by driving red and blue motional sideband transitions on individual ions with focused laser beams. While it is difficult to interact directly with the motional modes of the ion chain, there are many options that take advantage of the exquisite control over trapped-ion qubits to perform operations on modes mediated by an ion qubit. See also Ref.~\cite{sutherland2021a}, where different aspects of continuous-variable quantum computing and hybrid schemes have been discussed in the context of trapped ions.

To illustrate the potential of hybrid qubit and qumode quantum simulations, we consider the Jaynes-Cummings-Hubbard (JCH) model in this work, which is also referred to as the Jaynes-Cummings lattice model~\cite{Angelakis_2007, Schmidt_2009, Nunnenkamp:2011ww, Bertassoli:2024sri,Li:2022seq}. It is a 1+1 dimensional lattice model involving coupled spin and boson degrees of freedom per lattice site. Various aspects of the JCH model have been studied in the literature. For example, its phase diagram has been studied in Ref.~\cite{Greentree_2006, Nunnenkamp:2011ww} using mean-field theory, which is expected to exhibit a Mott insulator and superfluid phase. In addition, the JCH model shares features with one-dimensional quantum field theories (QFTs) such as the Schwinger model where fermionic matter is coupled to a U(1) gauge field~\cite{Schwinger:1962tp, Coleman:1975pw}, making our studies a natural starting point for the exploration of hybrid quantum simulations in the context of fundamental physics. See also Refs.~\cite{Casanova:2011wh,Lv:2017tmn,Davoudi:2021ney,Koch:2021wqd,Zache:2023cfj,Fromm:2024caq,Illa:2024kmf,Abel:2024kuv}.

\begin{figure}[t]
    \includegraphics[width=\linewidth]{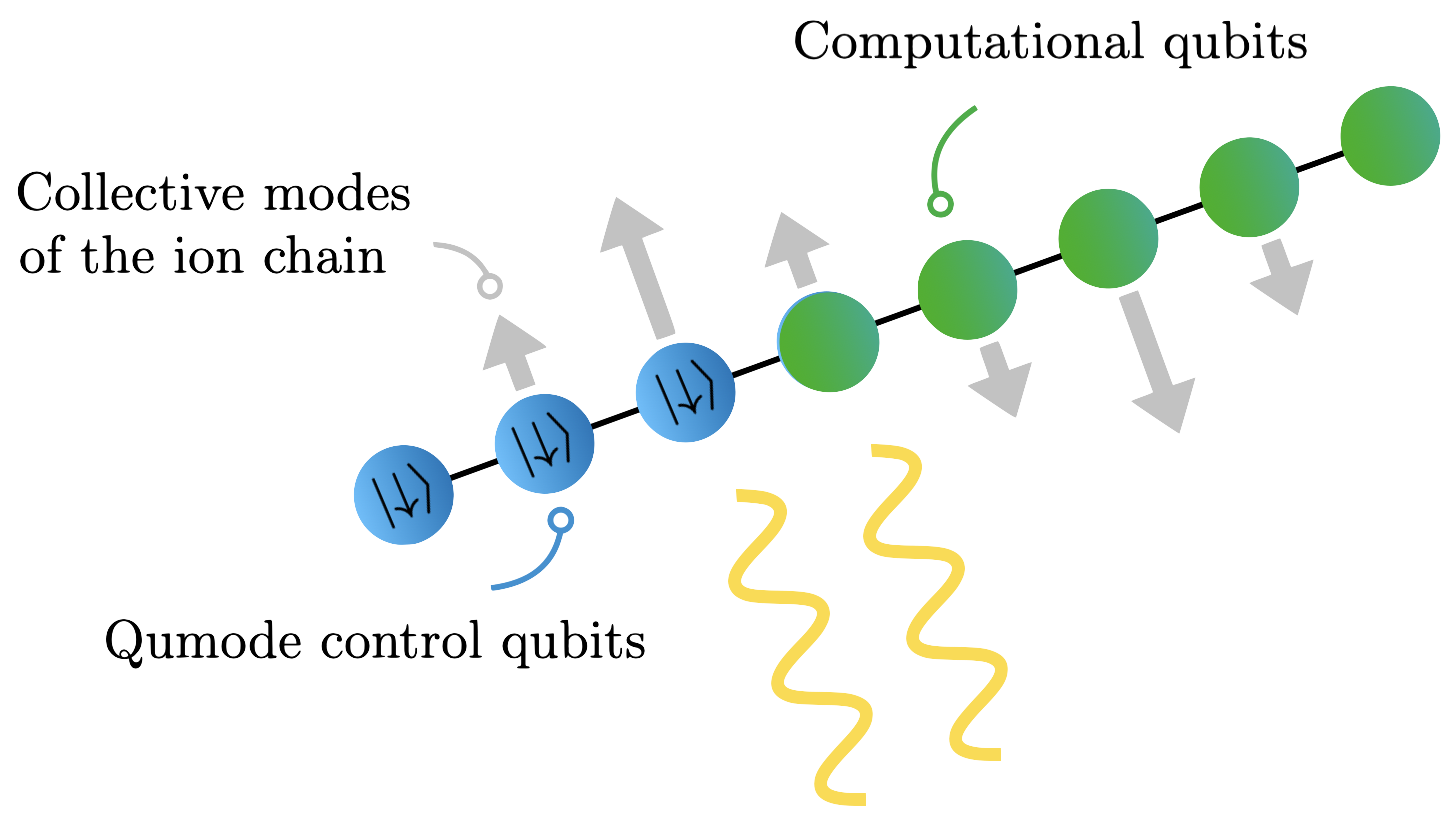}
    \caption{Illustration of the hybrid qubit and qumode setup using trapped-ions.~\label{fig:QubitQumode}}
\end{figure}

The remainder of this work is organized as follows. In section~\ref{sec:spin_boson_iontrap}, we discuss an example trapped-ion setup allowing access to both qubits and qumodes. We discuss different gate and measurement operations along with numerical simulations demonstrating the feasibility of our approach using current quantum hardware platforms. In section~\ref{sec:jch}, we introduce the JCH model and present numerical results for the real-time evolution and ground state preparation using a suitable variational algorithm. We conclude and present an outlook in section~\ref{sec:conclusion}.

\begin{figure*}[t!]
\centering
       \includegraphics[width=.9\linewidth]{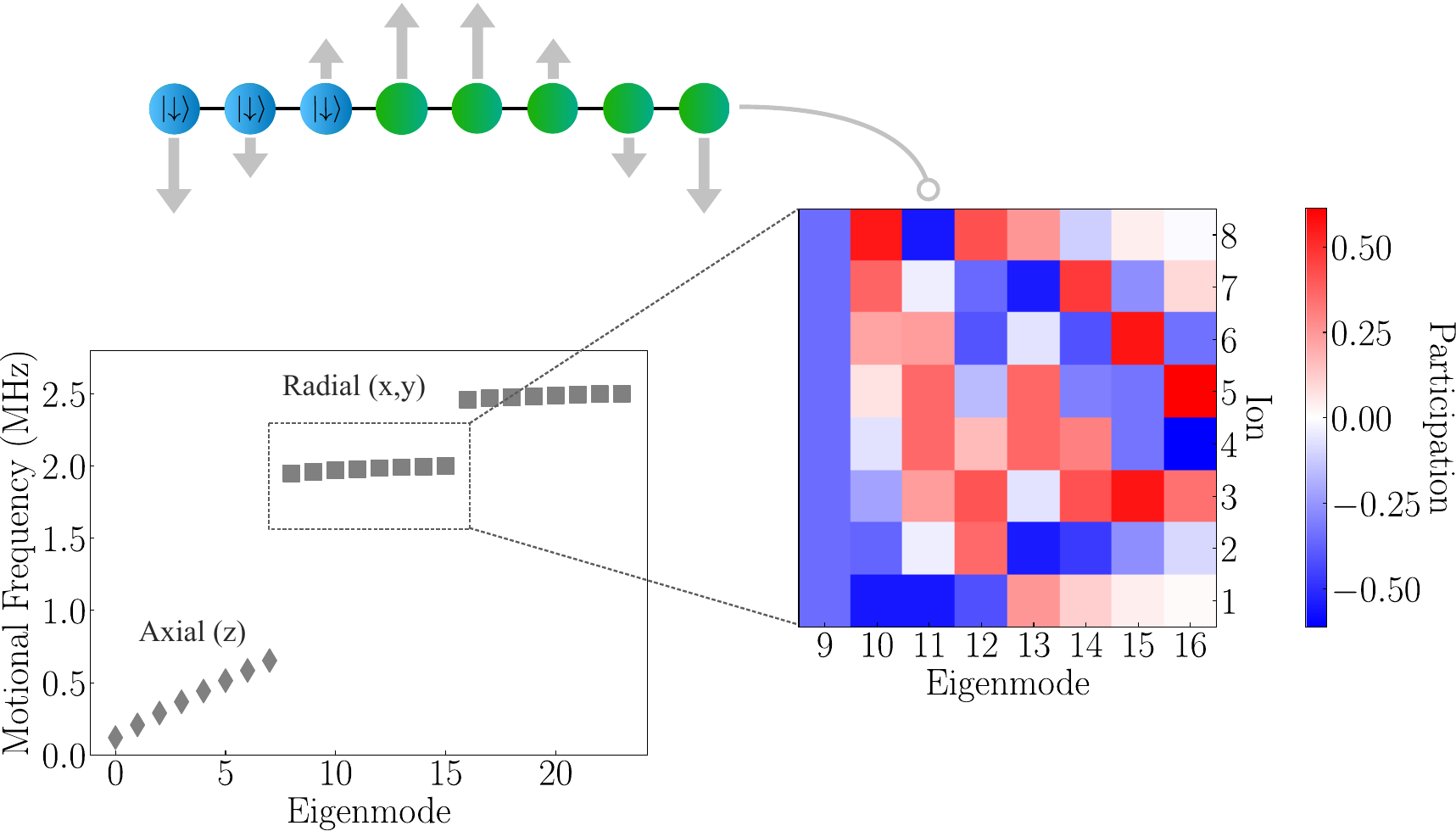}
       \caption{Bottom left: Frequencies of the 8 axial and 16 radial modes for the ion trap setup considered in this work. Right: The ion-mode participation matrix for the radial $x$ modes, the lower of the two radial bands. For example, the leftmost column represents the center-of-mass mode, where all ions are oscillating in phase. The second column corresponds to a rocking mode. Top left: The motion of ions in the third radial $x$ mode, the 11\textsuperscript{th} eigenmode, which is a bending mode. \label{fig:modesandparticipation}}
\end{figure*}

\section{Qubits \& qumodes with trapped ions~\label{sec:spin_boson_iontrap}}

In this section, we start by introducing the notation used throughout this work. We then discuss hybrid qubit and qumode gate and measurement operations for trapped ions. This will allow for the simultaneous use of continuous and discrete variables as computational resources. We present numerical results to demonstrate the feasibility on current quantum platforms.

\subsection{Notation}

We start by introducing the notation used throughout this work. We label the two states of a qubit as spin up, $\ket{\uparrow}$, and spin down, $\ket{\downarrow}$. The Pauli or spin operators are denoted by $\sigma^{x,y,z}$, and the spin raising and lowering operators are given by $\sigma^{\pm}=(\sigma^x\pm i \sigma^y)/2$. We have  $\ket{\uparrow} = \sigma^+ \ket{\downarrow}$ and $\ket{\downarrow} = \sigma^- \ket{\uparrow}$, as well as $\{\sigma^+,\sigma^-\}=\mathbb{I}$, where $\mathbb{I}$ is the identiy matrix. The state of a qumode is typically represented in terms of the Fock basis, the position and momentum basis, or the coherent state basis. Throughout this work, we will only work in the Fock and position/momentum basis. We denote the Fock states of a qumode or harmonic oscillator by $\ket{n}$, which can be written as
\begin{equation}
    \ket{n}=\frac{(\hat a^\dagger)^n}{\sqrt{n!}}\ket{0}\ .
\end{equation}
Here, $\ket{0}$ is the ground state of the qumode. The bosonic raising and lowering operators satisfy the commutation relation $[\hat a,\hat a^\dagger]=1$. The Fock states $\ket{n}$ are eigenstates of the number operator $\hat N=\hat a^\dagger\hat a$ with eigenvalue $n$.  We can write the position and momentum or quadrature operators $\hat X$ and $\hat P$, which satisfy the commutation relation $[\hat X,\hat P]=i$, in terms of the raising and lowering operators,
\begin{align}\label{eq:quadratures}
\hat X = \frac{1}{\sqrt{2}}(\hat a^\dagger + \hat a),& \quad \hat P = \frac{i}{\sqrt{2}}(\hat a^\dagger - \hat a)\ .
\end{align}
We denote the eigenstates of the position operator by $\ket{x}$ with $\hat X\ket{x}=x\ket{x}$ with eigenvalue $x$, and analogous relations hold for the momentum operator. Using the position basis, we can write the state of a qumode in terms of the continuous variable $x$ as
\begin{equation}\label{eq:position}
    \ket{\psi}=\int {\rm d}x\, \psi(x) \ket{x}\,,
\end{equation}
with the wave function $\psi(x)$. Note that we use $\hbar=1$ throughout this work. In circuit diagrams, single (triple) lines represent qubit (qumode) wires~\cite{Kay:2018huf}. See e.g., Fig.~\ref{fig:trotter-step}. 

\subsection{Trapped-ion setup and mode structure}

In this section, we will focus on a particular trapped ion implementation, $^{171}\text{Yb}^+$ ions in a linear Paul trap, for concreteness. However, the principles we discuss should apply equally well to other trapped-ion systems. The qubit levels, $\ket{\downarrow}$ and $\ket{\uparrow}$, are the two hyperfine levels $\ket{F=0, m_F=0}$ and $\ket{F=1, m_F=0}$, of the $^{2}\text{S}_{1/2}$ ground state of $^{171}\text{Yb}^+$, which have a demonstrated coherence time of several seconds \cite{olmschenk2007}.
In principle, multiple electronic or hyperfine states can be employed in a single ion to realize multiple qubits~\cite{allcock2021, yang2022, debry2023}, which would be useful to explore in the context of hybrid computing platforms in future work.

A qumode is idealized as a harmonic oscillator with an infinite number of equally spaced states. Using trapped ions, qumodes can be realized in terms of phonon modes, i.e., the vibrational modes arising from the harmonic motion of the ions in the trap. Generally, trapped-ion experiments can exhibit one of two classes of modes depending on the configuration of the ions in the trap: Local phonon modes of individual ions, when the motion of the ions can be described as independent harmonic oscillators~\cite{brown2011}, and collective phonon modes of the ion chain when the ions are sufficiently tightly confined in the trap~\cite{porras2008, ivanov2009, bermudez2010}. Here, we focus on collective modes as they exhibit full connectivity to all the qubits of the ion chain. 
Compared to the qubit degrees of freedom, the ion motional modes typically exhibit shorter coherence times due to electric field noise that can lead to heating and decoherence of the motional modes~\cite{wineland1975, brownnutt2015, Chen_2021}. With the exception of the center-of-mass mode, collective modes are less sensitive to decoherence arising from the coupling of the motional mode to currents in trap electrodes~\cite{king1998}. This coupling is suppressed for all modes except for the center-of-mass modes by a factor of the ion spacing divided by the ion-electrode distance~\cite{wineland1998experimental, cirac1995}.
We note that several proposals for realizing continuous variable quantum computing have been proposed using local ion modes~\cite{porras2004,deng2008, serafini2009, shen2014}. In general, $N$ ions can give access to at most $3N$ phonon modes, which in a linear trap will comprise $N$ axial modes and $2N$ radial modes. Higher numbers of modes of the ion chain can be increasingly difficult to resolve. This upper limit on the resolution of modes depends on the details of the experimental setup. Compared to qubits, the coherence time of qumodes is relatively short, ${\cal O}(ms)$ \cite{hou2024}. We will compare the coherence time to the different gate operations in the following section.

\begin{table*}
    \centering
    \begin{tabular}{c|c|c|c|c|c|c}
        Type & Operation & Short & Operator  & Estimated gate time & Estimated fidelity& Ref.\\ \hline\hline
        \multirow{3}{*}{Qubit gates} & Pauli operators & & $\sigma^i$ & 2 \us & 99.999\%&~\cite{loschnauer2024}\\
        & Rotation & ${\rm R}_i(\theta)$ & $e^{i\theta\sigma^i/2}$ & 2 \us & 99.999\%&~\cite{loschnauer2024}\\
        & Controlled NOT & CNOT & $e^{i\frac{\pi}{4}(\mathbb{I}_1-\sigma^z_{1})(\mathbb{I}_2-\sigma^x_{2})}$ & 30 \us & 99.9\%&~\cite{gaebler2016}\\ \hline  
      \multirow{6}{*}{Qumode gates} & Rotation & ${\rm R}(\theta)$ & $e^{i\theta \hat a^\dagger \hat a}$ & 200 \us$^{*}$ & 99\%$^{*}$&~\cite{shen2018}\\
       & Displacement & ${\rm D}(z)$ & $e^{z \hat a^\dagger-z^* \hat a}$ & 10 \us& 99\%&~\cite{McCormick_2019}\\
       & Single-mode squeezing & ${\rm S}(z)$ & $e^{(z^* \hat a\hat a-z \hat a^{\dagger}\hat a^\dagger)/2}$  & 3 \us & 98\%&~\cite{burd2019}\\
       & Beam splitter & ${\rm BS}(z)$ & $e^{z \hat a^\dagger \hat b- z^* \hat a \hat b^\dagger}$ & 250 \us & 99\%&~\cite{chen2023scalable} \\
       & Kerr & ${\rm K}(z)$ & $e^{i\theta (\hat a^\dagger \hat a)^2}$ & 10 ms$^{*}$ & 95\%$^{*}$&~\cite{stobinska2011} \\
       & Cross-Kerr & ${\rm CK}(z)$ & $e^{i\theta \hat a^\dagger \hat a\, \hat b^\dagger \hat b}$ & 800 \us & 97\%&~\cite{ding2017cross} \\ \hline
        \multirow{6}{*}{Hybrid gates} & Red sideband  & ${\rm RSB}(z)$&$  e^{i z \hat a \sigma^+ + i z^* \hat a^\dagger \sigma^-}$ & 200 \us & 99.9\%&~\cite{chen2023scalable}\\
         & Blue sideband & ${\rm BSB}(z)$&$e^{i z \hat a^\dagger \sigma^+ + i z^* \hat a \sigma^-}$ & 200 \us  & 99.9\%&~\cite{chen2023scalable} \\
         &Controlled rotation & ${\rm CR}(\theta)$&$e^{i\theta\sigma^z \hat a^\dagger\hat a}$& 200 \us$^{*}$ & 99\%$^{*}$&~\cite{sutherland2021a}\\ 
         &Controlled displacement & ${\rm CD}(z)$&$e^{\sigma^z(z \hat a^\dagger-z^* \hat a)}$& 800 \us & 95\%$^{*}$&~\cite{srinivas2019}\\ 
         &Controlled squeezing & ${\rm CS}(z)$&$e^{\sigma^z(z^* \hat a\hat a-z \hat a^{\dagger}\hat a^\dagger)/2}$& 120 \us$^{*}$ & 99\%$^{*}$&~\cite{sutherland2021a}\\ 
         &Controlled beam splitter & ${\rm CBS}(z)$&$e^{\sigma^z (z \hat a^\dagger \hat b- z^* \hat a \hat b^\dagger)}$ & 250 \us  & 99\%&~\cite{chen2023scalable} \\ \hline
    \multirow{5}{*}{Measurements}   & Qubit Pauli strings  & & $\sigma^i$ & 145 \us & 99.99\%&~\cite{myerson2008}\\
      & Average phonon number & & $\hat{N}$ & 200 \us & 97\%&~\cite{meekhof1996}\\
      & Qumode PNR & & $\ket{n}\bra{n}$ & 400 \us & 99\%&~\cite{Lv2017pnr}\\
      & Qumode homodyne & & $\hat X,\hat P$ & 200 \us & 95\%$^{*}$&~\cite{Lv2017pnr}\\
      & Hybrid & & $\sigma^i \hat X$, $\sigma^i \hat P$ & 200 \us$^\dagger$ & 95\%$^\dagger$\\\hline
    \end{tabular}
    \caption{Summary of the qubit and qumode gates, as well as measurement operations considered in this work along with their estimated gate times and fidelities for trapped $^{171}\text{Yb}^+$ ions in a linear Paul trap. Here, $\hat a,\hat a^\dagger$ and $\hat b, \hat b^\dagger$ refer to bosonic annihilation and creation operators acting on different qumodes (collective motional modes of the ion chain), and the identity $\mathbb{I}_j$ and Pauli operators $\sigma^i_j$ act on qubit $j$ with $i=x,y,z$. For parametrized gates, we introduce the complex-valued variable $z=\theta e^{i\phi}$ with $\theta\geq 0$, $\phi\in[0,2\pi)$.
    We estimate gate times and fidelities from references listed in the last column, where for experimental demonstrations we either include a measured gate time or infer one from a reported interaction strength. Note that red and blue sideband gates refer to the adiabatic versions of the traditionally understood coherent operations in trapped ions. Gate times are for $\theta=\pi/2$ and $\left|z\right|=1$. $^{*}$Estimate based on theoretical prediction. ${}^{\dagger}$Hybrid measurement using the upper bound on qubit and qumode homodyne measurements. 
    ~\label{tab:gates}}
\end{table*}

As illustrated in Fig.~\ref{fig:QubitQumode}, we consider a trapped-ion setup that gives access to both qubits and qumodes as computational resources. We consider $N=8$ ions that, in principle, allow for access to $24$ qumodes. 
However, in practice, several collective motional modes will be reserved to mediate two-qubit gates.
The qumodes are initialized to the $\ket{n=0}$ motional state by laser-cooling to the Doppler limit of about 1 mK, followed by resolved sideband cooling each mode $m$ to $\langle{n_m\rangle} < 0.01$ on a narrow transition with a linewidth less than the motional frequency \cite{resolvesideband}.
The qubits on each ion can be initialized to the $\ket{\downarrow}$ state by optical pumping \cite{opticalpumping}.

The gray arrows in Fig.~\ref{fig:QubitQumode} illustrate an exemplary radial mode. Since typical setups of current ion traps do not allow for direct interactions with the motional degrees of freedom of the chain, we need qubits to control the qumodes. Therefore, we reserve a certain number of control qubits, shown as blue circles in Fig.~\ref{fig:QubitQumode}. In principle, we only need a single control qubit. However, the performance of continuous variable gates can be improved with the help of multiple control qubits. For example, to perform the beam splitter operation, a continuous variable gate discussed in more detail below, the control qubits need to remain in the $\ket{\downarrow}$ state. The remaining qubits shown as green circles in Fig.~\ref{fig:QubitQumode} can be used as computational resources. Ideally, we set aside as few control qubits as possible, thus retaining as many computational qubits as possible. We explore this aspect in more detail below. Since all qubits share the collective motional degrees of freedom, we can realize hybrid gates that couple the computational qubits to the qumodes.

In the pseudopotential approximation, the ion trap provides a three-dimensional harmonic oscillator potential 
\begin{equation}
V_\text{trap}(\vec{x})= \frac{1}{2}\omega_x^2 x^2+\frac{1}{2}\omega_y^2 y^2+\frac{1}{2}\omega_z^2 z^2 \,,
\end{equation}
with the trap frequencies being $\{\omega_x, \omega_y, \omega_z \} = 2\pi \times \{2.0 ,2.5, 0.12\}$ MHz. The discrepancy between the trap frequencies causes the eigenfrequency spectrum of the system with multiple ions to be gapped, which is experimentally advantageous for manipulating the ions. For eight $^{171}\text{Yb}^+$ ions, the mode spectrum and mode participation are shown in Fig.~\ref{fig:modesandparticipation}.
The participation vector $b_{j,m}$ describes how strongly the mode $m$ couples to the qubit hosted on ion $j$.
We calculate the vector of 24 spatial coordinates of equilibrium positions $\left\{r_i\right\}$ of the ions in the trap by minimizing the total potential $V_\text{tot}(\left\{r_i\right\}) = V_\text{trap}(\left\{r_i\right\}) + V_\text{Coulomb}(\left\{r_i\right\})$ of the ion chain, where $V_\text{Coulomb}$ is the mutual Coulomb repulsion of the ions.
We then calculate the mode frequencies and participation vectors by linearizing the net force on each ion around its equilibrium position by calculating the Hessian $H_{ij} = \partial^2 V_\text{tot} / \partial r_i\partial r_j$ of the total potential. The eigenvalues of $H_{ij}$ are the square of the mode frequencies, and the eigenvectors are the mode participation vectors.

\subsection{Single-qumode gates}

Qumode rotation gates are implemented as controlled rotation gates (see section ~\ref{sec:hybrid_gates}), except that we use a control qubit in the $\ket{\downarrow}$ state. Before and after the controlled rotation, a $\sigma^x$ gate is applied to the control qubit. As a result, the action of the operator on the hybrid Hilbert space is only that of the rotation gate, and the controlled qubit is returned to the $\ket{\downarrow}$ state. We have
\begin{align}
     \sigma^x \text{CR}(\theta)\sigma^x \ket{\downarrow} \ket{\psi} = & \, \sigma^x \exp[i\theta \sigma^z \hat{a}^\dagger \hat{a}]\ket{\uparrow}\ket{\psi} \nonumber \\
    = & \, \ket{\downarrow} \otimes \text{R}(\theta)\ket{\psi}
\end{align}
where $\ket{\psi}$ is the qumode state.

Squeezing of motional states with trapped ions using laser fields was first demonstrated in Ref.~\cite{meekhof1996}. The generation of squeezed states is more robustly performed through the method proposed in Ref.~\cite{alonso2013}, which implements fast-switching electrodes inside the trap that oscillate the internal electric potential with strong control. Specifically, an ion in a motional ground state for a given quadratic potential is transported to a new potential with characteristic frequency $\omega_1$, which causes the curvature of the Gaussian wave function to flip. The curvature of the potential is then decreased to $\omega_2 = \lambda \omega_1$, which causes the wave function to breathe. The potential is then returned to frequency $\omega_1$, squeezing the state with a change in variance of $\lambda$.
Recent experiments have demonstrated up to 5.9 dB of single-mode squeezing in a single-ion surface-electrode trap~\cite{metzner2024}, and 5.4 dB of squeezing in the center-of-mass mode of a 2D ion crystal in a Penning trap via parametric modulation~\cite{affolter2023}. In both cases, the squeezed motional states remained coherent over timescales of 10 ms or longer.

It was demonstrated in Ref.~\cite{Leibfried:2003zz} that state displacement can be achieved by superimposing two standing wave electric fields that drive Raman transitions between different $\ket{n}$ levels. Choosing the detuning of the lasers to be $\omega_z$ gives an effective interaction Hamiltonian of 
\begin{equation}
\hat{H}_I(t) = \Omega \exp\left[i \eta ( \hat{a} e^{-i \omega_z t} + \hat{a}^\dagger e^{i \omega_z t} ) \right] + {\rm h.c.} 
\end{equation}
When working in the Lamb-Dicke regime ($\eta \ll 1$), the evolution operator corresponding to this Hamiltonian gives the qumode displacement gate in good approximation. 

To construct Kerr gates, we start with the interaction Hamiltonian of a laser incident on a single ion in a Paul trap, 
\begin{equation}
    H_I = \frac{\Omega}{2} \sigma^+ \exp\left[i \eta ( \hat{a} e^{-i \omega t} + \hat{a}^\dagger e^{i \omega t} ) \right] + {\rm h.c.} 
\end{equation}
where $\omega$ is the trap frequency and $\eta$ is the Lamb Dicke parameter. Expanding to fourth order in $\eta$ yields 
\begin{equation}\label{kerr}
    \hat H_I = \frac{\Omega}{2} \sigma^x \left[ 1- \frac{\eta^2}{2}+ \frac{\eta^4}{4} -\eta^2 \hat{N} + \frac{\eta^4}{4}\hat{N}^2 \right].
\end{equation}
There are higher-order interactions, but they are suppressed by additional factors of $\eta^2$ \cite{stobinska2011}. Since the action of $H_I$ is nontrivial on the qubit Hilbert space, we need to utilize a control qubit that is kept in the state $\ket{\downarrow}$. By first applying a rotation gate to the qubit  ${\rm R}_x\left(\pi /2 \right)$, the qubit is left in an eigenstate of the Hamiltonian, $\frac{1}{\sqrt2}\left(\ket{\uparrow}+\ket{\downarrow}\right)$, with eigenvalue $1$. The action of the Hamiltonian is then only nontrivial on the qumode Hilbert space. The term $\propto \eta^2 \hat{N}$  can be compensated by applying a rotation gate ${\rm R}\left(\frac{\Omega}{2}\eta^2 t \right)$, where $t$ is the desired time step of the gate. Inverting the qubit rotation after the time evolution of Eq.~(\ref{kerr}) returns the control qubit to its original state and recovers the action of the Kerr gate, 
\begin{equation}
    U(t)= \exp[i\frac{\Omega \eta^4}{8}\hat{N}^2 t]\,,
\end{equation}
up to a constant phase that we can ignore. This gate relies on the inherent nonlinearity of the trap potential, which is kept small in trapped-ion setups for practical considerations. This manifests as $\eta^4$ dependence in the gate. This makes the gate time $\propto \eta^{-4}$ and causes this gate to be the slowest in the set of gates we consider for the hybrid qubit-qumode system, see Tab.~\ref{tab:gates}. However, coherence times longer than the estimated gate time for the Kerr gate have been demonstrated in many-ion chains \cite{meth2023simulating}. This gate has not been performed experimentally, and we leave the technicalities of its implementation for future work.

\begin{figure}[t]
    \centering
    \includegraphics[width=\linewidth]{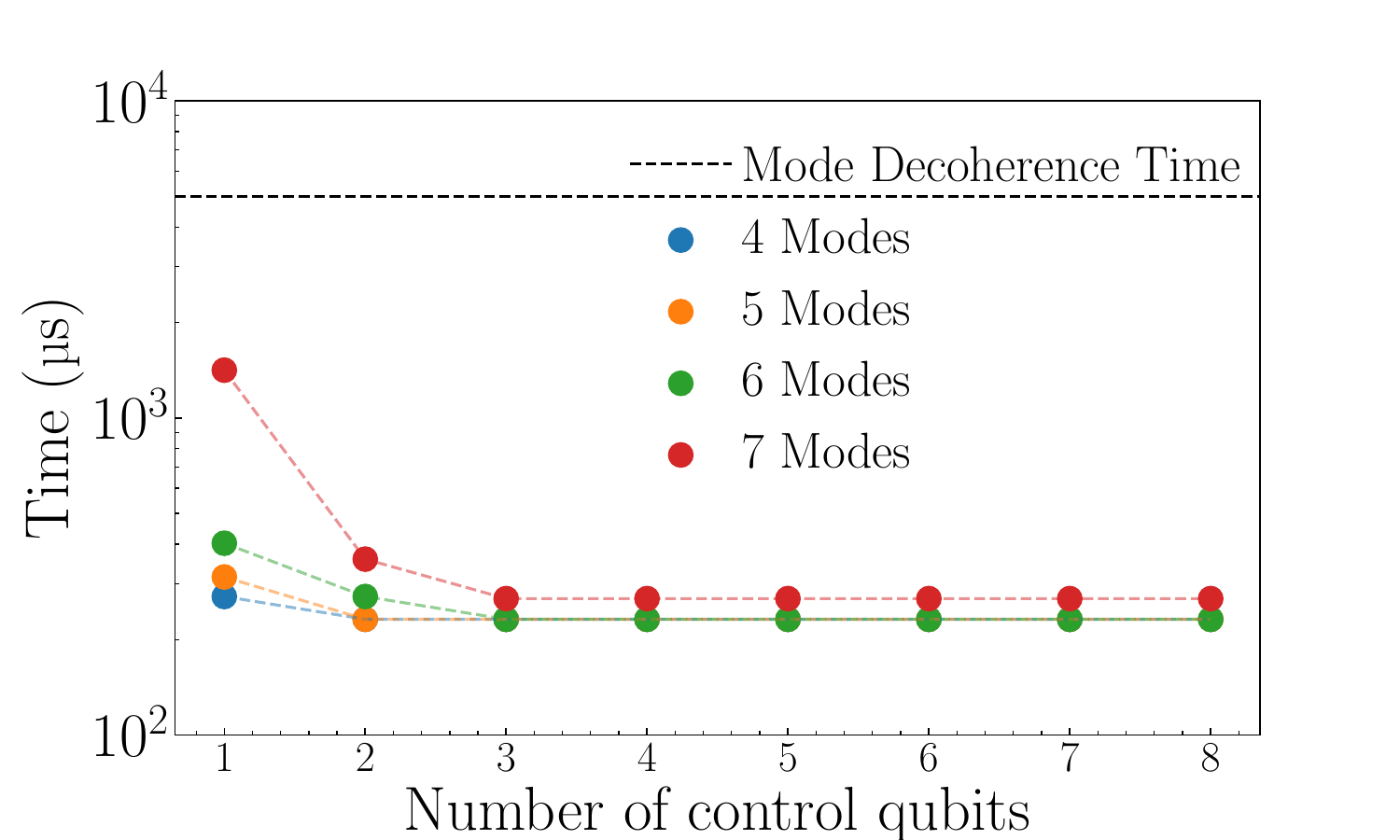}
    \caption{The maximum gate time of the set of beam splitter gate times covering nearest-neighbor mode couplings that correspond to having minimal total gate times for accessing a subset of the ions compared to the assumed coherence time of a qumode, 5 ms.\label{fig:max_gatetime}}
\end{figure}

\subsection{Two-qumode gates}

The participation of all ions in the collective modes of the system allows for the realization of programmable beam splitter gates between two qumodes with arbitrary phase and rotation~\cite{chen2023scalable}. A single ion, labeled $j$, is driven off of resonance with two tones detuned by $\Delta_\text{bs}$ from the red sidebands (RSB) of the two modes we want the beam splitter to act on, labeled $m$ and $n$. The distinct Rabi frequencies of these drives, $\Omega_{j,m}$ and $\Omega_{j,n}$, satisfy $\eta_{j,m} \Omega_{j,m} \approx \eta_{j,n} \Omega_{j,n} \approx \Delta_\text{bs}/2 $, where $\eta_{j,m}$ is related to the mode participation of each ion, $b_{j,m}$, and the Lamb Dicke parameter of each ion $\eta_j$, by $\eta_{j,m}= \eta_j b_{j,m}$~\cite{Katz:2022gra}. These drives acting at the same time result in the effective time evolution operator, 
\begin{align}
    U^{(m,n)}_j(t) &= \exp\left[ i\theta_\text{bs}(t) \,\sigma^z_j (\hat a^\dagger_m \hat a_n e^{i \phi} +\hat a_m \hat a^\dagger_n e^{-i \phi}) \right] \,, \\
     \theta_\text{bs}(t) &= \frac{\eta_{j,m}\Omega_{j,m}\eta_{j,n}\Omega_{j,n} t }{4 \Delta_\text{bs}}\,, 
\end{align}
where $\phi$ is the phase difference between the two tones, and $t$ is the time the amount of pulses are on. In principle, one control qubit kept in the state $\ket{\downarrow}$ can host all of the beam splitter gates that couple to any two modes, given that the ion participates in all modes. However, this is not true for all ions, and even a small participation of ions in different modes presents experimental concerns. The gate time for a beam splitter is a one-quarter period of the oscillation between state, $T_\text{bs} = \pi/2 \omega_\text{bs}$, where the beam splitter frequency, $\omega_{\rm bs}$, is given as 
\begin{eqnarray}\label{eq:BSfreq}
    \omega_\text{bs} = \frac{\eta_{j,m}\Omega_{j,m}\eta_{j,n}\Omega_{j,n} }{4 \Delta_\text{bs}} \ .
\end{eqnarray}
Low participation of an ion in a particular mode corresponds to long gate times for any beam splitter that is hosted on that ion that couples to that mode.

Therefore, it is pragmatic to optimize which ions host the beam splitter gates in order to minimize the sum of gate times of beam splitters that cover all of the mode-mode couplings that we wish to have. Any configuration of couplings that can be mapped to the target Hamiltonian is a valid configuration. For example, if one wishes to simulate a system with $N$ sites with periodic boundary conditions, one possible choice of mode couplings within a single band using $N$ ions is $(1,2),\; ... \;,\,(N-1,N),\,(N,1) $, which is trivially mapped to a local Hilbert space. From Eq.~(\ref{eq:BSfreq}), minimizing the sum of beam splitter gate times amounts to maximizing the product of the elements of the participation matrix for the control qubit and the two coupled modes, $b_{i,n},\,b_{j,m}$. For example, observe the participation matrix in Fig.~\ref{fig:modesandparticipation}. Ions three and six participate weakly in mode 13, making them poor choices to host couplings with this mode. In general, reserving more ions as control qubits will allow for more combinations of mode-mode couplings. This comes at the cost of computational qubits in the system. The more combinations of modes available (the more control qubits used) the better the optimized set of gates will perform, up to a point where the most optimized configuration can be found, beyond which there is no need to include more control qubits. 

This optimization scheme is demonstrated in Fig.~\ref{fig:max_gatetime}. We consider a chain of eight ions with modes from a single radial band. For a given number of control qubits, we optimize the sum of beam splitter gate times for a configuration of couplings that can be mapped to nearest neighbor interactions and report the highest gate time in that set. This is repeated for different numbers of modes, excluding the center of mass mode because of its shorter coherence time. We leave $\Delta$ and $\Omega$ constant since they can be individually tuned and kept equal for every ion. Going from one to two control qubits greatly speeds up the gate times, but going from two to three does offers much less improvement, if any at all, and there is no need to use more than three control qubits.
When performing beam splitters between seven modes, only two control qubits are needed to have all the gate times well below the typical qumode coherence time. This ensures all the beam splitters can be performed within the coherence time of the qumode. 
As a matter of practice, most experimental setups are designed to operate with a fixed number of trapped ions. Given this constraint, one can effectively trade off between the number of computational qubits available as a resource and the achievable gate times.

Cross Kerr gates, like Kerr gates, are realized by exploiting the nonlinearity of the Coulomb interaction between the ions in the trap that give rise to couplings between radial and axial motional modes~\cite{marquet2003,roos2008, nie2009}. While this coupling is normally quite weak, it can be enhanced in two modes that are near a parametric resonance ($\omega_a = 2 \omega_b$), such as a pair of carefully chosen axial and radial modes. Cross Kerr interactions have been performed for a three-ion chain ~\cite{ding2017cross}.

Estimates of gate times and fidelities taken from experimental demonstration in the literature are summarized in Tab. \ref{tab:gates}. As the length of the ion chain increases, the mode spectrum becomes denser, complicating the ability to address individual modes. While larger ion chains necessitate more complex experimental setups for the purposes of imaging and addressing individual ions, this scaling does not inherently limit the implementation of the gates discussed here.
 
\subsection{Hybrid qubit-qumode gates}\label{sec:hybrid_gates}

Adiabatic red and blue sideband gates (distinct from red/blue sidebands, the modes of the ions), also commonly known as Jaynes Cummings and anti-Jaynes Cummings interaction, respectively, are the simplest nontrivial operators that act on both modes and qubits. These gates are realized by applying a stimulated Raman adiabatic passage, which amounts to a sinusoidal modulation of the Rabi frequency of the laser incident on the ion, $\Omega\to \Omega(t) = \Omega \sin(\pi t/T) $ where $T$ is the total transfer time for the gate $\sim 7 \pi/\eta \Omega$ \cite{Chen_2021, um2016phonon}. This enables state transitions $\ket{\uparrow,n} \to \ket{\downarrow, n+1}$ and $\ket{\downarrow,n} \to \ket{\uparrow, n-1}$ for red and blue sidebands gates, respectively.

An implementation of the controlled rotation and controlled squeezing gates is proposed in Ref.~\cite{sutherland2021}. The controlled rotation is realized similarly to the controlled beam splitter gates. An ion, labeled $j$, is driven by two lasers. They are driven out of phase by $\phi = \pi/2$, and both are equally detuned by $\Delta_\text{r}$ far from the red and blue sidebands of the mode, labeled $m$, that is acting on. The Rabi frequencies satisfy the same relation with the Lamb Dicke parameters and the detuning as with the beam splitter gate. However, a difference is that the Rabi frequencies are chosen such that the two-photon Rabi frequency is purely imaginary. This gives an effective time evolution operator of 
\begin{eqnarray}
     U^m_j(t) &=& \exp[i \theta_\text{r}(t) \sigma^z_j \left( \hat{a}^{\dagger}_m\hat{a}_m \right)]\,,\\
    \theta_\text{r}(t) &=&\frac{4\eta_{j,m}^2\Omega_{j,m}^2 t }{\Delta_\text{r}}\,,
\end{eqnarray}
where $t$ is the amount of time the ion is driven for. This evolution operator corresponds to the gate CR($\theta$) with $\theta=\theta_{\rm r}(t)$. 

The controlled displacement gate relies on applying a spin-dependent force. Generating this force requires creating an electric field with a spatial gradient on the order of the mode wave function. Such a gradient is usually achieved using a laser with a wavelength that is much smaller than the size of the wave packet or by generating a strong microwave field gradient with electrodes close to the ions. The gradient causes the qubit level $\ket{\uparrow}$ to be displaced in energy relative to $\ket{\downarrow}$. This can be done using an optical lattice~\cite{wineland1998experimental} or microwave field gradient~\cite{srinivas2019}. Modulating this spin-dependent force near a motional mode frequency excites a coherent displacement in that motional mode that depends on the qubit state.

Controlled squeezing is done similarly to the controlled rotation. The key differences are: the detuning $\Delta_\text{sq}$ of the two lasers are equal in magnitude but with opposite sign, the phase between the two tones, and $\phi$ is now a free parameter. This gives rise to an effective time evolution operator
\begin{eqnarray}
    U^m_j(t) &=& \exp[i \theta_\text{sq}(t) \sigma^z_j \left( \hat{a}^{\dagger2}_m e^{i\phi }- \hat{a}^2_m e^{-i \phi} \right)]\,,\\
    \theta_\text{sq}(t) &=&\frac{2\eta_{j,m}^2\Omega_{j,m}^2 t }{\Delta_\text{sq}} \,.
\end{eqnarray}
This time evolution corresponds to the gate CS($z$) with $z= i\theta_\text{sq}(t) e^{i \phi} $. 

The beam splitter gate acts on a control qubit kept in the $\ket{\downarrow}$ state and two modes. A controlled beam splitter gate is instead applied directly to the computational qubit. If the coupling between two specific modes for a given ion is small, a SWAP gate can be applied to the control qubit and the desired computational qubit before and after performing the beam splitter operation.

Other hybrid operations can be synthesized from the gates described here. We leave a more detailed exploration for future work.

\subsection{Phonon number, homodyne and hybrid measurements}

\begin{figure}
    \centering
     \includegraphics[width=0.98\linewidth]{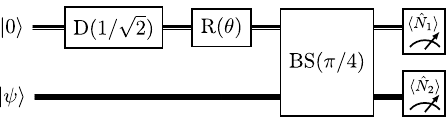}
    \caption{Circuit for a homodyne measurement of the quadrature operators $\hat X$\;(rotation $\theta=0$) and $\hat P$ (rotation $\theta=\pi/2$) using an ancillary qumode initialized in the $\ket{0}$ state.\label{fig:homodyne}}
\end{figure}

We can perform a homodyne measurement using the circuit diagram shown in Fig.~\ref{fig:homodyne}. Per homodyne measurement, we need to introduce an additional qumode. We then apply a displacement of the mode by $1/\sqrt{2}$ and rotate it by an angle of $\theta=0$ ($\theta=\pi/2$) for a position (momentum) measurement. Next, we interfere with both states using a 50-50 beam splitter and measure the average occupation number of both modes. For example, the homodyne measurement of $\hat X$ is then given by~\cite{Braunstein:2005zz}
\begin{equation}
    \langle\psi|\hat X|\psi\rangle = \langle\psi| \hat N_2 |\psi\rangle - \langle\psi| \hat N_1 |\psi\rangle \,.
\end{equation}

Trapped ion systems generally are not optimized to provide high-fidelity control over the motional degrees of freedom.
In order to perform a phonon number resolved (PNR) measurement, it is necessary to map the qumodes onto a qubit state.
Then the qubit's state $\ket{\uparrow,\downarrow}$ can be obtained via fluorescence measurements. We distinguish between two types of measurements, a projective single-shot measurement in the Fock basis and an average measurement of the phonon occupation number of a qumode.
In contrast, PNR measurements in photonic systems use detectors that are natively sensitive to photon number, such as transition edge sensors (TES) that employ a calorimetric effect~\cite{Cahall:17}, or by processing arrival times across a photon wave packet in superconducting nanowire single-photon detectors (SNSPDs)~\cite{Miller2003}. While these detectors work best for one to several photons, they can be extended to photon numbers $\sim$100 by multiplexing the signal across multiple detectors using beam splitters~\cite{Eaton:2022vjq}. 

Phonon state detection is commonly performed in trapped-ion systems by measuring the Rabi oscillation frequency of a blue sideband transition, which depends on the phonon number. The probability $P_n$ of being in the Fock state $\ket{n}$ can be inferred from the Fourier transform of the probability of the qubit to be in the $\ket{\downarrow}$ state, 
\begin{equation}
    P_\downarrow(t) = \frac{1}{2}\Big( 1+ \sum_{n=0}^\infty P_n \cos(2 \Omega_{n,n+1} t) \Big) \,,
\end{equation}
where $\Omega_{n,n+1} \approx \Omega\, \eta \,\sqrt{n+1}$ in the Lamb-Dicke regime \cite{wineland1998experimental}. The probability $P_\downarrow(t)$ is measured with resonant fluorescence. This method applies to individual ions and collective modes of the chain, but it is not a projective measurement, as it relies on using the average of many experimental shots. Additionally, this method has been efficiently generalized to simultaneously determine the Fock state distributions of multiple modes using fluorescence from multiple ions~\cite{Jia:2022qxf}.

There are two schemes for single-shot phonon number measurements demonstrated in trapped ion systems. Because these allow for high-fidelity estimates of the phonon number in a single experimental sequence, they are suitable for use in active feedback schemes or for mid-circuit measurements. The first method involves using red and blue sideband pulses to map the state of a single phonon mode onto one or more qubit states. An example of this is to initialize a single qubit in the $\ket{\uparrow}$ state and then perform a single adiabatic RSB pulse. The RSB pulse transfers all phonon population in the mode with $n\geq1$ to the $\ket{\downarrow}$ qubit state~\cite{Lv2017pnr}. Then, a single fluorescence measurement of the qubit will determine if the qubit and mode are in the $\ket{\uparrow}\ket{0}$ state (which is bright) or the $\ket{\downarrow}\ket{n\geq1}$ states (which are dark). If the mode is in $n
\geq1$ this is a quantum non-demolition measurement, and the precise value of $n$ can be done by performing a carrier $\pi$-pulse to rotate the qubit back to $\ket{\uparrow}$ and repeating the measurement sequence with another adiabatic RSB pulse until there is a bright fluorescence measurement.

The second scheme for single-shot measurements takes advantage of the cross-Kerr coupling between motional modes. This gives rise to a resolvable frequency shift on a ``readout'' mode that depends on the occupation of the ``target'' mode. Measuring the readout mode frequency is done by performing BSB pulses, with a successful qubit spin flip detected using resonant fluorescence~\cite{ding2017cross}. This measurement can be done simultaneously with a homodyne measurement, enabling the hybrid measurements displayed in table~\ref{tab:gates}, the measurement time of which is the longer of the two concurrent composite processes.

\begin{figure}[t!]
    \includegraphics[width=\linewidth]{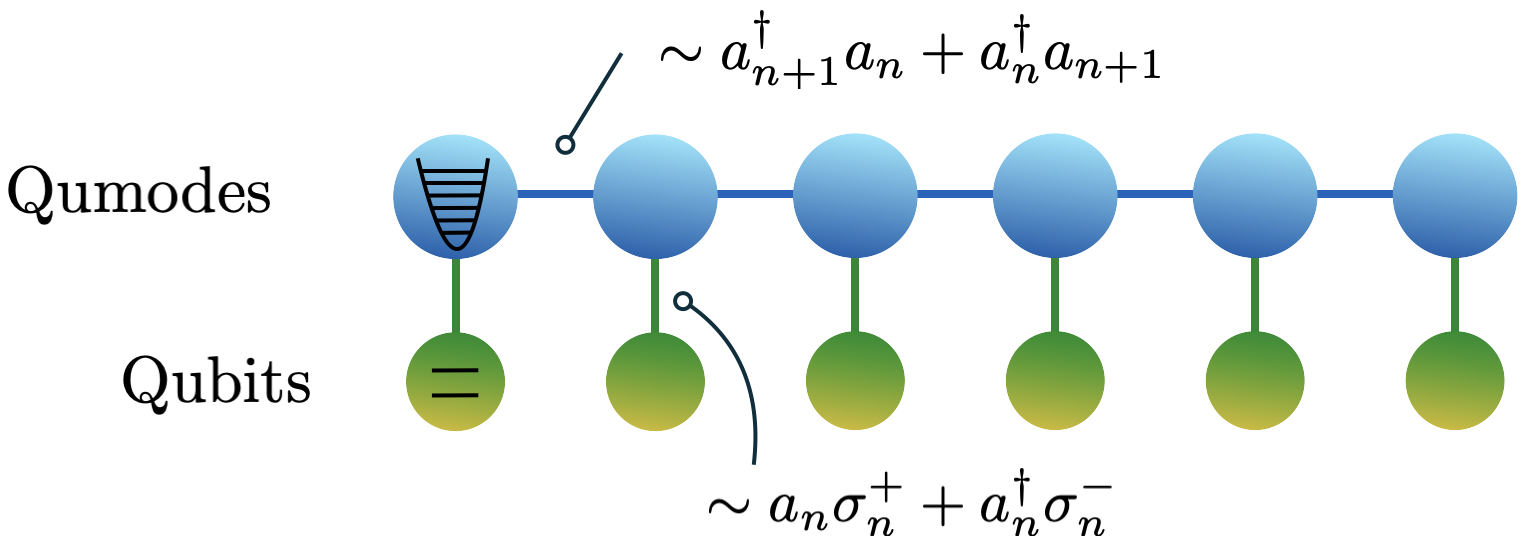}
    \caption{Illustration of the Jaynes-Cummings-Hubbard model for 6 lattice sites. Qumode (qubit) degrees of freedom are shown in blue (green).~\label{fig:JCH}}
\end{figure}

\section{Toward hybrid quantum simulations~\label{sec:jch}}

In this section, we explore an application of the hybrid quantum computing scheme using both qubits and qumodes. We start by introducing the Hamiltonian of the JCH model, which involves both spin and boson degrees of freedom. Using classical simulations, we study its real-time evolution, and we develop a variational quantum algorithm to prepare the ground state of the model where the total number of excitations is either variable or fixed.

\subsection{The Jaynes-Cummings-Hubbard model}\label{sec:lattice}

The Hamiltonian of the 1+1 dimensional JCH model with $M$ lattice sites is given by~\cite{Schmidt_2009}
\begin{eqnarray}\label{eq:JCH1}
    \hat{H} &= & \omega_c \sum_{n=1}^M \hat a_n^{\dagger}\hat a_n+\omega_a\sum_{n=1}^M\sigma_n^{+}\sigma_n^{-} \nonumber\\
    &-& \kappa \sum_{n=1}^M\left(\hat a_{n+1}^{\dagger} \hat a_n+\hat a_n^{\dagger} \hat a_{n+1}\right) \nonumber\\
    &+& \eta \sum_{n=1}^N\left(\hat a_n \sigma_n^{+}+\hat a_n^{\dagger} \sigma_n^{-}\right)\ .
\end{eqnarray}
Here, $\hat a_n^\dagger,\hat a_n$ denote the raising and lowering operators of the harmonic oscillator or qumode at lattice site $n$. Similarly, $\sigma_n^{\pm}$ denote the corresponding spin or qubit raising and lowering operators. The two terms in the first line of Eq.~(\ref{eq:JCH1}) account for the energies of the excitations of the qumodes and qubits that are distributed across the lattice. We define the difference between the respective energy levels or the detuning as 
\begin{eqnarray}\label{eq:detuning}
    \Delta=\omega_c-\omega_a\,.
\end{eqnarray}
The second line in Eq.~(\ref{eq:JCH1}) is a kinetic or nearest-neighbor hopping term between qumodes with coupling strength $\kappa$. The third line describes an onsite interaction between qubits and qumodes with strength $\eta$. See Fig.~\ref{fig:JCH} for a schematic illustration of the JCH model where qubits (qumodes) are represented by green (blue) circles and the nearest-neighbor and onsite interactions are indicated by blue and green lines, respectively. Throughout this work, we employ open boundary conditions. 

\begin{figure}
    \centering
    \includegraphics[width=.9\linewidth]{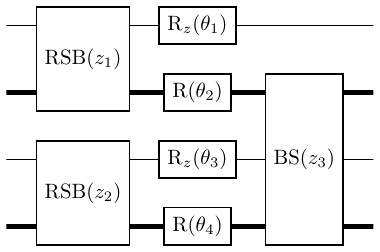}
    \caption{Illustration of a single Trotter step for the time evolution of the two-site JCH model, which also corresponds to a single layer of the variational algorithm employed here. Single (triple) lines represent qubit (qumode) wires. The beam splitter gate acts only on neighboring qumodes.}
    \label{fig:trotter-step}
\end{figure}

\begin{figure*}
    \centering
    \subfigure[$\kappa = 0.2,\ \eta=1 $ with initial state $|0100\rangle\otimes\ket{\uparrow\uparrow\uparrow\downarrow}$ ]
    {\includegraphics[width=.8\linewidth]{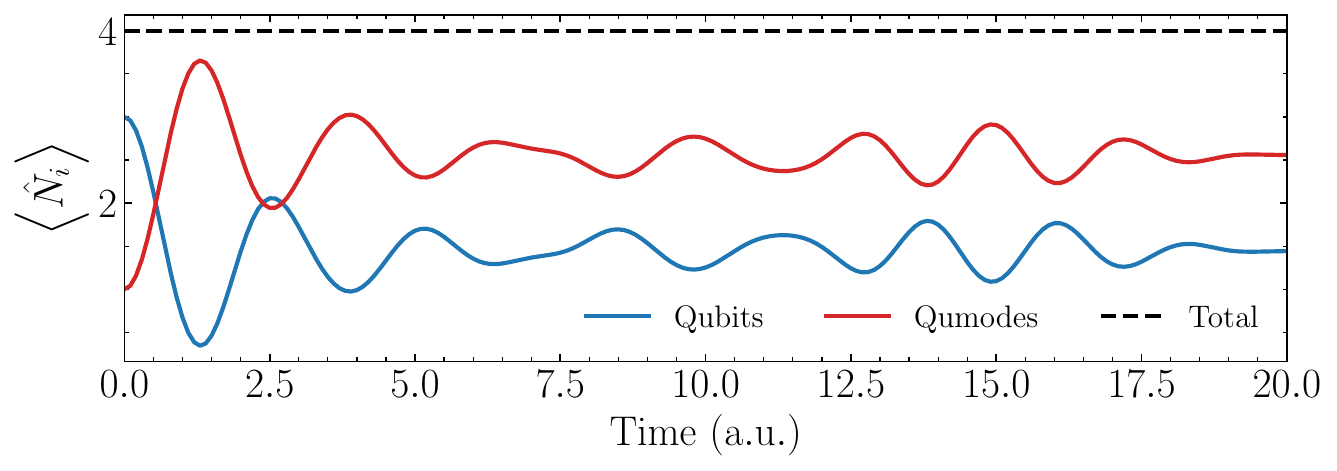}\label{fig:timeevo-numb}}
    \subfigure[$\kappa = 0.2,\ \eta=1$ with initial state $|0100\rangle\otimes\ket{\downarrow\downarrow\downarrow\uparrow}$]
    {\includegraphics[width=0.49\linewidth]{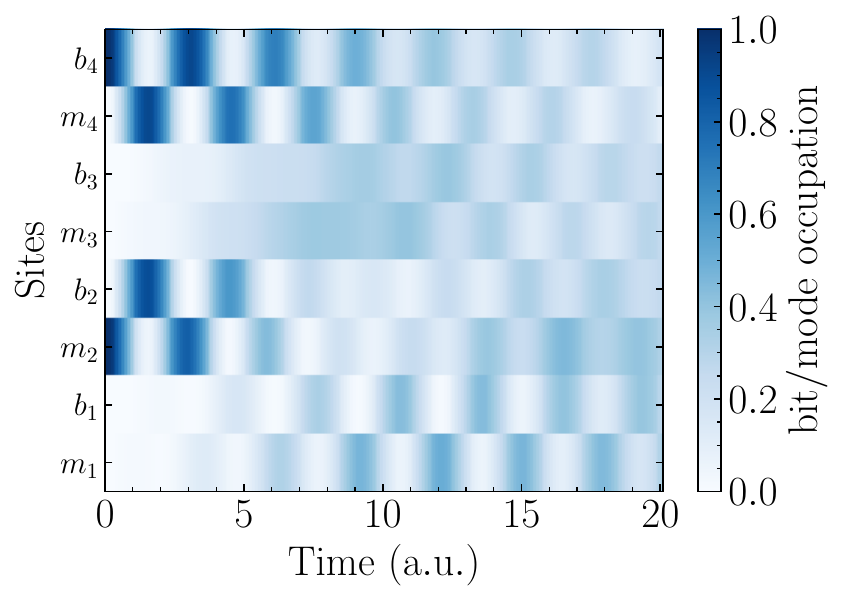}\label{fig:timeevo-k0.2}}
    \subfigure[$\kappa = 1,\ \eta=0.2$ with initial state $|0100\rangle\otimes\ket{\downarrow\downarrow\downarrow\uparrow}$]
    {\includegraphics[width=0.49\linewidth]{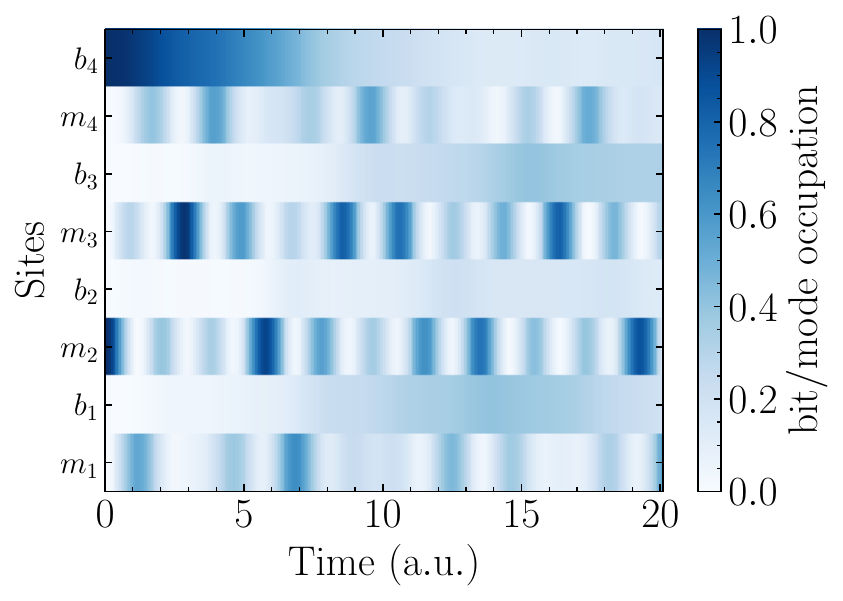}\label{fig:timeevo-k1}}
    \caption{Real-time evolution studies of the JCH model. The upper panel shows the evolution of the expectation value of the number operator for qubits (solid blue), qumodes (solid red), and the sum of both (dashed black) summed over all lattice sites. The lower two panels show the qubit and qumode occupation number for each lattice site as a function of time for two different sets of parameters. On the $y$-axis, we label the lattice site $i$ for each qubit and qumode as $b_i,m_i$, respectively.~\label{fig:time_evo}}
\end{figure*}

Each term of the Hamiltonian in Eq.~(\ref{eq:JCH1}) is written in terms of qubit and qumode raising and lowering operators. This form of the Hamiltonian is suitable for identifying the relevant gates, see table~\ref{tab:gates}, for the simulations discussed below. Instead, the measurement operations in table~\ref{tab:gates} are written in terms of Pauli, quadrature, and number operators. We therefore rewrite the Hamiltonian in Eq.~(\ref{eq:JCH1}) in terms of Pauli, quadrature, and number operators as follows
\begin{eqnarray}\label{eq:JCH2}
    \hat{H}&=&\,\omega_c\sum_{n=1}^M \hat{N}_n^m+\omega_a\sum_{n=1}^M \hat N_n^b \nonumber \\
        &-&\kappa \sum_{n=1}^M(\hat{X}_{n+1}\hat{X}_n + \hat{P}_{n+1}\hat{P}_{n}) \nonumber\\
        &+&\frac{\eta}{\sqrt{2}} \sum_{n=1}^M(\sigma^x_n \hat{X}_n - \sigma^y_n \hat{P}_n) \ . 
\end{eqnarray}
Here, $\hat X_n,\hat P_n$ represent the quadrature operators at lattice site $n$. We also introduced the number operators 
\begin{eqnarray}\label{eq:numberoperators}
    \hat{N}_{n}^m = \hat{a}^\dagger_n \hat{a}_n\quad , \quad \hat{N}_{n}^b = \sigma^+_n \sigma^-_n = \frac{1}{2}(\mathbb{I}_n + \sigma^z_n) \ ,
\end{eqnarray}
for the qumode and qubit register as indicated by the respective superscripts, and $\mathbb{I}_n,\sigma^{\{x,y,z\}}_n$ are the identity and Pauli operators at each lattice site, respectively. Using the form of the Hamiltonian in Eq.~(\ref{eq:JCH2}), we can identify the relevant qubit, qumode, and hybrid measurement operations listed in table~\ref{tab:gates}.

The number operator that measures the total number of excitations across the entire lattice is given by
\begin{eqnarray}\label{eq:Ntot}
    \hat N_{\rm tot}=\sum_{n=1}^M (\hat N_n^m+\hat N_n^b) \,.
\end{eqnarray}
The total number of excitations is conserved since the Hamiltonian commutes with the total number operator
\begin{equation}\label{eq:Nconserved}
    [\hat H,\hat N_{\rm tot}]=0\,.
\end{equation}
For the classical simulation described in the next section, the infinite-dimensional Hilbert space of the qumodes needs to be truncated. The Fock space truncation of the qumodes at level $\Lambda$ leads to the fact that the raising operator annihilates the state with the highest Fock space occupation number $\hat a^\dagger\ket{\Lambda}=0$. As a result, the commutation relation between the raising and lowering operators is modified as
\begin{equation}\label{eq:aadagger}
[\hat a,\hat a^\dagger] = \mathbb{I} - (\Lambda +1) \op{\Lambda}{\Lambda} .
\end{equation}
The commutation relation of the quadrature operators is modified analogously. However, the total number operator still commutes with the Hamiltonian, see Eq.~(\ref{eq:Nconserved}), despite the finite truncation of the Hilbert space. This ensures that the total number of excitations is a good quantum number in our numerical simulations. A numerical improvement could be achieved by normal ordering the raising and lower operators. That is, every term $\hat a\hat a^\dagger$ is replaced with $\hat a^\dagger\hat a+1$. In this case, the commutation relation between the quadrature operators is retained and the effect of the finite truncation of the qumode Hilbert space is reduced~\cite{Stavenger:2022wzz}. 

\subsection{Real-time evolution~\label{sec:timeevo}}

In this section, we are going to study real-time evolution of the JCH model. We employ a first-order Trotter decomposition~\cite{Boghosian:1996qd, Lloyd1996UniversalQS,Childs_2018} where the Hamiltonian in Eq.~(\ref{eq:JCH1}) is written as
\begin{eqnarray}
    \hat H=\sum_j \hat H_j\,,
\end{eqnarray}
where each term $\hat H_j$ can be written in terms of qubit, qumode or hybrid operators. Using a first-order Trotter decomposition, the unitary time evolution operator can be written as
\begin{eqnarray}\label{eq:U1}
    U_1(t)=\prod_j e^{-i t \hat H_j}\ .
\end{eqnarray}
By evolving in terms of $K$ small time steps with $\Delta t=t/K$, we can closely approximate the time evolution operator $e^{-it\hat H}$. We leave a more quantitative assessment of the error bound for the trotterized time evolution involving hybrid gates for future work~\cite{Childs_2018,Kalajdzievski_2018,Crane:2024tlj}. The different contributions to the trotterized time evolution operator in Eq.~(\ref{eq:U1}) can be expressed in terms of qumode rotation gates ${\rm R}(\theta)$, qubit rotation gates ${\rm R}_z(\theta)$, see Eq.~(\ref{eq:numberoperators}). In addition, we need the two-qumode beam splitter $\text{BS}(z)$, and the red sideband gate $\text{RSB}(z)$, which is a hybrid operation. All relevant gate operations can be found in table~\ref{tab:gates}. In Fig.~\ref{fig:trotter-step}, we show an illustration of the circuit for a single Trotter step for the two-site JCH model. The relevant parameters of the gates can be read off from Eq.~(\ref{eq:JCH1}). For example, in order to simulate a single Trotter step, a single term of the second line in Eq.~(\ref{eq:JCH1}), $\kappa (\hat a_{n+1}^{\dagger} \hat a_n+\hat a_n^{\dagger} \hat a_{n+1})$, is implemented by a beam splitter gate with parameter $z_3=\theta e^{i\phi}$, where $\theta=\kappa\Delta t$ and $\phi=\pi/2$. Similarly, the remaining terms in Eq.~(\ref{eq:JCH1}) are each implemented by the other gates depicted in Fig.~\ref{fig:trotter-step}. We note that for the model considered here, the different factors in the trotterized time evolution in Eq.~(\ref{eq:U1}) map directly to gates discussed in the previous section. However, more general hybrid gates involving Pauli strings and polynomials of the quadrature operators can be constructed from the gates listed in table~\ref{tab:gates}. See Ref.~\cite{Liu:2024mbr} for more details. These derived gates are relevant for more general models or gauge theories with fermions. 

For our numerical simulations, we use an extended version of \textsc{PennyLane} (version 0.35.1)~\cite{Bergholm:2018cyq}, where we include hybrid gates with an adjustable cutoff value $\Lambda$. We study the time evolution for a four-site JCH model. We choose the Fock space cutoff of each qumode as $\Lambda=4$. We note that for the JCH model considered in this work, the result for the real-time evolution is independent of the cutoff $\Lambda$ as long as it is sufficiently high to accommodate the initial total number of excitations in a single qumode. The reason is that the total number of excitations is preserved and no coherent or squeezed states are generated that would require an infinite sum over all Fock states $\ket{n}$ of a qumode. For the time evolution, we choose the time interval of each Trotter step as $\Delta t=0.1$, and we evolve for 200 steps. We verified that the Trotter errors are negligible for the time intervals shown in this section. We study the time evolution using different model parameters and initial states. Throughout this section, we choose the detuning between the qubit and qumode energy levels as $\Delta=0.5$, see Eq.~(\ref{eq:detuning}), with $\omega_c=1$.

\begin{figure}
    \centering
    \includegraphics[width=\linewidth]{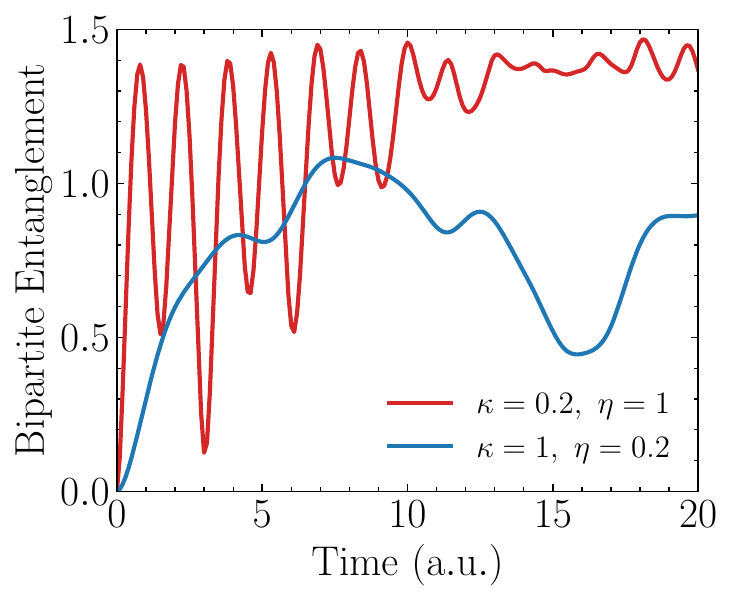}
    \caption{Time evolution of the bipartite entanglement between the qubit and qumode registers. We show the result for two different choices of the model parameters that correspond to Figs.~\ref{fig:timeevo-k0.2} and \ref{fig:timeevo-k1}.\label{fig:entanglement}}
\end{figure}

Fig.~\ref{fig:timeevo-numb} shows the expectation value of the number operators of the qumodes and qubits $\sum_i \hat{N}_i^{b,m}$, and the total $\hat{N}_{\rm tot}$, see Eq.~(\ref{eq:Ntot}), as a function of time for the model parameters $\kappa=0.2$ and $\eta=1$. As an initial state, we choose $|m\rangle\otimes|b\rangle=|0100\rangle\otimes\ket{\uparrow\uparrow\uparrow\downarrow}$, where the qubit and qumode registers are factorized. Here, $|m\rangle$, $|b\rangle$ denote the qumode and qubit register, respectively. As expected, the total number of excitations across the entire lattice is conserved, see Eq.~(\ref{eq:Nconserved}). However, we observe oscillations for the total number of excitations in the qubit and qumode sector. While the initial total number of excitations in the qubit subspace is larger, we observe an inversion at late times. Next, in Figs.~\ref{fig:timeevo-k1} and~\ref{fig:timeevo-k0.2}, we show the qubit and qumode occupation numbers for each lattice sites for two different sets of model parameters as indicated in the figures. For both simulations, we choose the initial state as $\ket{0100}\otimes\ket{\downarrow\downarrow\downarrow\uparrow}$. The $y$-axis of the two figures labels the lattice site $i$ for each qubit and qumode as $b_i,m_i$, respectively. In Fig.~\ref{fig:timeevo-k0.2}, due to the relatively large qubit-qumode coupling strength $\eta$, we observe initial on-site oscillations that decay only gradually with time and eventually spread across the entire lattice. Instead, in Fig.~\ref{fig:timeevo-k1}, we consider a relatively strong nearest-neighbor coupling between qumodes $\kappa$ compared to the qubit-qumode coupling. We observe that the initial qumode excitation at site $m_2$ moves across the lattice and backscatters at the endpoints. In comparison, the qubit excitation at lattice site $b_4$ only decays slowly.

To further illustrate the relation between the qubit and qumode registers of the JCH model, we consider the bipartite entanglement between the two and study its evolution as a function of time. The reduced density of the qubit register is obtained by tracing over the qumode degrees of freedom $\rho_b={\rm Tr}_m[\rho]$, where $\rho$ denotes the density matrix of the entire lattice. We approximate the von Neumann entropy of the reduced density matrix $\rho_b$ by the Shannon entropy as it is more directly accessible on quantum computers
\begin{eqnarray}\label{eq:entropy}
    S = -\tr[\rho_b\log\rho_b] \simeq -\sum_i p_i \log p_i\ .
\end{eqnarray}
The Shannon entropy on the right side of the equation can be constructed by measuring the qubits in the computational basis. The probabilities of each measurement outcome $i$ is denoted by $p_i$. Fig.~\ref{fig:entanglement} shows the time evolution of the bipartite entanglement between the qubit and qumode registers for two sets of model parameters, which correspond to Fig.~\ref{fig:timeevo-k0.2} and~\ref{fig:timeevo-k1}. At early times, the entanglement increases rapidly for large values of the qubit-qumode coupling strength $\eta$, which is shown in red. In addition, we observe faster oscillations and larger maximum values of the entanglement compared to the case when the qubit-qumode coupling strength is relatively small.

\subsection{Ground state preparation using variational algorithms~\label{sec:vqe}}

In this section, we explore the preparation of ground states using variational quantum algorithms (VQAs)~\cite{Peruzzo2014, Kandala:2017vok,Gard_2020,Tilly:2021jem, Bharti2022,Crane:2024tlj}. These algorithms generally employ a parametrized quantum circuit to approximate the ground state of a given Hamiltonian. The parameters are obtained using a classical optimization. Both gradient and non-gradient-based optimization techniques have been employed in the literature. On a quantum computer, gradients can be obtained using parameter shift rules~\cite{Mitarai:2018voy, Schuld:2018aiz}. For continuous variable quantum computing, suitable ansatze have been developed in Refs.~\cite{Killoran:2019yfa, Bangar:2023akc}. Here, a combination of Gaussian and non-Gaussian gates closely resembles classical neural networks where a linear or affine transformation is followed by a non-linearity. For example, compared to adiabatic state preparation algorithms~\cite{Farhi:2000ikn}, VQAs allow for significantly shorter circuits, making them well-suited for the near to intermediate-term future. However, the optimization can be challenging~\cite{Bittel:2021ley} and VQAs often exhibit so-called barren plateaus where the gradient and its variance vanish exponentially in terms of the number of qubits or layers~\cite{McClean_2018}. Mitigation strategies have been developed in the context of qubit-based algorithms~\cite{Grimsley:2018wnd, Larocca:2021ksf, Ragone:2023qbn, Diaz:2023uuo, Fontana:2023wnj}. Other state preparation algorithms have been proposed in the literature. See for example Refs.~\cite{10.1063/1.5027484,Gilyen:2019,Motta:2020,Dong:2022mmq,Stetcu:2022nhy,Rrapaj:2024shg}. We leave an exploration of these methods in the context of hybrid quantum computing using qubits and qumodes for future work and, instead, focus here on proof-of-concept studies. 

\begin{figure}[t]
    \centering
    \includegraphics[width=.9\linewidth]{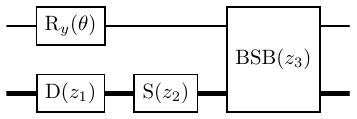}
    \caption{Additional parametrized gates that are included in our variational ansatz per lattice site and layer for a variable number of total excitations.}
    \label{fig:init}
\end{figure}

\begin{figure*}
    \centering
    \subfigure[Reconstructed ground state fidelity]{\includegraphics[width=0.4\linewidth]{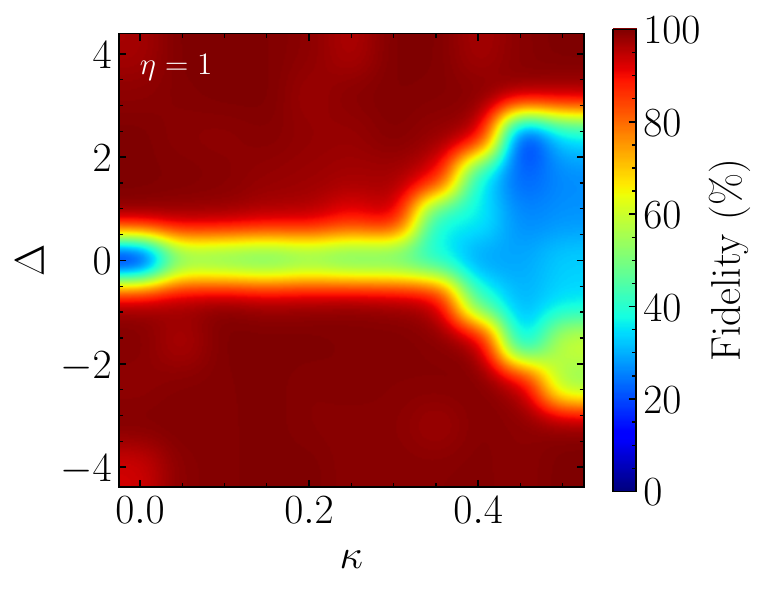}\label{fig:fid}}\quad\quad\quad\quad
    \subfigure[Energy gap of the JCH Hamiltonian using exact diagonalization]{\includegraphics[width=0.4\linewidth]{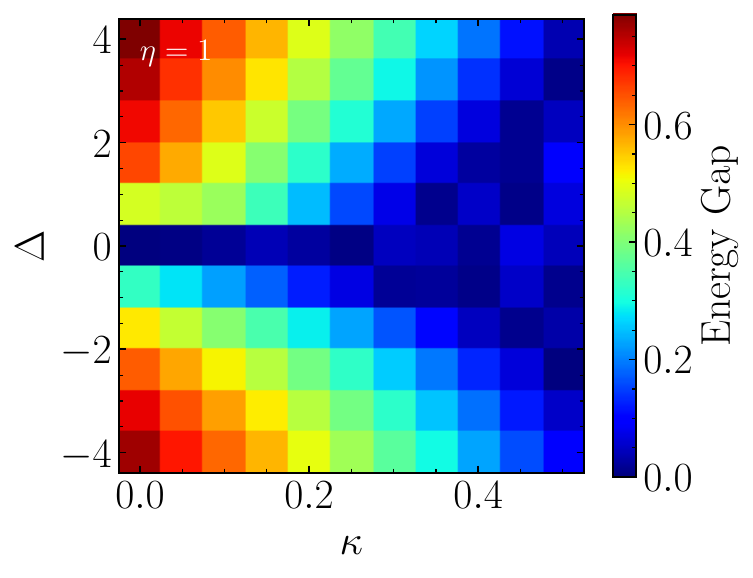}\label{fig:egap}}
    \subfigure[Total number of excitations of the ground state using exact diagonalization]{\includegraphics[width=0.4\linewidth]{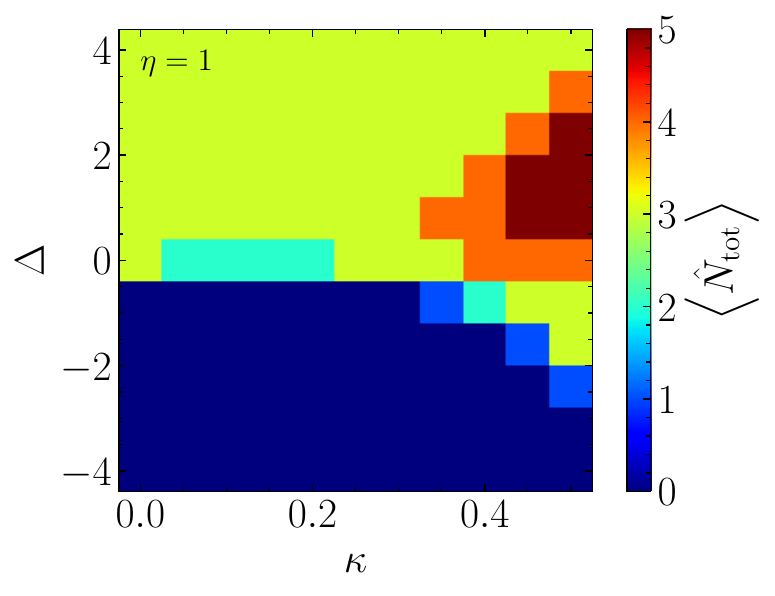}\label{fig:numb_ed}}\quad\quad\quad\quad
    \subfigure[Total number of excitations of the ground state using the VQA]{\includegraphics[width=0.4\linewidth]{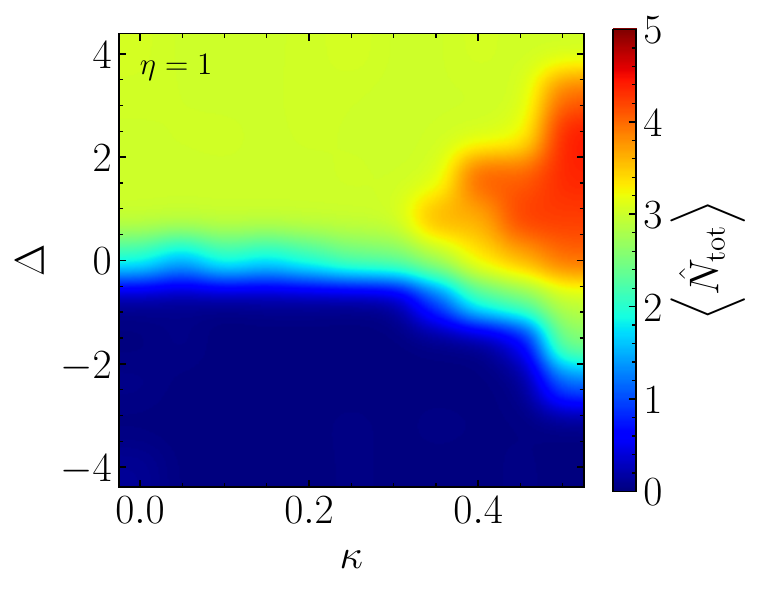}\label{fig:numb_vqa}}
    \caption{Ground state preparation of the JCH model using a qubit and qumode-based VQA. Each panel shows results for a grid of different model parameters allowing for a variable number of total excitations $N_{\rm tot}$. (a) Fidelity of the variationally prepared ground state. (b) Energy gap between the ground and first excited state of the JCH model using exact diagonalization. (c) Expectation value of the total number operator of the JCH ground state using exact diagonalization. (d) The reconstructed total number of excitations of the JCH ground state using the VQA. \label{fig:VQAresults}}
\end{figure*}

VQAs employ an ansatz consisting of parametrized circuits. We denote the corresponding unitary operator by $U(\Theta)$, where $\Theta$ denotes the set of variational parameters. The ground state of a given Hamiltonian is approximated by minimizing the expectation value
\begin{eqnarray}\label{eq:VQAoptimize}
\min_{\Theta} \langle\phi|U^\dagger(\Theta) \hat{H} U(\Theta) |\phi \rangle\ .
\end{eqnarray}
using a classical optimization loop. Here, $\ket{\phi}$ denotes a suitable initial state of the qubits and qumodes. A frequently employed method is to construct an ansatz $U(\Theta)$ using the gates that appear in the trotterized version of the time evolution operator $U_1(t)$, see Eq.~(\ref{eq:U1})~\cite{Gard_2020}. A single layer of the VQA ansatz consists of the gates of one Trotter step, see Fig.~\ref{fig:trotter-step}. Since the $\text{RSB}(z)$ gate is non-Gaussian, whereas all other gates in the Hamiltonian-based ansatz~\cite{Gard_2020} are Gaussian, this approach closely mimics the structure proposed for continuous-variable quantum neural networks in Ref.~\cite{Killoran:2019yfa}. The measurement operations needed to obtain the expectation value of the Hamiltonian in Eq.~(\ref{eq:VQAoptimize}) require the measurement of Pauli strings on the qubit side as well as homodyne measurements of the quadrature operators, and hybrid measurements. See the discussion around Eq.~(\ref{eq:JCH2}) and the list of measurement operations given in table~\ref{tab:gates}.

\begin{figure*}
    \centering
    \subfigure[Reconstructed ground state fidelity for $N_{\rm tot}=1$]{\includegraphics[width=0.4\linewidth]{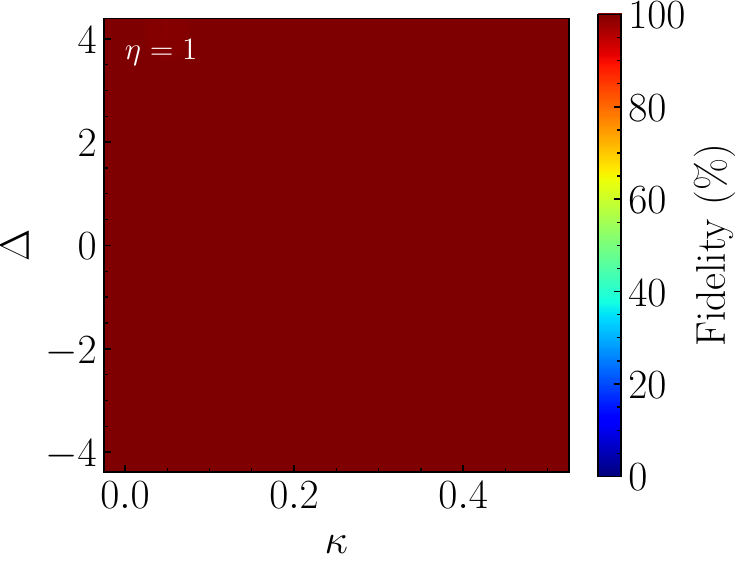}\label{fig:fidN1}}
    \quad\quad\quad\quad
    \subfigure[Energy gap for $N_{\rm tot}=1$ sector of the Hilbert space]{\includegraphics[width=0.38\linewidth]{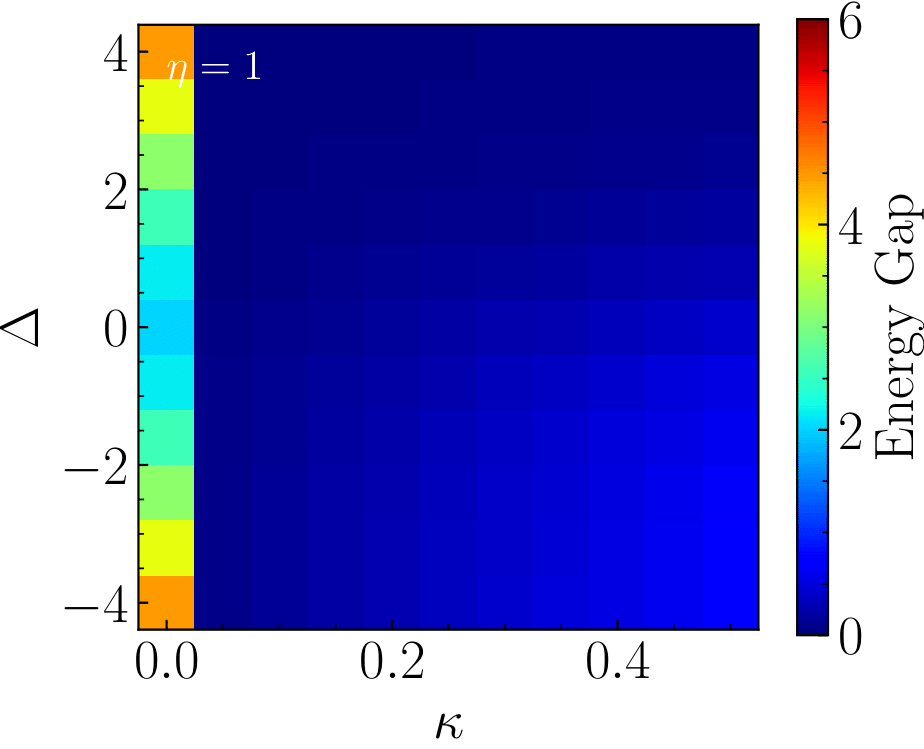}\label{fig:gapN1}}
    \subfigure[Reconstructed ground state fidelity for $N_{\rm tot}=3$]{\includegraphics[width=0.4\linewidth]{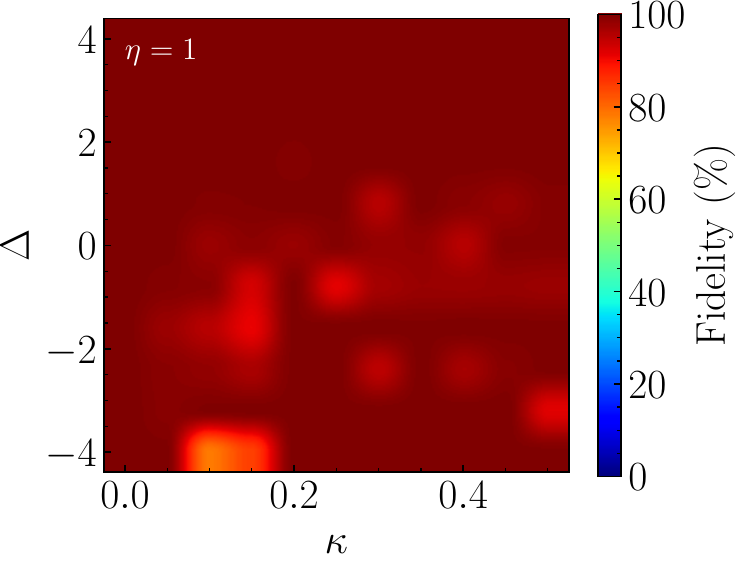}\label{fig:fidN3}}\quad\quad\quad\quad
    \subfigure[Energy gap for $N_{\rm tot}=3$ sector of the Hilbert space]{\includegraphics[width=0.4\linewidth]{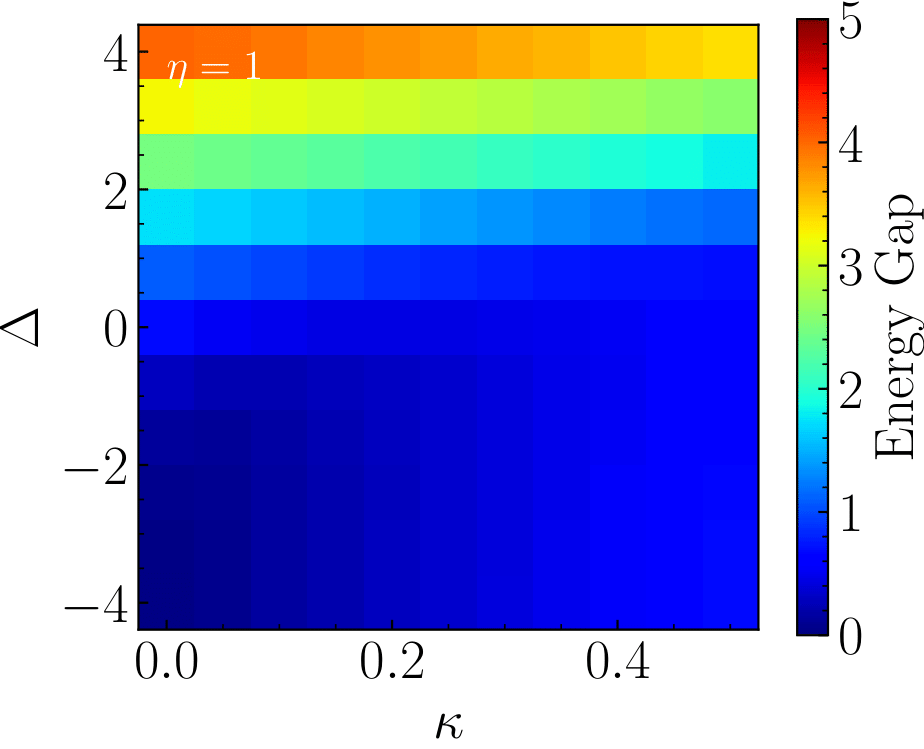}\label{fig:gapN3}}
    \caption{Ground state preparation of the JCH model using a qubit and qumode-based VQA. Different than the results in Fig.~\ref{fig:VQAresults}, we limit ourselves here to a fixed total number of excitations $N_{\rm tot}$ as indicated below each panel. (a) Fidelity of the variationally prepared ground state for $N_{\rm tot}=1$, (b) Energy gap between the ground and excited state of the JCH model for the $N_{\rm tot}=1$ subspace. (c), (d) Same as (a), (b) but for $N_{\rm tot}=3$.~\label{fig:VQAresultsN}}
\end{figure*}

The total number of excitations of the ground state of the JCH Hamiltonian depends on the choice of model parameters. In this section, we explore the ground state preparation with both a variable number of total excitations $N_{\rm tot}$ and when we restrict ourselves to the subspace where $N_{\rm tot}$ is fixed. For the latter case, the Hilbert space is significantly smaller and the gap between the ground state and the first excited state of this subspace is relatively large compared to the case with variable $N_{\rm tot}$. While the Hamiltonian-based variational ansatz described above is sufficient for fixed $N_{\rm tot}$ as it conserves the total number of excitations across the lattice, it is not sufficient to prepare the ground state of the JCH model for a variable total number of excitations. In this case, we extend the variational ansatz in Fig.~\ref{fig:trotter-step} by including additional parametrized gates that can change the total number of excitations. For every site of the JCH model, we include a single-qubit rotation ${\rm R}_y(\theta)$, a qumode displacement gate ${\rm D}(z)$, single-qumode squeezing ${\rm S}(z)$ and a non-Gaussian blue sideband gate $\text{BSB}(z)$, see table~\ref{tab:gates}. Each of these gates can change the total number of excitations. By including the corresponding parameters in the optimization procedure of the VQA, we can generate approximately the correct number of excitations of the ground state when $N_{\rm tot}$ is not fixed. Fig.~\ref{fig:init} shows a circuit diagram of the additional gates included in our VQA per lattice site and layer for the JCH model with variable $N_{\rm tot}$.

For the numerical simulations presented in this section, we employ again the extended \textsc{PennyLane} (version 0.35.1)~\cite{Bergholm:2018cyq} implementation with the \textsc{Jax} extension (version 0.4.23)~\cite{jax2018github} to evaluate the gradients. The optimization is performed using \textsc{SciPy}'s Sequential Least Squares Programming method (version 1.10.0)~\cite{2020SciPy-NMeth}. To explore the feasibility of our VQA, we consider the following ranges of the JCH model parameters: Detuning $\Delta \in [-4,4]$ with $\omega_c=1$, nearest-neighbor qumode coupling $\kappa\in [0,0.5]$, and qubit-qumode coupling strength $\eta=1$. See also Ref.~\cite{Cai_2021}.

We start by considering the ground state preparation with variable $N_{\rm tot}$. To assess how well the VQA approximates the ground state, we calculate the fidelity between the variationally prepared state and the result using exact diagonalization. For two states $\ket{\psi}$, $\ket{\chi}$ the fidelity is defined as the overlap $F(\ket{\psi},\ket{\chi})=\lVert\langle \psi| \chi\rangle \rVert^2$. For each set of model parameters, we perform the variational optimization and increase the number of layers up to at most 10 until we reach a fidelity of $F\gtrsim 90\%$. Note that here we include the variational circuits in Figs.~\ref{fig:trotter-step} and~\ref{fig:init}. Each qubit and qumode is initialized in the $\ket{0}$ state. We present our numerical results in Fig.~\ref{fig:VQAresults}. Each panel shows a two-dimensional histogram for the range of model parameters listed above. The results for the fidelity are shown in Fig.~\ref{fig:fid}. Over a considerable range of the parameter space, we achieve good results with $F\gtrsim 95\%$ except for a band centered around $\Delta\approx 0$. The band where the fidelity of the ground state reconstruction is low, widens almost linearly for $\kappa\gtrsim 0.35$. In order to better understand this behavior, we plot the energy gap of the JCH Hamiltonian in Fig.~\ref{fig:egap} for the same model parameters using exact diagonalization. We observe a correlation between the parameter regions where the gap is small and where the achieved fidelity of the VQA is low. The performance of VQAs is known to degrade for Hamiltonians with a small value of the energy gap~\cite{Peruzzo2014}. Similar considerations apply for example to adiabatic state preparation algorithms~\cite{Farhi:2000ikn}.

We now consider the expectation value of the total number operator of the variationally prepared ground state. The results using exact diagonalization are shown in Fig.~\ref{fig:numb_ed}. We observe that the total number of excitations for the chosen model parameters is $\expval*{\hat N_{\rm tot}}\leq 5$. As discussed above, since the number of excitations of the ground state is not known apriori, it needs to be reconstructed by the variational algorithm. We note that obtaining the correct number of excitations is not directly part of the objective function of the VQA. Instead, the variational parameters are obtained by minimizing the expectation value of the Hamiltonian, see Eq.~(\ref{eq:VQAoptimize}). The results of the VQA for $\expval*{\hat N_{\rm tot}}$ are shown in Fig.~\ref{fig:numb_vqa}. Compared to the result using exact diagonalization, we observe that the VQA can reconstruct this observable for the chosen model parameters even in regions with low fidelity.

Next, we consider the ground state preparation for fixed values of the total number of excitations $N_{\rm tot}$. Here, we only employ the Hamiltonian-based ansatz~\cite{Gard_2020} as illustrated in Fig.~\ref{fig:trotter-step}. For various applications in fundamental particle and nuclear physics this is the relevant case since it allows us to isolate sectors of the Hilbert space with fixed quantum numbers. By analyzing the Hamiltonian in Eq.~(\ref{eq:JCH2}), we observe that the spectrum can be degenerate for the parameter range we consider. Generally, the level of degeneracy depends on the choice of parameters. As an example, we consider the case of $\kappa=0$ where the model decouples into $M$ copies of the single-site JCH model. For the $N_{\rm tot}=1$ subspace, the lowest energy is obtained by the configuration where a single excitation is at any of the $M$ decoupled lattice sites.  As a result, the Hilbert space is at least $M$-fold degenerate. In contrast, the $N_{\rm tot}=3$ sector does not have this particular degeneracy. The variational algorithm cannot differentiate between the states in a degenerate subspace. Since we cannot compare to a single state when quantifying the fidelity of the variational state preparation, we extend our definition of fidelity to quantify the overlap of a variationally prepared state with a set of degenerate states that span the subspace of the Hilbert space associated with the ground state energy. The fidelity introduced above can be written as the expectation value of the projector onto the state we are comparing to, $F(\ket{\phi}, \ket{\psi}) = \bra{\psi}(\ket{\phi}\bra{\phi})\ket{\psi}$. Here $\ket{\phi}$ and $\ket{\psi}$ are the exact eigenstate and variational state, respectively. The natural generalization to a degenerate subspace is to include more states in the projection operator. For a given set of degenerate states, $\{\ket{\phi_i}\}$, we define the fidelity as
\begin{align}
F(\{\ket{\phi_i}\}, \ket{\psi}) & = \bra{\psi} \Big(\sum_{i} \ket{\phi_i} \bra{\phi_i} \Big) \ket{\psi} \nonumber \\
& = \sum_i F(\ket{\phi_i}, \ket{\psi})\,,
\end{align}
i.e. it is the sum of the fidelities between, or the modulus square of the overlap of the variationally prepared state with all the states in the degenerate subspace. 

\begin{figure*}
    \centering
    \subfigure[][Average Negativity\label{fig:NegativityN3_a}]{\includegraphics[width=0.3\textwidth]{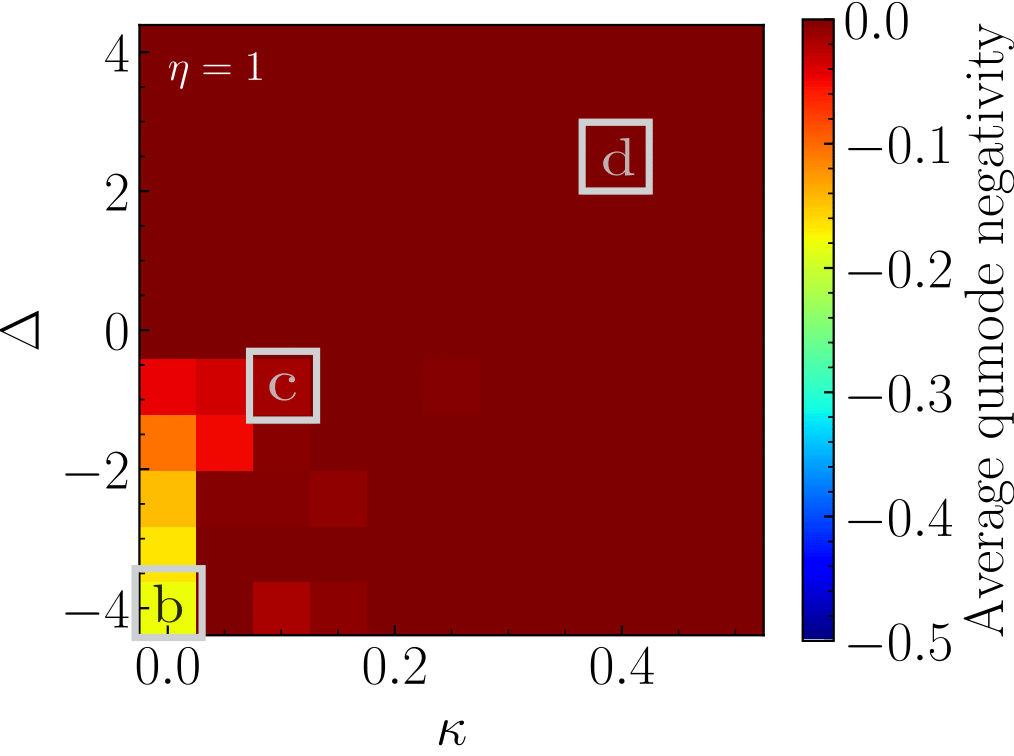}}
    \subfigure[][$\kappa=0, \Delta=-4.0$\label{fig:NegativityN3_b}]{\includegraphics[width=0.22\textwidth]{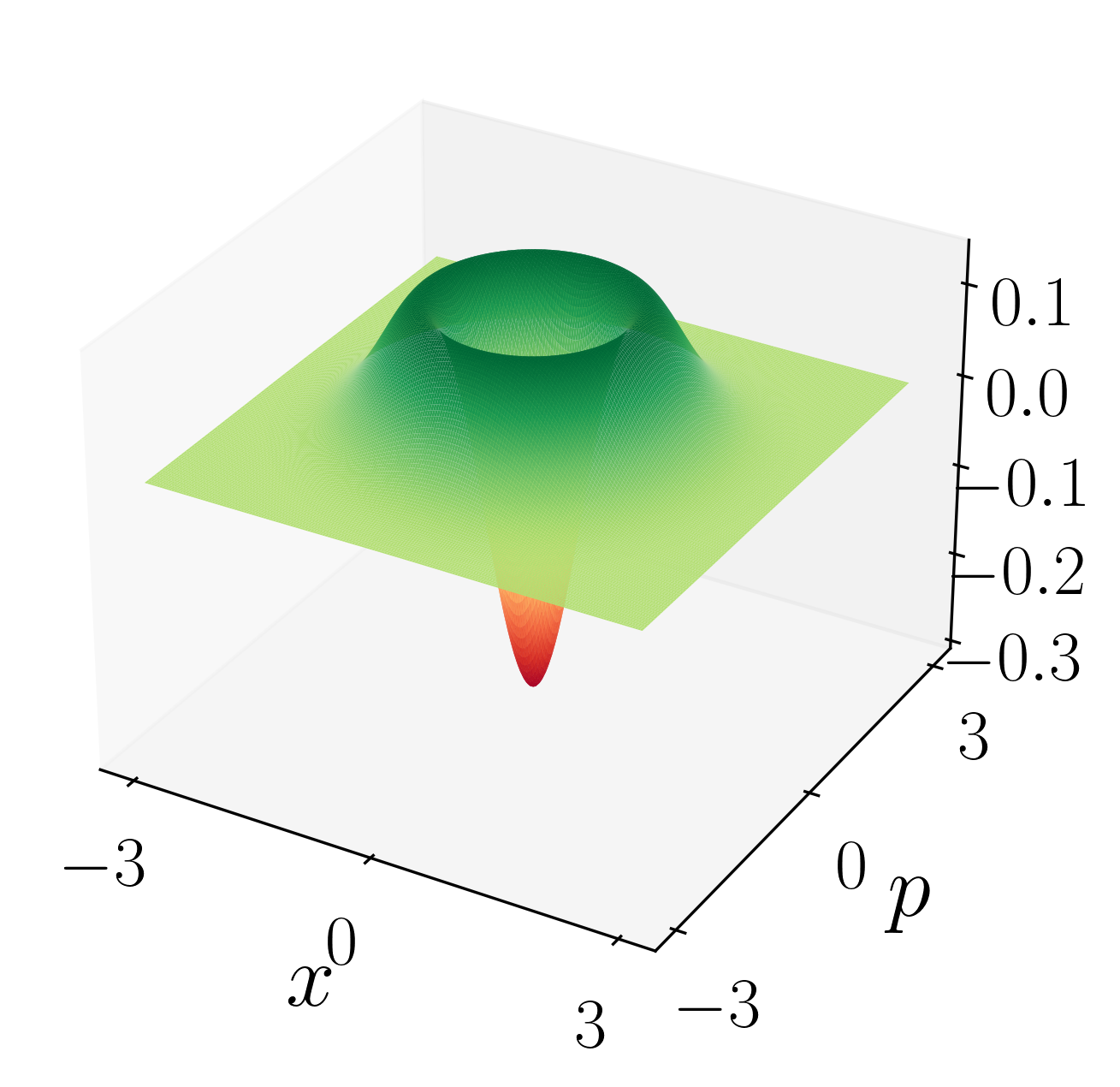}}
    \subfigure[][$\kappa=0.1, \Delta=-0.8$\label{fig:NegativityN3_c}]{\includegraphics[width=0.22\textwidth]{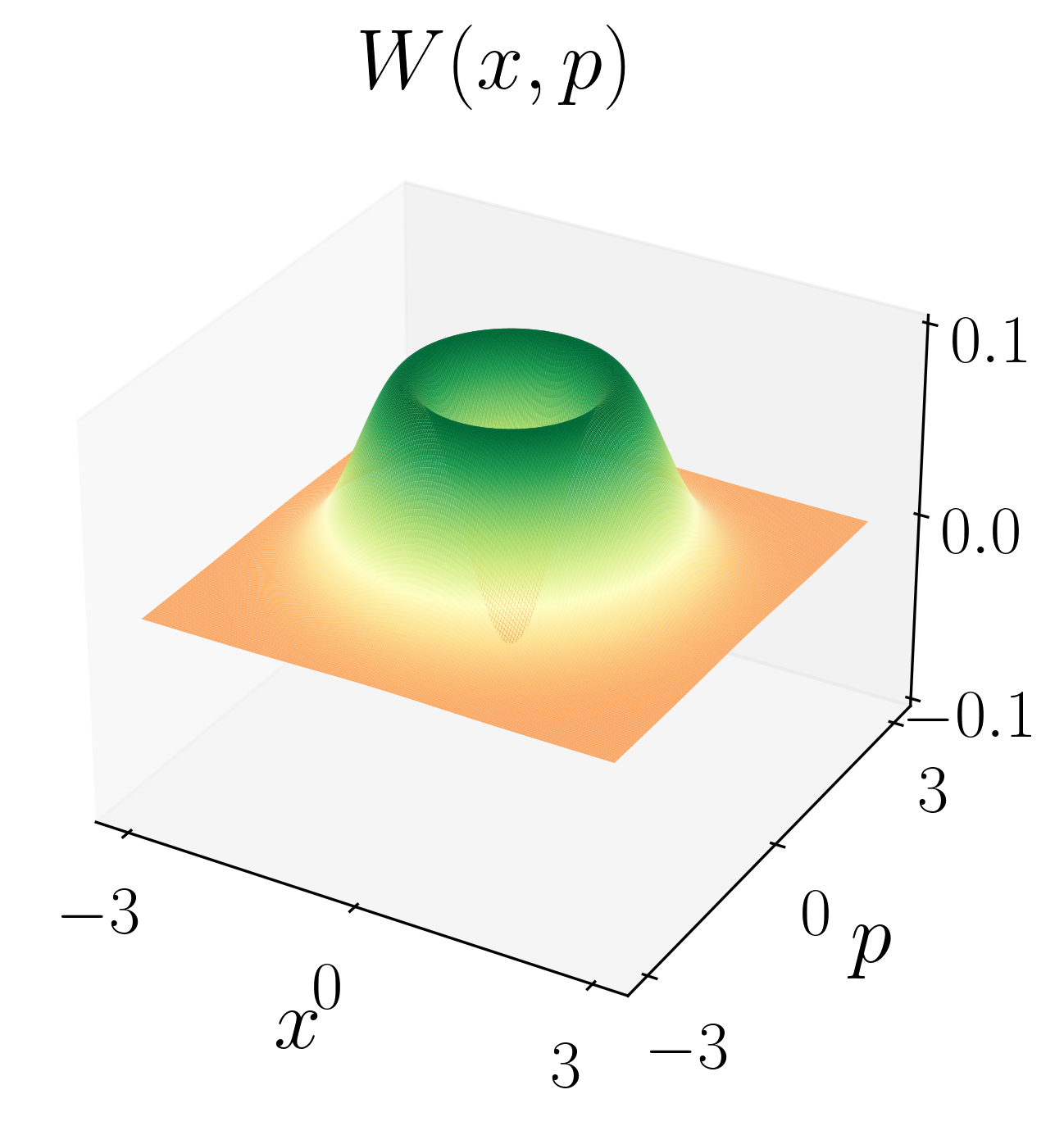}}
    \subfigure[][$\kappa=0.4, \Delta=2.4$\label{fig:NegativityN3_d}]{\includegraphics[width=0.22\textwidth]{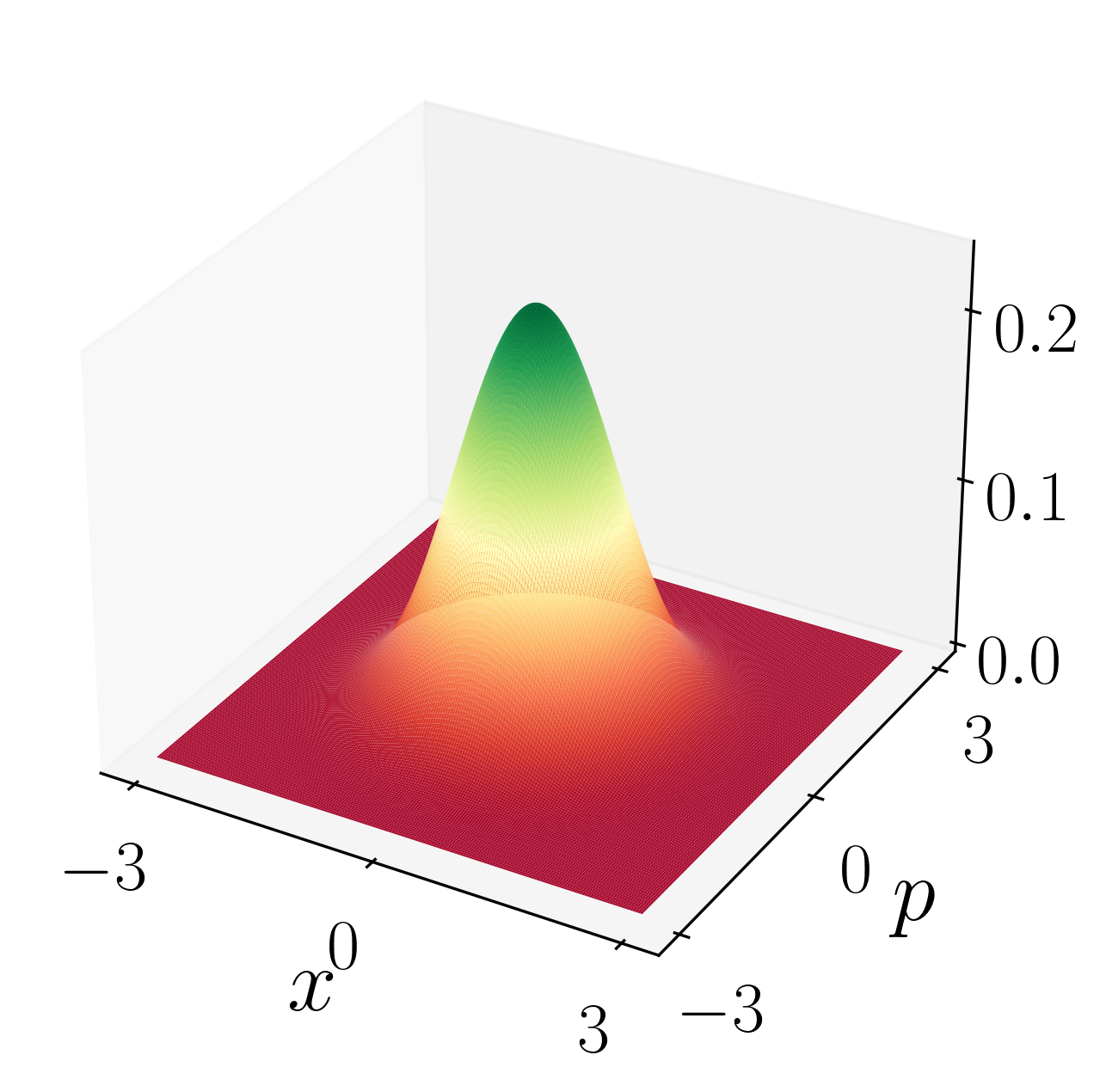}}
    \caption{a) Average negativity of the qumodes for the 3-site JCH model ground state with $N_{\rm tot}=3$ for different values of the model parameters. b)-d) The Wigner functions for the central mode for three representative points of the model parameter space as indicated in a).}
    \label{fig:NegativityN3}
\end{figure*}

We consider the ground state preparation for a total number of excitations of $N_{\rm tot}=1$ and 3. For the initial state of the variational algorithm, we choose $\ket{\phi}=\ket{000}\otimes\ket{\uparrow\downarrow\downarrow}$ and $\ket{000}\otimes\ket{\uparrow\uparrow\uparrow}$ for $N_{\rm tot}=1$ and 3, respectively. These states can be obtained by applying $\sigma^x$ gates to the qubits. We show our numerical results in Fig.~\ref{fig:VQAresultsN} for the same model parameters considered in Fig. ~\ref{fig:VQAresults}. The obtained fidelities for the two different values of $N_{\rm tot}$ are shown in Figs.~\ref{fig:fidN1} and~\ref{fig:fidN3}. Overall, the VQA achieves very high values for the fidelities over the entire parameter range. In Figs.~\ref{fig:gapN1} and~\ref{fig:gapN3}, we show the energy gap for the two values of fixed $N_{\rm tot}$. Specifically, we plot the energy difference between the lowest two, possibly degenerate, energy eigenvalues in the fixed particle number sector of the Hilbert space. In general, the gap is significantly larger compared to the case with a variable total number of excitations, see Fig.~\ref{fig:egap}. While the gap for $N_{\rm tot}=1$ is smaller compared to the case with $N_{\rm tot}=3$, the Hilbert space is sufficiently small such that the achieved fidelity is close to 100\% for the entire parameter. The smallest gap observed in the considered region of parameters is $\sim3.7\cdot 10^{-3}$ for $N_{\rm tot}=1$ and $\sim2.2 \cdot 10^{-2}$ for $N_{\rm tot} =3$. While the VQA also works well for $N_{\rm tot}=3$, we do observe that the fidelity slightly deteriorates in the parameter regions where the energy gap is small. This can likely be improved by including more layers in the ansatz.

\subsection{Qumode Wigner functions} 

The state of qumodes is often characterized in terms of their Wigner functions. In this section, we investigate the structure of the Wigner functions of the qumode ground states of the JCH model. The Wigner function is given in terms of the following integral over the position space wave function, see Eq.~(\ref{eq:position}) above,
\begin{eqnarray}
    W(x, p)=\frac{1}{\pi} \int \mathrm{d} y\,  \psi^*(x+y) \psi(x-y) e^{2 i p y}  \,.
\end{eqnarray}
The Wigner function is a quasi-probability distribution since it is normalized to unity upon integration over $x,p$. However, it can be negative, which distinguishes it from classical probability distributions. The Wigner function was introduced as a quantum analog to classical phase space distributions. In the context of fundamental physics and hadron structure see for example Refs.~\cite{Belitsky:2003nz,Hagiwara:2016kam}. Generally, the more negative the Wigner function is, the ``more quantum'' a state is. The negativity $\delta$ is defined as the integral over the volume of the negative part of the Wigner function~\cite{negativity1}
\begin{eqnarray}
\delta = \frac{1}{2}\int {\rm d}x{\rm d}p \left(W(x,p) - | W(x,p)| \right) \,,
\end{eqnarray}
which represents a measure of the non-classicality or ``quantumness'' of a given state. Typically, states that exhibit a large negativity are relatively difficult to prepare. In Fig.~\ref{fig:NegativityN3}, we show the average negativity of the qumodes corresponding to the lowest energy state of the JCH model with $N_{\rm tot}=3$ for the same range of model parameters as in Fig.~\ref{fig:VQAresultsN}. For a wide range of the parameter space, the Wigner function is Gaussian resulting in a vanishing negativity $\delta=0$. Only in the lower left region of the plot, $\Delta \lesssim 0$, $\kappa \lesssim 0.2$, do the Wigner functions exhibit a nonzero negativity. The Wigner functions of the central qumode of the lattice system are shown in Figs. \ref{fig:NegativityN3_b}-\ref{fig:NegativityN3_d} for three sets of the model parameters: $(\kappa, 
\Delta) = (0, -4)$, $(0.1, -0.8)$, and $(0.4, 2.4)$ as indicated in Fig.~\ref{fig:NegativityN3_a}. The rightmost panel shows a Wigner function that closely resembles a 2d Gaussian surface and it is positive definite. The Wigner functions shown in the middle and leftmost panels display a nontrivial behavior where their surfaces dip to negative values at the center. This dip is more pronounced in the leftmost panel indicating that it corresponds to the ``most quantum'' state of the three exemplary Wigner functions of the JCH ground states shown here. The negativity is introduced by non-Gaussian gates, which is the RSB gate in our case.

\section{Conclusions and outlook~\label{sec:conclusion}}

In this work, we explored the feasibility of realizing a hybrid quantum computing approach using both qubits and qumodes with trapped-ion platforms. Qubits can be realized using for example the hyperfine states of the trapped-ions allowing for high gate fidelities and long coherence times. While trapped-ion-based qumodes have remained relatively unexplored, they can be realized using the collective vibrational modes of the ion chain. Given their relatively short coherence time, qumode operations currently represent the main bottleneck for the hybrid qubit and qumode quantum computing paradigm explored in this work. Using representative device parameters of existing trapped-ion platforms, we performed simulations to estimate the times and fidelities of a representative set of hybrid gate operations. In addition, we explored different measurement protocols for continuous variable and hybrid operators. Our results indicate that it is already possible to carry our small-scale simulations using existing trapped-ion platforms such as the QSCOUT testbed~\cite{QSCOUT}, motivating further developments in this direction. In addition, we expect that it is possible to further extend the current list of hybrid gates considered in this work.

The use of hybrid quantum computing involving both discrete and continuous variables may extend the reach of quantum simulations relevant to, for example, condensed matter and fundamental particle and nuclear physics. As a first step in this direction, we considered quantum simulations of the Jaynes-Cummings-Hubbard (JCH) model, which is a representative example of a lattice model that can be naturally mapped to qubits and qumodes. It consists of a one-dimensional lattice involving interacting spin and boson degrees of freedom. We explored the trotterized time evolution of this system using the hybrid gate set described above and studied for example the entanglement between the spin and boson degrees of freedom. In addition, we developed suitable variational algorithms that mimic the structure of neural networks to prepare the ground state of the JCH model. We considered both the case where the total number of excitations $N_{\rm tot}$ is variable and fixed, i.e. limiting the state preparation to a certain subspace of the total Hilbert space. The JCH model shares similarities with quantum field theories involving fermion, scalar, and gauge field degrees of freedom making it a natural starting point for further explorations of hybrid quantum simulations in the context of fundamental physics, which will be addressed in future work.

\section*{Data availability}

All data relevant to this work can be obtained from the authors upon reasonable request.

\section*{Acknowledgements}

We would like to thank Robert Edwards, Raghav Jha, Or Katz, Dima Kharzeev, Jianwei Qiu, Tommaso Rainaldi, Enrique Rico, and Feng Yuan for helpful discussions. JYA and FR were supported by the U.S. Department of Energy, Office of Science, Contract No.~DE-AC05-06OR23177, under which Jefferson Science Associates, LLC operates Jefferson Lab. JYA, JM, and FR were supported in part by the DOE, Office of Science, Office of Nuclear Physics, Early Career Program under contract No. DE-SC0024358/DE-SC0025881.

\bibliographystyle{utphys}
\bibliography{bibliography}

\providecommand{\href}[2]{#2}\begingroup\raggedright\begin{thebibliography}{100}

\bibitem{bullock2005}
S.~S. Bullock, D.~P. O'Leary, and G.~K. Brennen, ``Asymptotically optimal quantum circuits for $d$-level systems,'' \href{http://dx.doi.org/10.1103/PhysRevLett.94.230502}{{\em Phys. Rev. Lett.} {\bfseries 94} (Jun, 2005) 230502}. \url{https://link.aps.org/doi/10.1103/PhysRevLett.94.230502}.

\bibitem{rico2018}
E.~Rico, M.~Dalmonte, P.~Zoller, D.~Banerjee, M.~B{\"o}gli, P.~Stebler, and U.-J. Wiese, ``So(3) ``nuclear physics'' with ultracold gases,'' \href{http://dx.doi.org/https://doi.org/10.1016/j.aop.2018.03.020}{{\em Annals of Physics} {\bfseries 393} (2018) 466--483}. \url{https://www.sciencedirect.com/science/article/pii/S0003491618300757}.

\bibitem{Gottesman:2000di}
D.~Gottesman, A.~Kitaev, and J.~Preskill, ``{Encoding a qubit in an oscillator},'' \href{http://dx.doi.org/10.1103/PhysRevA.64.012310}{{\em Phys. Rev. A} {\bfseries 64} (2001) 012310}, \href{http://arxiv.org/abs/quant-ph/0008040}{{\ttfamily arXiv:quant-ph/0008040}}.

\bibitem{Lloyd:1998jk}
S.~Lloyd and S.~L. Braunstein, ``{Quantum computation over continuous variables},'' \href{http://dx.doi.org/10.1103/PhysRevLett.82.1784}{{\em Phys. Rev. Lett.} {\bfseries 82} (1999) 1784--1787}, \href{http://arxiv.org/abs/quant-ph/9810082}{{\ttfamily arXiv:quant-ph/9810082}}.

\bibitem{bartlett2002}
S.~D. Bartlett and B.~C. Sanders, ``Universal continuous-variable quantum computation: Requirement of optical nonlinearity for photon counting,'' \href{http://dx.doi.org/10.1103/PhysRevA.65.042304}{{\em Phys. Rev. A} {\bfseries 65} (Mar, 2002) 042304}. \url{https://link.aps.org/doi/10.1103/PhysRevA.65.042304}.

\bibitem{menicucci2008}
N.~C. Menicucci, S.~T. Flammia, and O.~Pfister, ``One-way quantum computing in the optical frequency comb,'' \href{http://dx.doi.org/10.1103/PhysRevLett.101.130501}{{\em Phys. Rev. Lett.} {\bfseries 101} (Sep, 2008) 130501}. \url{https://link.aps.org/doi/10.1103/PhysRevLett.101.130501}.

\bibitem{tasca2011}
D.~S. Tasca, R.~M. Gomes, F.~Toscano, P.~H. Souto~Ribeiro, and S.~P. Walborn, ``Continuous-variable quantum computation with spatial degrees of freedom of photons,'' \href{http://dx.doi.org/10.1103/PhysRevA.83.052325}{{\em Phys. Rev. A} {\bfseries 83} (May, 2011) 052325}. \url{https://link.aps.org/doi/10.1103/PhysRevA.83.052325}.

\bibitem{heeres2017}
R.~W. Heeres, P.~Reinhold, N.~Ofek, L.~Frunzio, L.~Jiang, M.~H. Devoret, and R.~J. Schoelkopf, ``Implementing a universal gate set on a logical qubit encoded in an oscillator,'' \href{http://dx.doi.org/10.1038/s41467-017-00045-1}{{\em Nature Communications} {\bfseries 8} no.~1, (2017) 94}. \url{https://doi.org/10.1038/s41467-017-00045-1}.

\bibitem{Stavenger:2022wzz}
T.~J. Stavenger, E.~Crane, K.~C. Smith, C.~T. Kang, S.~M. Girvin, and N.~Wiebe, \href{http://dx.doi.org/10.1109/HPEC55821.2022.9926318}{``{C2QA - Bosonic Qiskit},''} in {\em {26th IEEE High Performance Extreme Computing}}.
\newblock 9, 2022.
\newblock \href{http://arxiv.org/abs/2209.11153}{{\ttfamily arXiv:2209.11153 [quant-ph]}}.

\bibitem{Wang:2020ghj}
C.~S. Wang {\em et~al.}, ``{Efficient Multiphoton Sampling of Molecular Vibronic Spectra on a Superconducting Bosonic Processor},'' \href{http://dx.doi.org/10.1103/PhysRevX.10.021060}{{\em Phys. Rev. X} {\bfseries 10} no.~2, (2020) 021060}.

\bibitem{lau2012}
H.-K. Lau and D.~F.~V. James, ``Proposal for a scalable universal bosonic simulator using individually trapped ions,'' \href{http://dx.doi.org/10.1103/PhysRevA.85.062329}{{\em Phys. Rev. A} {\bfseries 85} (Jun, 2012) 062329}. \url{https://link.aps.org/doi/10.1103/PhysRevA.85.062329}.

\bibitem{fluhmann2019}
C.~Fl{\"u}hmann, T.~L. Nguyen, M.~Marinelli, V.~Negnevitsky, K.~Mehta, and J.~P. Home, ``Encoding a qubit in a trapped-ion mechanical oscillator,'' \href{http://dx.doi.org/10.1038/s41586-019-0960-6}{{\em Nature} {\bfseries 566} no.~7745, (2019) 513--517}. \url{https://doi.org/10.1038/s41586-019-0960-6}.

\bibitem{sutherland2021a}
R.~T. Sutherland and R.~Srinivas, ``Universal hybrid quantum computing in trapped ions,'' \href{http://dx.doi.org/10.1103/PhysRevA.104.032609}{{\em Phys. Rev. A} {\bfseries 104} (Sep, 2021) 032609}. \url{https://link.aps.org/doi/10.1103/PhysRevA.104.032609}.

\bibitem{neeve2022}
B.~de~Neeve, T.-L. Nguyen, T.~Behrle, and J.~P. Home, ``Error correction of a logical grid state qubit by dissipative pumping,'' \href{http://dx.doi.org/10.1038/s41567-021-01487-7}{{\em Nature Physics} {\bfseries 18} no.~3, (2022) 296--300}. \url{https://doi.org/10.1038/s41567-021-01487-7}.

\bibitem{matsos2024}
V.~G. Matsos, C.~H. Valahu, T.~Navickas, A.~D. Rao, M.~J. Millican, X.~C. Kolesnikow, M.~J. Biercuk, and T.~R. Tan, ``Robust and deterministic preparation of bosonic logical states in a trapped ion,'' \href{http://dx.doi.org/10.1103/PhysRevLett.133.050602}{{\em Phys. Rev. Lett.} {\bfseries 133} (Jul, 2024) 050602}. \url{https://link.aps.org/doi/10.1103/PhysRevLett.133.050602}.

\bibitem{bouchoule1999}
I.~Bouchoule, H.~Perrin, A.~Kuhn, M.~Morinaga, and C.~Salomon, ``Neutral atoms prepared in fock states of a one-dimensional harmonic potential,'' \href{http://dx.doi.org/10.1103/PhysRevA.59.R8}{{\em Phys. Rev. A} {\bfseries 59} (Jan, 1999) R8--R11}. \url{https://link.aps.org/doi/10.1103/PhysRevA.59.R8}.

\bibitem{kendell2024}
H.~C.~P. Kendell, G.~Ferranti, and C.~A. Weidner, ``{Deterministic generation of highly squeezed GKP states in ultracold atoms},'' \href{http://dx.doi.org/10.1063/5.0197119}{{\em APL Quantum} {\bfseries 1} no.~2, (05, 2024) 026109}, \href{http://arxiv.org/abs/https://pubs.aip.org/aip/apq/article-\\ pdf/doi/10.1063/5.0197119/19934138/026109\_1\_5.\\0197119.pdf}{{\ttfamily https://pubs.aip.org/aip/apq/article-\\ pdf/doi/10.1063/5.0197119/19934138/026109\_1\_5.\\0197119.pdf}}. \url{https://doi.org/10.1063/5.0197119}.

\bibitem{shaw2024}
A.~L. Shaw, P.~Scholl, R.~Finkelstein, R.~B.-S. Tsai, J.~Choi, and M.~Endres, ``Erasure-cooling, control, and hyper-entanglement of motion in optical tweezers,'' 2024.
\newblock \url{https://arxiv.org/abs/2311.15580}.

\bibitem{bohnmann2024}
L.~H. Bohnmann, D.~F. Locher, J.~Zeiher, and M.~M{\"u}ller, ``Bosonic quantum error correction with neutral atoms in optical dipole traps,'' 2024.
\newblock \url{https://arxiv.org/abs/2408.14251}.

\bibitem{Hu:2019zbe}
L.~Hu {\em et~al.}, ``{Quantum error correction and universal gate set operation on a binomial bosonic logical qubit},'' \href{http://dx.doi.org/10.1038/s41567-018-0414-3}{{\em Nature Phys.} {\bfseries 15} no.~5, (2019) 503--508}.

\bibitem{Campagne-Ibarcq:2019nmy}
P.~Campagne-Ibarcq {\em et~al.}, ``{Quantum error correction of a qubit encoded in grid states of an oscillator},'' \href{http://dx.doi.org/10.1038/s41586-020-2603-3}{{\em Nature} {\bfseries 584} (2020) 368--372}, \href{http://arxiv.org/abs/1907.12487}{{\ttfamily arXiv:1907.12487 [quant-ph]}}.

\bibitem{Cai_2021}
K.~Cai, P.~Parajuli, G.~Long, C.~W. Wong, and L.~Tian, ``Robust preparation of many-body ground states in jaynes--cummings lattices,'' \href{http://dx.doi.org/10.1038/s41534-021-00433-y}{{\em npj Quantum Information} {\bfseries 7} no.~1, (June, 2021) }. \url{http://dx.doi.org/10.1038/s41534-021-00433-y}.

\bibitem{Kogut:1974ag}
J.~B. Kogut and L.~Susskind, ``{Hamiltonian Formulation of Wilson's Lattice Gauge Theories},'' \href{http://dx.doi.org/10.1103/PhysRevD.11.395}{{\em Phys. Rev. D} {\bfseries 11} (1975) 395--408}.

\bibitem{Jordan:1928wi}
P.~Jordan and E.~P. Wigner, ``{About the Pauli exclusion principle},'' \href{http://dx.doi.org/10.1007/BF01331938}{{\em Z. Phys.} {\bfseries 47} (1928) 631--651}.

\bibitem{Chandrasekharan:1996ih}
S.~Chandrasekharan and U.~J. Wiese, ``{Quantum link models: A Discrete approach to gauge theories},'' \href{http://dx.doi.org/10.1016/S0550-3213(97)00006-0}{{\em Nucl. Phys. B} {\bfseries 492} (1997) 455--474}, \href{http://arxiv.org/abs/hep-lat/9609042}{{\ttfamily arXiv:hep-lat/9609042}}.

\bibitem{Jordan:2012xnu}
S.~P. Jordan, K.~S.~M. Lee, and J.~Preskill, ``{Quantum Algorithms for Quantum Field Theories},'' \href{http://dx.doi.org/10.1126/science.1217069}{{\em Science} {\bfseries 336} (2012) 1130--1133}, \href{http://arxiv.org/abs/1111.3633}{{\ttfamily arXiv:1111.3633 [quant-ph]}}.

\bibitem{Ercolessi:2017jbi}
E.~Ercolessi, P.~Facchi, G.~Magnifico, S.~Pascazio, and F.~V. Pepe, ``{Phase Transitions in $Z_{n}$ Gauge Models: Towards Quantum Simulations of the Schwinger-Weyl QED},'' \href{http://dx.doi.org/10.1103/PhysRevD.98.074503}{{\em Phys. Rev. D} {\bfseries 98} no.~7, (2018) 074503}, \href{http://arxiv.org/abs/1705.11047}{{\ttfamily arXiv:1705.11047 [quant-ph]}}.

\bibitem{Banerjee:2012pg}
D.~Banerjee, M.~Dalmonte, M.~Muller, E.~Rico, P.~Stebler, U.~J. Wiese, and P.~Zoller, ``{Atomic Quantum Simulation of Dynamical Gauge Fields coupled to Fermionic Matter: From String Breaking to Evolution after a Quench},'' \href{http://dx.doi.org/10.1103/PhysRevLett.109.175302}{{\em Phys. Rev. Lett.} {\bfseries 109} (2012) 175302}, \href{http://arxiv.org/abs/1205.6366}{{\ttfamily arXiv:1205.6366 [cond-mat.quant-gas]}}.

\bibitem{Zohar:2015hwa}
E.~Zohar, J.~I. Cirac, and B.~Reznik, ``{Quantum Simulations of Lattice Gauge Theories using Ultracold Atoms in Optical Lattices},'' \href{http://dx.doi.org/10.1088/0034-4885/79/1/014401}{{\em Rept. Prog. Phys.} {\bfseries 79} no.~1, (2016) 014401}, \href{http://arxiv.org/abs/1503.02312}{{\ttfamily arXiv:1503.02312 [quant-ph]}}.

\bibitem{Hauke:2013jga}
P.~Hauke, D.~Marcos, M.~Dalmonte, and P.~Zoller, ``{Quantum simulation of a lattice Schwinger model in a chain of trapped ions},'' \href{http://dx.doi.org/10.1103/PhysRevX.3.041018}{{\em Phys. Rev. X} {\bfseries 3} no.~4, (2013) 041018}, \href{http://arxiv.org/abs/1306.2162}{{\ttfamily arXiv:1306.2162 [cond-mat.quant-gas]}}.

\bibitem{deJong:2021wsd}
W.~A. de~Jong, K.~Lee, J.~Mulligan, M.~P\l{}osko\'n, F.~Ringer, and X.~Yao, ``{Quantum simulation of nonequilibrium dynamics and thermalization in the Schwinger model},'' \href{http://dx.doi.org/10.1103/PhysRevD.106.054508}{{\em Phys. Rev. D} {\bfseries 106} no.~5, (2022) 054508}, \href{http://arxiv.org/abs/2106.08394}{{\ttfamily arXiv:2106.08394 [quant-ph]}}.

\bibitem{Jordan:2017lea}
S.~P. Jordan, H.~Krovi, K.~S.~M. Lee, and J.~Preskill, ``{BQP-completeness of Scattering in Scalar Quantum Field Theory},'' \href{http://dx.doi.org/10.22331/q-2018-01-08-44}{{\em Quantum} {\bfseries 2} (2018) 44}, \href{http://arxiv.org/abs/1703.00454}{{\ttfamily arXiv:1703.00454 [quant-ph]}}.

\bibitem{Klco:2018kyo}
N.~Klco, E.~F. Dumitrescu, A.~J. McCaskey, T.~D. Morris, R.~C. Pooser, M.~Sanz, E.~Solano, P.~Lougovski, and M.~J. Savage, ``{Quantum-classical computation of Schwinger model dynamics using quantum computers},'' \href{http://dx.doi.org/10.1103/PhysRevA.98.032331}{{\em Phys. Rev. A} {\bfseries 98} no.~3, (2018) 032331}, \href{http://arxiv.org/abs/1803.03326}{{\ttfamily arXiv:1803.03326 [quant-ph]}}.

\bibitem{Tong:2021rfv}
Y.~Tong, V.~V. Albert, J.~R. McClean, J.~Preskill, and Y.~Su, ``{Provably accurate simulation of gauge theories and bosonic systems},'' \href{http://dx.doi.org/10.22331/q-2022-09-22-816}{{\em Quantum} {\bfseries 6} (2022) 816}, \href{http://arxiv.org/abs/2110.06942}{{\ttfamily arXiv:2110.06942 [quant-ph]}}.

\bibitem{Marshall:2015mna}
K.~Marshall, R.~Pooser, G.~Siopsis, and C.~Weedbrook, ``{Quantum simulation of quantum field theory using continuous variables},'' \href{http://dx.doi.org/10.1103/PhysRevA.92.063825}{{\em Phys. Rev. A} {\bfseries 92} no.~6, (2015) 063825}, \href{http://arxiv.org/abs/1503.08121}{{\ttfamily arXiv:1503.08121 [quant-ph]}}.

\bibitem{Bauer:2021gek}
C.~W. Bauer and D.~M. Grabowska, ``{Efficient representation for simulating U(1) gauge theories on digital quantum computers at all values of the coupling},'' \href{http://dx.doi.org/10.1103/PhysRevD.107.L031503}{{\em Phys. Rev. D} {\bfseries 107} no.~3, (2023) L031503}, \href{http://arxiv.org/abs/2111.08015}{{\ttfamily arXiv:2111.08015 [hep-ph]}}.

\bibitem{Yeter-Aydeniz:2021mol}
K.~Yeter-Aydeniz, E.~Moschandreou, and G.~Siopsis, ``{Quantum imaginary-time evolution algorithm for quantum field theories with continuous variables},'' \href{http://dx.doi.org/10.1103/PhysRevA.105.012412}{{\em Phys. Rev. A} {\bfseries 105} no.~1, (2022) 012412}, \href{http://arxiv.org/abs/2107.00791}{{\ttfamily arXiv:2107.00791 [quant-ph]}}.

\bibitem{Barata:2020jtq}
J.~a. Barata, N.~Mueller, A.~Tarasov, and R.~Venugopalan, ``{Single-particle digitization strategy for quantum computation of a $\phi^4$ scalar field theory},'' \href{http://dx.doi.org/10.1103/PhysRevA.103.042410}{{\em Phys. Rev. A} {\bfseries 103} no.~4, (2021) 042410}, \href{http://arxiv.org/abs/2012.00020}{{\ttfamily arXiv:2012.00020 [hep-th]}}.

\bibitem{Spagnoli:2024mib}
L.~Spagnoli, A.~Roggero, and N.~Wiebe, ``{Fault-tolerant simulation of Lattice Gauge Theories with gauge covariant codes},'' \href{http://arxiv.org/abs/2405.19293}{{\ttfamily arXiv:2405.19293 [quant-ph]}}.

\bibitem{Honda:2021aum}
M.~Honda, E.~Itou, Y.~Kikuchi, L.~Nagano, and T.~Okuda, ``{Classically emulated digital quantum simulation for screening and confinement in the Schwinger model with a topological term},'' \href{http://dx.doi.org/10.1103/PhysRevD.105.014504}{{\em Phys. Rev. D} {\bfseries 105} no.~1, (2022) 014504}, \href{http://arxiv.org/abs/2105.03276}{{\ttfamily arXiv:2105.03276 [hep-lat]}}.

\bibitem{Yeter-Aydeniz:2017ubh}
K.~Yeter-Aydeniz and G.~Siopsis, ``{Quantum Computation of Scattering Amplitudes in Scalar Quantum Electrodynamics},'' \href{http://dx.doi.org/10.1103/PhysRevD.97.036004}{{\em Phys. Rev. D} {\bfseries 97} no.~3, (2018) 036004}, \href{http://arxiv.org/abs/1709.02355}{{\ttfamily arXiv:1709.02355 [quant-ph]}}.

\bibitem{Li:2021kcs}
{\bfseries QuNu} Collaboration, T.~Li, X.~Guo, W.~K. Lai, X.~Liu, E.~Wang, H.~Xing, D.-B. Zhang, and S.-L. Zhu, ``{Partonic collinear structure by quantum computing},'' \href{http://dx.doi.org/10.1103/PhysRevD.105.L111502}{{\em Phys. Rev. D} {\bfseries 105} no.~11, (2022) L111502}, \href{http://arxiv.org/abs/2106.03865}{{\ttfamily arXiv:2106.03865 [hep-ph]}}.

\bibitem{Cohen:2021imf}
{\bfseries NuQS} Collaboration, T.~D. Cohen, H.~Lamm, S.~Lawrence, and Y.~Yamauchi, ``{Quantum algorithms for transport coefficients in gauge theories},'' \href{http://dx.doi.org/10.1103/PhysRevD.104.094514}{{\em Phys. Rev. D} {\bfseries 104} no.~9, (2021) 094514}, \href{http://arxiv.org/abs/2104.02024}{{\ttfamily arXiv:2104.02024 [hep-lat]}}.

\bibitem{Jha:2023ecu}
R.~G. Jha, F.~Ringer, G.~Siopsis, and S.~Thompson, ``{Continuous-variable quantum computation of the O(3) model in 1+1 dimensions},'' \href{http://dx.doi.org/10.1103/PhysRevA.109.052412}{{\em Phys. Rev. A} {\bfseries 109} no.~5, (2024) 052412}, \href{http://arxiv.org/abs/2310.12512}{{\ttfamily arXiv:2310.12512 [quant-ph]}}.

\bibitem{Florio:2023dke}
A.~Florio, D.~Frenklakh, K.~Ikeda, D.~Kharzeev, V.~Korepin, S.~Shi, and K.~Yu, ``{Real-Time Nonperturbative Dynamics of Jet Production in Schwinger Model: Quantum Entanglement and Vacuum Modification},'' \href{http://dx.doi.org/10.1103/PhysRevLett.131.021902}{{\em Phys. Rev. Lett.} {\bfseries 131} no.~2, (2023) 021902}, \href{http://arxiv.org/abs/2301.11991}{{\ttfamily arXiv:2301.11991 [hep-ph]}}.

\bibitem{Belyansky:2023rgh}
R.~Belyansky, S.~Whitsitt, N.~Mueller, A.~Fahimniya, E.~R. Bennewitz, Z.~Davoudi, and A.~V. Gorshkov, ``{High-Energy Collision of Quarks and Mesons in the Schwinger Model: From Tensor Networks to Circuit QED},'' \href{http://dx.doi.org/10.1103/PhysRevLett.132.091903}{{\em Phys. Rev. Lett.} {\bfseries 132} no.~9, (2024) 091903}, \href{http://arxiv.org/abs/2307.02522}{{\ttfamily arXiv:2307.02522 [quant-ph]}}.

\bibitem{Briceno:2023xcm}
R.~A. Brice\~no, R.~G. Edwards, M.~Eaton, C.~Gonz\'alez-Arciniegas, O.~Pfister, and G.~Siopsis, ``{Toward coherent quantum computation of scattering amplitudes with a measurement-based photonic quantum processor},'' \href{http://arxiv.org/abs/2312.12613}{{\ttfamily arXiv:2312.12613 [quant-ph]}}.

\bibitem{Davoudi:2022xmb}
Z.~Davoudi, A.~F. Shaw, and J.~R. Stryker, ``{General quantum algorithms for Hamiltonian simulation with applications to a non-Abelian lattice gauge theory},'' \href{http://dx.doi.org/10.22331/q-2023-12-20-1213}{{\em Quantum} {\bfseries 7} (2023) 1213}, \href{http://arxiv.org/abs/2212.14030}{{\ttfamily arXiv:2212.14030 [hep-lat]}}.

\bibitem{Funcke:2023lli}
L.~Funcke, K.~Jansen, and S.~K\"uhn, ``{Exploring the CP-violating Dashen phase in the Schwinger model with tensor networks},'' \href{http://dx.doi.org/10.1103/PhysRevD.108.014504}{{\em Phys. Rev. D} {\bfseries 108} no.~1, (2023) 014504}, \href{http://arxiv.org/abs/2303.03799}{{\ttfamily arXiv:2303.03799 [hep-lat]}}.

\bibitem{Araz:2022tbd}
J.~Y. Araz, S.~Schenk, and M.~Spannowsky, ``{Toward a quantum simulation of nonlinear sigma models with a topological term},'' \href{http://dx.doi.org/10.1103/PhysRevA.107.032619}{{\em Phys. Rev. A} {\bfseries 107} no.~3, (2023) 032619}, \href{http://arxiv.org/abs/2210.03679}{{\ttfamily arXiv:2210.03679 [quant-ph]}}.

\bibitem{Araz:2023ngh}
J.~Y. Araz, M.~Spannowsky, and M.~Wingate, ``{Exploring thermal equilibria of the Fermi-Hubbard model with variational quantum algorithms},'' \href{http://dx.doi.org/10.1103/PhysRevA.109.062422}{{\em Phys. Rev. A} {\bfseries 109} no.~6, (2024) 062422}, \href{http://arxiv.org/abs/2312.09292}{{\ttfamily arXiv:2312.09292 [quant-ph]}}.

\bibitem{Asaduzzaman:2023wtd}
M.~Asaduzzaman, R.~G. Jha, and B.~Sambasivam, ``{Sachdev-Ye-Kitaev model on a noisy quantum computer},'' \href{http://dx.doi.org/10.1103/PhysRevD.109.105002}{{\em Phys. Rev. D} {\bfseries 109} no.~10, (2024) 105002}, \href{http://arxiv.org/abs/2311.17991}{{\ttfamily arXiv:2311.17991 [quant-ph]}}.

\bibitem{Fromm:2023npm}
M.~Fromm, O.~Philipsen, M.~Spannowsky, and C.~Winterowd, ``{Simulating Z\_{2} lattice gauge theory with the variational quantum thermalizer},'' \href{http://dx.doi.org/10.1140/epjqt/s40507-024-00232-2}{{\em EPJ Quant. Technol.} {\bfseries 11} no.~1, (2024) 20}, \href{http://arxiv.org/abs/2306.06057}{{\ttfamily arXiv:2306.06057 [hep-lat]}}.

\bibitem{Florio:2024aix}
A.~Florio, D.~Frenklakh, K.~Ikeda, D.~E. Kharzeev, V.~Korepin, S.~Shi, and K.~Yu, ``{Quantum simulation of entanglement and hadronization in jet production: lessons from the massive Schwinger model},'' \href{http://arxiv.org/abs/2404.00087}{{\ttfamily arXiv:2404.00087 [hep-ph]}}.

\bibitem{Grieninger:2024cdl}
S.~Grieninger, K.~Ikeda, and I.~Zahed, ``{Quasi-parton distributions in massive QED2: Towards quantum computation},'' \href{http://arxiv.org/abs/2404.05112}{{\ttfamily arXiv:2404.05112 [hep-ph]}}.

\bibitem{Lee:2024jnt}
K.~Lee, F.~Turro, and X.~Yao, ``{Quantum Computing for Energy Correlators},'' \href{http://arxiv.org/abs/2409.13830}{{\ttfamily arXiv:2409.13830 [hep-ph]}}.

\bibitem{Briceno:2020rar}
R.~A. Brice\~no, J.~V. Guerrero, M.~T. Hansen, and A.~M. Sturzu, ``{Role of boundary conditions in quantum computations of scattering observables},'' \href{http://dx.doi.org/10.1103/PhysRevD.103.014506}{{\em Phys. Rev. D} {\bfseries 103} no.~1, (2021) 014506}, \href{http://arxiv.org/abs/2007.01155}{{\ttfamily arXiv:2007.01155 [hep-lat]}}.

\bibitem{Barata:2024apg}
J.~a. Barata and S.~Mukherjee, ``{Probing Celestial Energy and Charge Correlations through Real-Time Quantum Simulations: Insights from the Schwinger Model},'' \href{http://arxiv.org/abs/2409.13816}{{\ttfamily arXiv:2409.13816 [hep-ph]}}.

\bibitem{Neill:2024klc}
D.~Neill, H.~Liu, J.~Martin, and A.~Roggero, ``{Scattering Neutrinos, Spin Models, and Permutations},'' \href{http://arxiv.org/abs/2406.18677}{{\ttfamily arXiv:2406.18677 [hep-ph]}}.

\bibitem{Farrell:2024fit}
R.~C. Farrell, M.~Illa, A.~N. Ciavarella, and M.~J. Savage, ``{Quantum simulations of hadron dynamics in the Schwinger model using 112 qubits},'' \href{http://dx.doi.org/10.1103/PhysRevD.109.114510}{{\em Phys. Rev. D} {\bfseries 109} no.~11, (2024) 114510}, \href{http://arxiv.org/abs/2401.08044}{{\ttfamily arXiv:2401.08044 [quant-ph]}}.

\bibitem{Gustafson:2024bww}
E.~Gustafson {\em et~al.}, ``{Surrogate Constructed Scalable Circuits ADAPT-VQE in the Schwinger model},'' \href{http://arxiv.org/abs/2408.12641}{{\ttfamily arXiv:2408.12641 [quant-ph]}}.

\bibitem{Hardy:2024ric}
A.~Hardy {\em et~al.}, ``{Optimized Quantum Simulation Algorithms for Scalar Quantum Field Theories},'' \href{http://arxiv.org/abs/2407.13819}{{\ttfamily arXiv:2407.13819 [quant-ph]}}.

\bibitem{porras2004}
D.~Porras and J.~I. Cirac, ``Bose-einstein condensation and strong-correlation behavior of phonons in ion traps,'' \href{http://dx.doi.org/10.1103/PhysRevLett.93.263602}{{\em Phys. Rev. Lett.} {\bfseries 93} (Dec, 2004) 263602}. \url{https://link.aps.org/doi/10.1103/PhysRevLett.93.263602}.

\bibitem{serafini2009}
A.~Serafini, A.~Retzker, and M.~B. Plenio, ``Manipulating the quantum information of the radial modes of trapped ions: linear phononics, entanglement generation, quantum state transmission and non-locality tests,'' \href{http://dx.doi.org/10.1088/1367-2630/11/2/023007}{{\em New Journal of Physics} {\bfseries 11} no.~2, (Feb, 2009) 023007}. \url{https://dx.doi.org/10.1088/1367-2630/11/2/023007}.

\bibitem{debnath2018}
S.~Debnath, N.~M. Linke, S.-T. Wang, C.~Figgatt, K.~A. Landsman, L.-M. Duan, and C.~Monroe, ``Observation of hopping and blockade of bosons in a trapped ion spin chain,'' \href{http://dx.doi.org/10.1103/PhysRevLett.120.073001}{{\em Phys. Rev. Lett.} {\bfseries 120} (Feb, 2018) 073001}. \url{https://link.aps.org/doi/10.1103/PhysRevLett.120.073001}.

\bibitem{Chen_2021}
W.~Chen, J.~Gan, J.-N. Zhang, D.~Matuskevich, and K.~Kim, ``Quantum computation and simulation with vibrational modes of trapped ions,'' \href{http://dx.doi.org/10.1088/1674-1056/ac01e3}{{\em Chinese Physics B} {\bfseries 30} no.~6, (Jun, 2021) 060311}. \url{https://doi.org/10.1088%2F1674-1056%2Fac01e3}.

\bibitem{Katz:2022gra}
O.~Katz and C.~Monroe, ``{Programmable Quantum Simulations of Bosonic Systems with Trapped Ions},'' \href{http://dx.doi.org/10.1103/PhysRevLett.131.033604}{{\em Phys. Rev. Lett.} {\bfseries 131} no.~3, (2023) 033604}, \href{http://arxiv.org/abs/2207.13653}{{\ttfamily arXiv:2207.13653 [quant-ph]}}.

\bibitem{chen2023scalable}
W.~Chen, Y.~Lu, S.~Zhang, K.~Zhang, G.~Huang, M.~Qiao, X.~Su, J.~Zhang, J.-N. Zhang, L.~Banchi, {\em et~al.}, ``Scalable and programmable phononic network with trapped ions,'' {\em Nature Physics} {\bfseries 19} no.~6, (2023) 877--883.

\bibitem{cirac1995}
J.~I. Cirac and P.~Zoller, ``Quantum computations with cold trapped ions,'' \href{http://dx.doi.org/10.1103/PhysRevLett.74.4091}{{\em Phys. Rev. Lett.} {\bfseries 74} (May, 1995) 4091--4094}. \url{https://link.aps.org/doi/10.1103/PhysRevLett.74.4091}.

\bibitem{molmer1999}
K.~M\o{}lmer and A.~S\o{}rensen, ``Multiparticle entanglement of hot trapped ions,'' \href{http://dx.doi.org/10.1103/PhysRevLett.82.1835}{{\em Phys. Rev. Lett.} {\bfseries 82} (Mar, 1999) 1835--1838}. \url{https://link.aps.org/doi/10.1103/PhysRevLett.82.1835}.

\bibitem{sorensen1999}
A.~S\o{}rensen and K.~M\o{}lmer, ``Quantum computation with ions in thermal motion,'' \href{http://dx.doi.org/10.1103/PhysRevLett.82.1971}{{\em Phys. Rev. Lett.} {\bfseries 82} (Mar, 1999) 1971--1974}. \url{https://link.aps.org/doi/10.1103/PhysRevLett.82.1971}.

\bibitem{Macridin:2023xny}
A.~Macridin, A.~C.~Y. Li, and P.~Spentzouris, ``{Qumode transfer between continuous- and discrete-variable devices},'' \href{http://dx.doi.org/10.1103/PhysRevA.109.032419}{{\em Phys. Rev. A} {\bfseries 109} no.~3, (2024) 032419}, \href{http://arxiv.org/abs/2305.03179}{{\ttfamily arXiv:2305.03179 [quant-ph]}}.

\bibitem{Liu:2024mbr}
Y.~Liu {\em et~al.}, ``{Hybrid Oscillator-Qubit Quantum Processors: Instruction Set Architectures, Abstract Machine Models, and Applications},'' \href{http://arxiv.org/abs/2407.10381}{{\ttfamily arXiv:2407.10381 [quant-ph]}}.

\bibitem{Crane:2024tlj}
E.~Crane {\em et~al.}, ``{Hybrid Oscillator-Qubit Quantum Processors: Simulating Fermions, Bosons, and Gauge Fields},'' \href{http://arxiv.org/abs/2409.03747}{{\ttfamily arXiv:2409.03747 [quant-ph]}}.

\bibitem{romanencko2020}
A.~Romanenko, R.~Pilipenko, S.~Zorzetti, D.~Frolov, M.~Awida, S.~Belomestnykh, S.~Posen, and A.~Grassellino, ``Three-dimensional superconducting resonators at $t<20$ mk with photon lifetimes up to $\tau=2$ s,'' \href{http://dx.doi.org/10.1103/PhysRevApplied.13.034032}{{\em Phys. Rev. Appl.} {\bfseries 13} (Mar, 2020) 034032}. \url{https://link.aps.org/doi/10.1103/PhysRevApplied.13.034032}.

\bibitem{schreier2008}
J.~A. Schreier, A.~A. Houck, J.~Koch, D.~I. Schuster, B.~R. Johnson, J.~M. Chow, J.~M. Gambetta, J.~Majer, L.~Frunzio, M.~H. Devoret, S.~M. Girvin, and R.~J. Schoelkopf, ``Suppressing charge noise decoherence in superconducting charge qubits,'' \href{http://dx.doi.org/10.1103/PhysRevB.77.180502}{{\em Phys. Rev. B} {\bfseries 77} (May, 2008) 180502}. \url{https://link.aps.org/doi/10.1103/PhysRevB.77.180502}.

\bibitem{wang2021}
P.~Wang, C.-Y. Luan, M.~Qiao, M.~Um, J.~Zhang, Y.~Wang, X.~Yuan, M.~Gu, J.~Zhang, and K.~Kim, ``Single ion qubit with estimated coherence time exceeding one hour,'' \href{http://dx.doi.org/10.1038/s41467-020-20330-w}{{\em Nature Communications} {\bfseries 12} no.~1, (2021) 233}. \url{https://doi.org/10.1038/s41467-020-20330-w}.

\bibitem{Jarlaud_2021}
V.~Jarlaud, P.~Hrmo, M.~K. Joshi, and R.~C. Thompson, ``Coherence properties of highly-excited motional states of a trapped ion,'' \href{http://dx.doi.org/10.1088/1361-6455/abc271}{{\em Journal of Physics B: Atomic, Molecular and Optical Physics} {\bfseries 54} no.~1, (Dec, 2020) 015501}. \url{https://dx.doi.org/10.1088/1361-6455/abc271}.

\bibitem{Casanova:2011wh}
J.~Casanova, L.~Lamata, I.~L. Egusquiza, R.~Gerritsma, C.~F. Roos, J.~J. Garcia-Ripoll, and E.~Solano, ``{Quantum Simulation of Quantum Field Theories in Trapped Ions},'' \href{http://dx.doi.org/10.1103/PhysRevLett.107.260501}{{\em Phys. Rev. Lett.} {\bfseries 107} (2011) 260501}, \href{http://arxiv.org/abs/1107.5233}{{\ttfamily arXiv:1107.5233 [quant-ph]}}.

\bibitem{Lv:2017tmn}
D.~Lv, S.~An, Z.~Liu, J.-N. Zhang, J.~S. Pedernales, L.~Lamata, E.~Solano, and K.~Kim, ``{Quantum Simulation of the Quantum Rabi Model in a Trapped Ion},'' \href{http://dx.doi.org/10.1103/PhysRevX.8.021027}{{\em Phys. Rev. X} {\bfseries 8} no.~2, (2018) 021027}, \href{http://arxiv.org/abs/1711.00582}{{\ttfamily arXiv:1711.00582 [quant-ph]}}.

\bibitem{Davoudi:2021ney}
Z.~Davoudi, N.~M. Linke, and G.~Pagano, ``{Toward simulating quantum field theories with controlled phonon-ion dynamics: A hybrid analog-digital approach},'' \href{http://dx.doi.org/10.1103/PhysRevResearch.3.043072}{{\em Phys. Rev. Res.} {\bfseries 3} no.~4, (2021) 043072}, \href{http://arxiv.org/abs/2104.09346}{{\ttfamily arXiv:2104.09346 [quant-ph]}}.

\bibitem{Angelakis_2007}
D.~G. Angelakis, M.~F. Santos, and S.~Bose, ``Photon-blockade-induced mott transitions and $xy$ spin models in coupled cavity arrays,'' \href{http://dx.doi.org/10.1103/physreva.76.031805}{{\em Physical Review A} {\bfseries 76} no.~3, (Sept., 2007) }. \url{http://dx.doi.org/10.1103/PhysRevA.76.031805}.

\bibitem{Schmidt_2009}
S.~Schmidt and G.~Blatter, ``Strong coupling theory for the jaynes-cummings-hubbard model,'' \href{http://dx.doi.org/10.1103/physrevlett.103.086403}{{\em Physical Review Letters} {\bfseries 103} no.~8, (Aug., 2009) }. \url{http://dx.doi.org/10.1103/PhysRevLett.103.086403}.

\bibitem{Nunnenkamp:2011ww}
A.~Nunnenkamp, J.~Koch, and S.~M. Girvin, ``{Synthetic gauge fields and homodyne transmission in Jaynes-Cummings lattices},'' \href{http://dx.doi.org/10.1088/1367-2630/13/9/095008}{{\em New J. Phys.} {\bfseries 13} (2011) 095008}, \href{http://arxiv.org/abs/1105.1817}{{\ttfamily arXiv:1105.1817 [cond-mat.mes-hall]}}.

\bibitem{Bertassoli:2024sri}
J.~L.~T. Bertassoli and A.~Vidiella-Barranco, ``{A note on the emission spectrum and trapping states in the Jaynes-Cummings model},'' \href{http://arxiv.org/abs/2406.10763}{{\ttfamily arXiv:2406.10763 [quant-ph]}}.

\bibitem{Li:2022seq}
B.~W. Li, Q.~X. Mei, Y.~K. Wu, M.~L. Cai, Y.~Wang, L.~Yao, Z.~C. Zhou, and L.~M. Duan, ``{Observation of Non-Markovian Spin Dynamics in a Jaynes-Cummings-Hubbard Model Using a Trapped-Ion Quantum Simulator},'' \href{http://dx.doi.org/10.1103/PhysRevLett.129.140501}{{\em Phys. Rev. Lett.} {\bfseries 129} no.~14, (2022) 140501}, \href{http://arxiv.org/abs/2205.15529}{{\ttfamily arXiv:2205.15529 [quant-ph]}}.

\bibitem{Greentree_2006}
A.~D. Greentree, C.~Tahan, J.~H. Cole, and L.~C.~L. Hollenberg, ``Quantum phase transitions of light,'' \href{http://dx.doi.org/10.1038/nphys466}{{\em Nature Physics} {\bfseries 2} no.~12, (Nov., 2006) 856--861}. \url{http://dx.doi.org/10.1038/nphys466}.

\bibitem{Schwinger:1962tp}
J.~S. Schwinger, ``{Gauge Invariance and Mass. 2.},'' \href{http://dx.doi.org/10.1103/PhysRev.128.2425}{{\em Phys. Rev.} {\bfseries 128} (1962) 2425--2429}.

\bibitem{Coleman:1975pw}
S.~R. Coleman, R.~Jackiw, and L.~Susskind, ``{Charge Shielding and Quark Confinement in the Massive Schwinger Model},'' \href{http://dx.doi.org/10.1016/0003-4916(75)90212-2}{{\em Annals Phys.} {\bfseries 93} (1975) 267}.

\bibitem{Koch:2021wqd}
J.~Koch, G.~R. Hunanyan, T.~Ockenfels, E.~Rico, E.~Solano, and M.~Weitz, ``{Quantum Rabi dynamics of trapped atoms far in the deep strong coupling regime},'' \href{http://dx.doi.org/10.1038/s41467-023-36611-z}{{\em Nature Commun.} {\bfseries 14} no.~1, (2023) 954}, \href{http://arxiv.org/abs/2112.12488}{{\ttfamily arXiv:2112.12488 [quant-ph]}}.

\bibitem{Zache:2023cfj}
T.~V. Zache, D.~Gonz\'alez-Cuadra, and P.~Zoller, ``{Fermion-qudit quantum processors for simulating lattice gauge theories with matter},'' \href{http://dx.doi.org/10.22331/q-2023-10-16-1140}{{\em Quantum} {\bfseries 7} (2023) 1140}, \href{http://arxiv.org/abs/2303.08683}{{\ttfamily arXiv:2303.08683 [quant-ph]}}.

\bibitem{Fromm:2024caq}
M.~Fromm, L.~Katschke, O.~Philipsen, and W.~Unger, ``{Quantum computational resources for lattice QCD in the strong-coupling limit},'' \href{http://arxiv.org/abs/2406.18721}{{\ttfamily arXiv:2406.18721 [hep-lat]}}.

\bibitem{Illa:2024kmf}
M.~Illa, C.~E.~P. Robin, and M.~J. Savage, ``{Qu8its for quantum simulations of lattice quantum chromodynamics},'' \href{http://dx.doi.org/10.1103/PhysRevD.110.014507}{{\em Phys. Rev. D} {\bfseries 110} no.~1, (2024) 014507}, \href{http://arxiv.org/abs/2403.14537}{{\ttfamily arXiv:2403.14537 [quant-ph]}}.

\bibitem{Abel:2024kuv}
S.~Abel, M.~Spannowsky, and S.~Williams, ``{Simulating quantum field theories on continuous-variable quantum computers},'' \href{http://dx.doi.org/10.1103/PhysRevA.110.012607}{{\em Phys. Rev. A} {\bfseries 110} no.~1, (2024) 012607}, \href{http://arxiv.org/abs/2403.10619}{{\ttfamily arXiv:2403.10619 [quant-ph]}}.

\bibitem{Kay:2018huf}
A.~Kay, ``{Tutorial on the Quantikz Package},'' \href{http://arxiv.org/abs/1809.03842}{{\ttfamily arXiv:1809.03842 [quant-ph]}}.

\bibitem{olmschenk2007}
S.~Olmschenk, K.~C. Younge, D.~L. Moehring, D.~N. Matsukevich, P.~Maunz, and C.~Monroe, ``Manipulation and detection of a trapped ${\mathrm{yb}}^{+}$ hyperfine qubit,'' \href{http://dx.doi.org/10.1103/PhysRevA.76.052314}{{\em Phys. Rev. A} {\bfseries 76} (Nov, 2007) 052314}. \url{https://link.aps.org/doi/10.1103/PhysRevA.76.052314}.

\bibitem{allcock2021}
D.~T.~C. Allcock, W.~C. Campbell, J.~Chiaverini, I.~L. Chuang, E.~R. Hudson, I.~D. Moore, A.~Ransford, C.~Roman, J.~M. Sage, and D.~J. Wineland, ``{omg blueprint for trapped ion quantum computing with metastable states},'' \href{http://dx.doi.org/10.1063/5.0069544}{{\em Applied Physics Letters} {\bfseries 119} no.~21, (11, 2021) 214002}, \href{http://arxiv.org/abs/https://pubs.aip.org/aip/apl/article-pdf/doi \\ /10.1063/5.0069544/19773783/214002\_1\_online.pdf}{{\ttfamily https://pubs.aip.org/aip/apl/article-pdf/doi \\ /10.1063/5.0069544/19773783/214002\_1\_online.pdf}}. \url{https://doi.org/10.1063/5.0069544}.

\bibitem{yang2022}
H.~X. Yang, J.~Y. Ma, Y.~K. Wu, Y.~Wang, M.~M. Cao, W.~X. Guo, Y.~Y. Huang, L.~Feng, Z.~C. Zhou, and L.~M. Duan, ``Realizing coherently convertible dual-type qubits with the same ion species,'' \href{http://dx.doi.org/10.1038/s41567-022-01661-5}{{\em Nature Physics} {\bfseries 18} no.~9, (2022) 1058--1061}. \url{https://doi.org/10.1038/s41567-022-01661-5}.

\bibitem{debry2023}
K.~DeBry, J.~Sinanan-Singh, C.~D. Bruzewicz, D.~Reens, M.~E. Kim, M.~P. Roychowdhury, R.~McConnell, I.~L. Chuang, and J.~Chiaverini, ``Experimental quantum channel discrimination using metastable states of a trapped ion,'' \href{http://dx.doi.org/10.1103/PhysRevLett.131.170602}{{\em Phys. Rev. Lett.} {\bfseries 131} (Oct, 2023) 170602}. \url{https://link.aps.org/doi/10.1103/PhysRevLett.131.170602}.

\bibitem{brown2011}
K.~R. Brown, C.~Ospelkaus, Y.~Colombe, A.~C. Wilson, D.~Leibfried, and D.~J. Wineland, ``Coupled quantized mechanical oscillators,'' \href{http://dx.doi.org/10.1038/nature09721}{{\em Nature} {\bfseries 471} no.~7337, (2011) 196--199}. \url{https://doi.org/10.1038/nature09721}.

\bibitem{porras2008}
D.~Porras, F.~Marquardt, J.~von Delft, and J.~I. Cirac, ``Mesoscopic spin-boson models of trapped ions,'' \href{http://dx.doi.org/10.1103/PhysRevA.78.010101}{{\em Phys. Rev. A} {\bfseries 78} (Jul, 2008) 010101}. \url{https://link.aps.org/doi/10.1103/PhysRevA.78.010101}.

\bibitem{ivanov2009}
P.~A. Ivanov, S.~S. Ivanov, N.~V. Vitanov, A.~Mering, M.~Fleischhauer, and K.~Singer, ``Simulation of a quantum phase transition of polaritons with trapped ions,'' \href{http://dx.doi.org/10.1103/PhysRevA.80.060301}{{\em Phys. Rev. A} {\bfseries 80} (Dec, 2009) 060301}. \url{https://link.aps.org/doi/10.1103/PhysRevA.80.060301}.

\bibitem{bermudez2010}
A.~Bermudez, M.~A. Martin-Delgado, and D.~Porras, ``The localization of phonons in ion traps with controlled quantum disorder,'' \href{http://dx.doi.org/10.1088/1367-2630/12/12/123016}{{\em New Journal of Physics} {\bfseries 12} no.~12, (Dec, 2010) 123016}. \url{https://dx.doi.org/10.1088/1367-2630/12/12/123016}.

\bibitem{wineland1975}
D.~J. Wineland and H.~G. Dehmelt, ``{Principles of the stored ion calorimeter},'' \href{http://dx.doi.org/10.1063/1.321602}{{\em Journal of Applied Physics} {\bfseries 46} no.~2, (02, 1975) 919--930}. \url{https://doi.org/10.1063/1.321602}.

\bibitem{brownnutt2015}
M.~Brownnutt, M.~Kumph, P.~Rabl, and R.~Blatt, ``Ion-trap measurements of electric-field noise near surfaces,'' \href{http://dx.doi.org/10.1103/RevModPhys.87.1419}{{\em Rev. Mod. Phys.} {\bfseries 87} (Dec, 2015) 1419--1482}. \url{https://link.aps.org/doi/10.1103/RevModPhys.87.1419}.

\bibitem{king1998}
B.~E. King, C.~S. Wood, C.~J. Myatt, Q.~A. Turchette, D.~Leibfried, W.~M. Itano, C.~Monroe, and D.~J. Wineland, ``Cooling the collective motion of trapped ions to initialize a quantum register,'' \href{http://dx.doi.org/10.1103/PhysRevLett.81.1525}{{\em Phys. Rev. Lett.} {\bfseries 81} (Aug, 1998) 1525--1528}. \url{https://link.aps.org/doi/10.1103/PhysRevLett.81.1525}.

\bibitem{wineland1998experimental}
D.~J. Wineland, C.~Monroe, W.~M. Itano, D.~Leibfried, B.~E. King, and D.~M. Meekhof, ``Experimental issues in coherent quantum-state manipulation of trapped atomic ions,'' {\em Journal of research of the National Institute of Standards and Technology} {\bfseries 103} no.~3, (1998) 259.

\bibitem{deng2008}
X.-L. Deng, D.~Porras, and J.~I. Cirac, ``Quantum phases of interacting phonons in ion traps,'' \href{http://dx.doi.org/10.1103/PhysRevA.77.033403}{{\em Phys. Rev. A} {\bfseries 77} (Mar, 2008) 033403}. \url{https://link.aps.org/doi/10.1103/PhysRevA.77.033403}.

\bibitem{shen2014}
C.~Shen, Z.~Zhang, and L.-M. Duan, ``Scalable implementation of boson sampling with trapped ions,'' \href{http://dx.doi.org/10.1103/PhysRevLett.112.050504}{{\em Phys. Rev. Lett.} {\bfseries 112} (Feb, 2014) 050504}. \url{https://link.aps.org/doi/10.1103/PhysRevLett.112.050504}.

\bibitem{hou2024}
P.-Y. Hou, J.~J. Wu, S.~D. Erickson, D.~C. Cole, G.~Zarantonello, A.~D. Brandt, S.~Geller, A.~Kwiatkowski, S.~Glancy, E.~Knill, A.~C. Wilson, D.~H. Slichter, and D.~Leibfried, ``Coherent coupling and non-destructive measurement of trapped-ion mechanical oscillators,'' \href{http://dx.doi.org/10.1038/s41567-024-02585-y}{{\em Nature Physics} (2024) }. \url{https://doi.org/10.1038/s41567-024-02585-y}.

\bibitem{loschnauer2024}
C.~M. Löschnauer, J.~M. Toba, A.~C. Hughes, S.~A. King, M.~A. Weber, R.~Srinivas, R.~Matt, R.~Nourshargh, D.~T.~C. Allcock, C.~J. Ballance, C.~Matthiesen, M.~Malinowski, and T.~P. Harty, ``Scalable, high-fidelity all-electronic control of trapped-ion qubits,'' 2024.
\newblock \url{https://arxiv.org/abs/2407.07694}.

\bibitem{gaebler2016}
J.~P. Gaebler, T.~R. Tan, Y.~Lin, Y.~Wan, R.~Bowler, A.~C. Keith, S.~Glancy, K.~Coakley, E.~Knill, D.~Leibfried, and D.~J. Wineland, ``High-fidelity universal gate set for ${^{9}\mathrm{Be}}^{+}$ ion qubits,'' \href{http://dx.doi.org/10.1103/PhysRevLett.117.060505}{{\em Phys. Rev. Lett.} {\bfseries 117} (Aug, 2016) 060505}. \url{https://link.aps.org/doi/10.1103/PhysRevLett.117.060505}.

\bibitem{shen2018}
Y.~Shen, Y.~Lu, K.~Zhang, J.~Zhang, S.~Zhang, J.~Huh, and K.~Kim, ``Quantum optical emulation of molecular vibronic spectroscopy using a trapped-ion device,'' \href{http://dx.doi.org/10.1039/C7SC04602B}{{\em Chem. Sci.} {\bfseries 9} (2018) 836--840}. \url{http://dx.doi.org/10.1039/C7SC04602B}.

\bibitem{McCormick_2019}
``Coherently displaced oscillator quantum states of a single trapped atom,'' \href{http://dx.doi.org/10.1088/2058-9565/ab0513}{{\em Quantum Science and Technology} {\bfseries 4} no.~2, (Mar, 2019) 024010}. \url{https://dx.doi.org/10.1088/2058-9565/ab0513}.

\bibitem{burd2019}
S.~C. Burd, R.~Srinivas, J.~J. Bollinger, A.~C. Wilson, D.~J. Wineland, D.~Leibfried, D.~H. Slichter, and D.~T.~C. Allcock, ``Quantum amplification of mechanical oscillator motion,'' \href{http://dx.doi.org/10.1126/science.aaw2884}{{\em Science} {\bfseries 364} no.~6446, (2019) 1163--1165}. \url{https://www.science.org/doi/abs/10.1126/science.aaw2884}.

\bibitem{stobinska2011}
M.~Stobi{\'n}ska, A.~Villar, and G.~Leuchs, ``Generation of kerr non-gaussian motional states of trapped ions,'' \href{http://dx.doi.org/10.1209/0295-5075/94/54002}{{\em EPL (Europhysics Letters)} {\bfseries 94} (05, 2011) 54002}.

\bibitem{ding2017cross}
S.~Ding, G.~Maslennikov, R.~Habl{\"u}tzel, and D.~Matsukevich, ``Cross-kerr nonlinearity for phonon counting,'' {\em Physical review letters} {\bfseries 119} no.~19, (2017) 193602.

\bibitem{srinivas2019}
R.~Srinivas, S.~C. Burd, R.~T. Sutherland, A.~C. Wilson, D.~J. Wineland, D.~Leibfried, D.~T.~C. Allcock, and D.~H. Slichter, ``Trapped-ion spin-motion coupling with microwaves and a near-motional oscillating magnetic field gradient,'' \href{http://dx.doi.org/10.1103/PhysRevLett.122.163201}{{\em Phys. Rev. Lett.} {\bfseries 122} (Apr, 2019) 163201}. \url{https://link.aps.org/doi/10.1103/PhysRevLett.122.163201}.

\bibitem{myerson2008}
A.~H. Myerson, D.~J. Szwer, S.~C. Webster, D.~T.~C. Allcock, M.~J. Curtis, G.~Imreh, J.~A. Sherman, D.~N. Stacey, A.~M. Steane, and D.~M. Lucas, ``High-fidelity readout of trapped-ion qubits,'' \href{http://dx.doi.org/10.1103/PhysRevLett.100.200502}{{\em Phys. Rev. Lett.} {\bfseries 100} (May, 2008) 200502}. \url{https://link.aps.org/doi/10.1103/PhysRevLett.100.200502}.

\bibitem{meekhof1996}
D.~M. Meekhof, C.~Monroe, B.~E. King, W.~M. Itano, and D.~J. Wineland, ``Generation of nonclassical motional states of a trapped atom,'' \href{http://dx.doi.org/10.1103/PhysRevLett.76.1796}{{\em Phys. Rev. Lett.} {\bfseries 76} (Mar, 1996) 1796--1799}. \url{https://link.aps.org/doi/10.1103/PhysRevLett.76.1796}.

\bibitem{Lv2017pnr}
D.~Lv, S.~An, M.~Um, J.~Zhang, J.-N. Zhang, M.~S. Kim, and K.~Kim, ``Reconstruction of the jaynes-cummings field state of ionic motion in a harmonic trap,'' \href{http://dx.doi.org/10.1103/PhysRevA.95.043813}{{\em Phys. Rev. A} {\bfseries 95} (Apr, 2017) 043813}. \url{https://link.aps.org/doi/10.1103/PhysRevA.95.043813}.

\bibitem{resolvesideband}
C.~Monroe, D.~M. Meekhof, B.~E. King, S.~R. Jefferts, W.~M. Itano, D.~J. Wineland, and P.~Gould, ``Resolved-sideband raman cooling of a bound atom to the 3d zero-point energy,'' \href{http://dx.doi.org/10.1103/PhysRevLett.75.4011}{{\em Phys. Rev. Lett.} {\bfseries 75} (Nov, 1995) 4011--4014}. \url{https://link.aps.org/doi/10.1103/PhysRevLett.75.4011}.

\bibitem{opticalpumping}
W.~HAPPER, ``Optical pumping,'' \href{http://dx.doi.org/10.1103/RevModPhys.44.169}{{\em Rev. Mod. Phys.} {\bfseries 44} (Apr, 1972) 169--249}. \url{https://link.aps.org/doi/10.1103/RevModPhys.44.169}.

\bibitem{alonso2013}
J.~Alonso, F.~M. Leupold, B.~C. Keitch, and J.~P. Home, ``Quantum control of the motional states of trapped ions through fast switching of trapping potentials,'' \href{http://dx.doi.org/10.1088/1367-2630/15/2/023001}{{\em New Journal of Physics} {\bfseries 15} no.~2, (Feb, 2013) 023001}. \url{https://dx.doi.org/10.1088/1367-2630/15/2/023001}.

\bibitem{metzner2024}
J.~Metzner, A.~Quinn, S.~Brudney, I.~D. Moore, S.~C. Burd, D.~J. Wineland, and D.~T.~C. Allcock, ``Two-mode squeezing and su(1,1) interferometry with trapped ions,'' \href{http://dx.doi.org/10.1103/PhysRevA.110.022613}{{\em Phys. Rev. A} {\bfseries 110} (Aug, 2024) 022613}. \url{https://link.aps.org/doi/10.1103/PhysRevA.110.022613}.

\bibitem{affolter2023}
M.~Affolter, W.~Ge, B.~Bullock, S.~C. Burd, K.~A. Gilmore, J.~F. Lilieholm, A.~L. Carter, and J.~J. Bollinger, ``Toward improved quantum simulations and sensing with trapped two-dimensional ion crystals via parametric amplification,'' \href{http://dx.doi.org/10.1103/PhysRevA.107.032425}{{\em Phys. Rev. A} {\bfseries 107} (Mar, 2023) 032425}. \url{https://link.aps.org/doi/10.1103/PhysRevA.107.032425}.

\bibitem{Leibfried:2003zz}
D.~Leibfried, R.~Blatt, C.~Monroe, and D.~Wineland, ``{Quantum dynamics of single trapped ions},'' \href{http://dx.doi.org/10.1103/RevModPhys.75.281}{{\em Rev. Mod. Phys.} {\bfseries 75} (2003) 281--324}.

\bibitem{meth2023simulating}
M.~Meth, J.~F. Haase, J.~Zhang, C.~Edmunds, L.~Postler, A.~Steiner, A.~J. Jena, L.~Dellantonio, R.~Blatt, P.~Zoller, {\em et~al.}, ``Simulating 2d lattice gauge theories on a qudit quantum computer,'' {\em arXiv preprint arXiv:2310.12110} (2023) .

\bibitem{marquet2003}
C.~Marquet, F.~Schmidt-Kaler, and D.~F.~V. James, ``Phonon--phonon interactions due to non-linear effects in a linear ion trap,'' \href{http://dx.doi.org/10.1007/s00340-003-1097-7}{{\em Applied Physics B} {\bfseries 76} no.~3, (2003) 199--208}. \url{https://doi.org/10.1007/s00340-003-1097-7}.

\bibitem{roos2008}
C.~F. Roos, T.~Monz, K.~Kim, M.~Riebe, H.~H\"affner, D.~F.~V. James, and R.~Blatt, ``Nonlinear coupling of continuous variables at the single quantum level,'' \href{http://dx.doi.org/10.1103/PhysRevA.77.040302}{{\em Phys. Rev. A} {\bfseries 77} (Apr, 2008) 040302}. \url{https://link.aps.org/doi/10.1103/PhysRevA.77.040302}.

\bibitem{nie2009}
X.~R. Nie, C.~F. Roos, and D.~F. James, ``Theory of cross phase modulation for the vibrational modes of trapped ions,'' \href{http://dx.doi.org/https://doi.org/10.1016/j.physleta.2008.11.045}{{\em Physics Letters A} {\bfseries 373} no.~4, (2009) 422--425}. \url{https://www.sciencedirect.com/science/article/pii/S0375960108017106}.

\bibitem{um2016phonon}
M.~Um, J.~Zhang, D.~Lv, Y.~Lu, S.~An, J.-N. Zhang, H.~Nha, M.~Kim, and K.~Kim, ``Phonon arithmetic in a trapped ion system,'' {\em Nature communications} {\bfseries 7} no.~1, (2016) 11410.

\bibitem{sutherland2021}
R.~T. Sutherland, S.~C. Burd, D.~H. Slichter, S.~B. Libby, and D.~Leibfried, ``Motional squeezing for trapped ion transport and separation,'' \href{http://dx.doi.org/10.1103/PhysRevLett.127.083201}{{\em Phys. Rev. Lett.} {\bfseries 127} (Aug, 2021) 083201}. \url{https://link.aps.org/doi/10.1103/PhysRevLett.127.083201}.

\bibitem{Braunstein:2005zz}
S.~L. Braunstein and P.~van Loock, ``{Quantum information with continuous variables},'' \href{http://dx.doi.org/10.1103/RevModPhys.77.513}{{\em Rev. Mod. Phys.} {\bfseries 77} (2005) 513--577}, \href{http://arxiv.org/abs/quant-ph/0410100}{{\ttfamily arXiv:quant-ph/0410100}}.

\bibitem{Cahall:17}
C.~Cahall, K.~L. Nicolich, N.~T. Islam, G.~P. Lafyatis, A.~J. Miller, D.~J. Gauthier, and J.~Kim, ``Multi-photon detection using a conventional superconducting nanowire single-photon detector,'' \href{http://dx.doi.org/10.1364/OPTICA.4.001534}{{\em Optica} {\bfseries 4} no.~12, (Dec, 2017) 1534--1535}. \url{https://opg.optica.org/optica/abstract.cfm?URI=optica-4-12-1534}.

\bibitem{Miller2003}
A.~J. Miller, S.~W. Nam, J.~M. Martinis, and A.~V. Sergienko, ``{Demonstration of a low-noise near-infrared photon counter with multiphoton discrimination},'' \href{http://dx.doi.org/10.1063/1.1596723}{{\em Applied Physics Letters} {\bfseries 83} no.~4, (07, 2003) 791--793}. \url{https://doi.org/10.1063/1.1596723}.

\bibitem{Eaton:2022vjq}
M.~Eaton, A.~Hossameldin, R.~J. Birrittella, P.~M. Alsing, C.~C. Gerry, H.~Dong, C.~Cuevas, and O.~Pfister, ``{Resolution of 100 photons and quantum generation of unbiased random numbers},'' \href{http://dx.doi.org/10.1038/s41566-022-01105-9}{{\em Nature Photon.} {\bfseries 17} no.~1, (2023) 106--111}, \href{http://arxiv.org/abs/2205.01221}{{\ttfamily arXiv:2205.01221 [quant-ph]}}.

\bibitem{Jia:2022qxf}
Z.~Jia, Y.~Wang, B.~Zhang, J.~Whitlow, C.~Fang, J.~Kim, and K.~R. Brown, ``{Determination of Multimode Motional Quantum States in a Trapped Ion System},'' \href{http://dx.doi.org/10.1103/PhysRevLett.129.103602}{{\em Phys. Rev. Lett.} {\bfseries 129} no.~10, (2022) 103602}, \href{http://arxiv.org/abs/2205.11444}{{\ttfamily arXiv:2205.11444 [quant-ph]}}.

\bibitem{Boghosian:1996qd}
B.~M. Boghosian and W.~Taylor, ``{Simulating quantum mechanics on a quantum computer},'' \href{http://dx.doi.org/10.1016/S0167-2789(98)00042-6}{{\em Physica D} {\bfseries 120} (1998) 30--42}, \href{http://arxiv.org/abs/quant-ph/9701019}{{\ttfamily arXiv:quant-ph/9701019}}.

\bibitem{Lloyd1996UniversalQS}
S.~Lloyd, ``Universal quantum simulators,'' {\em Science} {\bfseries 273} (1996) 1073 -- 1078. \url{https://api.semanticscholar.org/CorpusID:43496899}.

\bibitem{Childs_2018}
A.~M. Childs, D.~Maslov, Y.~Nam, N.~J. Ross, and Y.~Su, ``Toward the first quantum simulation with quantum speedup,'' \href{http://dx.doi.org/10.1073/pnas.1801723115}{{\em Proceedings of the National Academy of Sciences} {\bfseries 115} no.~38, (Sept., 2018) 9456--9461}. \url{http://dx.doi.org/10.1073/pnas.1801723115}.

\bibitem{Kalajdzievski_2018}
T.~Kalajdzievski, C.~Weedbrook, and P.~Rebentrost, ``Continuous-variable gate decomposition for the bose-hubbard model,'' \href{http://dx.doi.org/10.1103/physreva.97.062311}{{\em Physical Review A} {\bfseries 97} no.~6, (June, 2018) }. \url{http://dx.doi.org/10.1103/PhysRevA.97.062311}.

\bibitem{Bergholm:2018cyq}
V.~Bergholm {\em et~al.}, ``{PennyLane: Automatic differentiation of hybrid quantum-classical computations},'' \href{http://arxiv.org/abs/1811.04968}{{\ttfamily arXiv:1811.04968 [quant-ph]}}.

\bibitem{Peruzzo2014}
A.~Peruzzo, J.~McClean, P.~Shadbolt, M.-H. Yung, X.-Q. Zhou, P.~J. Love, A.~Aspuru-Guzik, and J.~L. O'Brien, ``A variational eigenvalue solver on a photonic quantum processor,'' \href{http://dx.doi.org/10.1038/ncomms5213}{{\em Nature Communications} {\bfseries 5} no.~1, (July, 2014) }. \url{http://dx.doi.org/10.1038/ncomms5213}.

\bibitem{Kandala:2017vok}
A.~Kandala, A.~Mezzacapo, K.~Temme, M.~Takita, M.~Brink, J.~M. Chow, and J.~M. Gambetta, ``{Hardware-efficient variational quantum eigensolver for small molecules and quantum magnets},'' \href{http://dx.doi.org/10.1038/nature23879}{{\em Nature} {\bfseries 549} no.~7671, (2017) 242--246}, \href{http://arxiv.org/abs/1704.05018}{{\ttfamily arXiv:1704.05018 [quant-ph]}}.

\bibitem{Gard_2020}
B.~T. Gard, L.~Zhu, G.~S. Barron, N.~J. Mayhall, S.~E. Economou, and E.~Barnes, ``Efficient symmetry-preserving state preparation circuits for the variational quantum eigensolver algorithm,'' \href{http://dx.doi.org/10.1038/s41534-019-0240-1}{{\em npj Quantum Information} {\bfseries 6} no.~1, (Jan., 2020) }. \url{http://dx.doi.org/10.1038/s41534-019-0240-1}.

\bibitem{Tilly:2021jem}
J.~Tilly {\em et~al.}, ``{The Variational Quantum Eigensolver: A review of methods and best practices},'' \href{http://dx.doi.org/10.1016/j.physrep.2022.08.003}{{\em Phys. Rept.} {\bfseries 986} (2022) 1--128}, \href{http://arxiv.org/abs/2111.05176}{{\ttfamily arXiv:2111.05176 [quant-ph]}}.

\bibitem{Bharti2022}
K.~Bharti, A.~Cervera-Lierta, T.~H. Kyaw, T.~Haug, S.~Alperin-Lea, A.~Anand, M.~Degroote, H.~Heimonen, J.~S. Kottmann, T.~Menke, W.-K. Mok, S.~Sim, L.-C. Kwek, and A.~Aspuru-Guzik, ``Noisy intermediate-scale quantum algorithms,'' \href{http://dx.doi.org/10.1103/RevModPhys.94.015004}{{\em Rev. Mod. Phys.} {\bfseries 94} (Feb, 2022) 015004}. \url{https://link.aps.org/doi/10.1103/RevModPhys.94.015004}.

\bibitem{Mitarai:2018voy}
K.~Mitarai, M.~Negoro, M.~Kitagawa, and K.~Fujii, ``{Quantum circuit learning},'' \href{http://dx.doi.org/10.1103/PhysRevA.98.032309}{{\em Phys. Rev. A} {\bfseries 98} no.~3, (2018) 032309}, \href{http://arxiv.org/abs/1803.00745}{{\ttfamily arXiv:1803.00745 [quant-ph]}}.

\bibitem{Schuld:2018aiz}
M.~Schuld, V.~Bergholm, C.~Gogolin, J.~Izaac, and N.~Killoran, ``{Evaluating analytic gradients on quantum hardware},'' \href{http://dx.doi.org/10.1103/PhysRevA.99.032331}{{\em Phys. Rev. A} {\bfseries 99} no.~3, (2019) 032331}, \href{http://arxiv.org/abs/1811.11184}{{\ttfamily arXiv:1811.11184 [quant-ph]}}.

\bibitem{Killoran:2019yfa}
N.~Killoran, T.~R. Bromley, J.~M. Arrazola, M.~Schuld, N.~Quesada, and S.~Lloyd, ``{Continuous-variable quantum neural networks},'' \href{http://dx.doi.org/10.1103/PhysRevResearch.1.033063}{{\em Phys. Rev. Res.} {\bfseries 1} no.~3, (2019) 033063}.

\bibitem{Bangar:2023akc}
S.~Bangar, L.~Sunny, K.~Yeter-Aydeniz, and G.~Siopsis, ``{Experimentally realizable continuous-variable quantum neural networks},'' \href{http://dx.doi.org/10.1103/PhysRevA.108.042414}{{\em Phys. Rev. A} {\bfseries 108} no.~4, (2023) 042414}, \href{http://arxiv.org/abs/2306.02525}{{\ttfamily arXiv:2306.02525 [quant-ph]}}.

\bibitem{Farhi:2000ikn}
E.~Farhi, J.~Goldstone, S.~Gutmann, and M.~Sipser, ``{Quantum Computation by Adiabatic Evolution},'' \href{http://arxiv.org/abs/quant-ph/0001106}{{\ttfamily arXiv:quant-ph/0001106}}.

\bibitem{Bittel:2021ley}
L.~Bittel and M.~Kliesch, ``{Training variational quantum algorithms is NP-hard},'' \href{http://dx.doi.org/10.1103/PhysRevLett.127.120502}{{\em Phys. Rev. Lett.} {\bfseries 127} (2021) 120502}, \href{http://arxiv.org/abs/2101.07267}{{\ttfamily arXiv:2101.07267 [quant-ph]}}.

\bibitem{McClean_2018}
J.~R. McClean, S.~Boixo, V.~N. Smelyanskiy, R.~Babbush, and H.~Neven, ``Barren plateaus in quantum neural network training landscapes,'' \href{http://dx.doi.org/10.1038/s41467-018-07090-4}{{\em Nature Communications} {\bfseries 9} no.~1, (Nov, 2018) }. \url{https://doi.org/10.1038%2Fs41467-018-07090-4}.

\bibitem{Grimsley:2018wnd}
H.~R. Grimsley, S.~E. Economou, E.~Barnes, and N.~J. Mayhall, ``{An adaptive variational algorithm for exact molecular simulations on a quantum computer},'' \href{http://dx.doi.org/10.1038/s41467-019-10988-2}{{\em Nature Commun.} {\bfseries 10} (2019) 3007}, \href{http://arxiv.org/abs/1812.11173}{{\ttfamily arXiv:1812.11173 [quant-ph]}}.

\bibitem{Larocca:2021ksf}
M.~Larocca, P.~Czarnik, K.~Sharma, G.~Muraleedharan, P.~J. Coles, and M.~Cerezo, ``{Diagnosing Barren Plateaus with Tools from Quantum Optimal Control},'' \href{http://dx.doi.org/10.22331/q-2022-09-29-824}{{\em Quantum} {\bfseries 6} (2022) 824}, \href{http://arxiv.org/abs/2105.14377}{{\ttfamily arXiv:2105.14377 [quant-ph]}}.

\bibitem{Ragone:2023qbn}
M.~Ragone, B.~N. Bakalov, F.~Sauvage, A.~F. Kemper, C.~O. Marrero, M.~Larocca, and M.~Cerezo, ``{A Unified Theory of Barren Plateaus for Deep Parametrized Quantum Circuits},'' \href{http://arxiv.org/abs/2309.09342}{{\ttfamily arXiv:2309.09342 [quant-ph]}}.

\bibitem{Diaz:2023uuo}
N.~L. Diaz, D.~Garc\'\i{}a-Mart\'\i{}n, S.~Kazi, M.~Larocca, and M.~Cerezo, ``{Showcasing a Barren Plateau Theory Beyond the Dynamical Lie Algebra},'' \href{http://arxiv.org/abs/2310.11505}{{\ttfamily arXiv:2310.11505 [quant-ph]}}.

\bibitem{Fontana:2023wnj}
E.~Fontana, D.~Herman, S.~Chakrabarti, N.~Kumar, R.~Yalovetzky, J.~Heredge, S.~H. Sureshbabu, and M.~Pistoia, ``{The Adjoint Is All You Need: Characterizing Barren Plateaus in Quantum Ans\"atze},'' \href{http://arxiv.org/abs/2309.07902}{{\ttfamily arXiv:2309.07902 [quant-ph]}}.

\bibitem{10.1063/1.5027484}
Y.~Ge, J.~Tura, and J.~I. Cirac, ``{Faster ground state preparation and high-precision ground energy estimation with fewer qubits},'' \href{http://dx.doi.org/10.1063/1.5027484}{{\em Journal of Mathematical Physics} {\bfseries 60} no.~2, (02, 2019) 022202}. \url{https://doi.org/10.1063/1.5027484}.

\bibitem{Gilyen:2019}
A.~Gily\'{e}n, Y.~Su, G.~H. Low, and N.~Wiebe, \href{http://dx.doi.org/10.1145/3313276.3316366}{``Quantum singular value transformation and beyond: exponential improvements for quantum matrix arithmetics,''} in {\em Proceedings of the 51st Annual ACM SIGACT Symposium on Theory of Computing}, STOC 2019, pp.~193--204.
\newblock Association for Computing Machinery, New York, NY, USA, 2019.
\newblock \url{https://doi.org/10.1145/3313276.3316366}.

\bibitem{Motta:2020}
M.~Motta, C.~Sun, A.~T.~K. Tan, M.~J. O'Rourke, E.~Ye, A.~J. Minnich, F.~G. S.~L. Brand{\~a}o, and G.~K.-L. Chan, ``Determining eigenstates and thermal states on a quantum computer using quantum imaginary time evolution,'' \href{http://dx.doi.org/10.1038/s41567-019-0704-4}{{\em Nature Physics} {\bfseries 16} no.~2, (Feb, 2020) 205--210}. \url{https://doi.org/10.1038/s41567-019-0704-4}.

\bibitem{Dong:2022mmq}
Y.~Dong, L.~Lin, and Y.~Tong, ``{Ground-State Preparation and Energy Estimation on Early Fault-Tolerant Quantum Computers via Quantum Eigenvalue Transformation of Unitary Matrices},'' \href{http://dx.doi.org/10.1103/PRXQuantum.3.040305}{{\em PRX Quantum} {\bfseries 3} no.~4, (2022) 040305}, \href{http://arxiv.org/abs/2204.05955}{{\ttfamily arXiv:2204.05955 [quant-ph]}}.

\bibitem{Stetcu:2022nhy}
I.~Stetcu, A.~Baroni, and J.~Carlson, ``{Projection algorithm for state preparation on quantum computers},'' \href{http://dx.doi.org/10.1103/PhysRevC.108.L031306}{{\em Phys. Rev. C} {\bfseries 108} no.~3, (2023) L031306}, \href{http://arxiv.org/abs/2211.10545}{{\ttfamily arXiv:2211.10545 [quant-ph]}}.

\bibitem{Rrapaj:2024shg}
E.~Rrapaj and E.~Rule, ``{Exact block encoding of imaginary time evolution with universal quantum neural networks},'' \href{http://arxiv.org/abs/2403.17273}{{\ttfamily arXiv:2403.17273 [quant-ph]}}.

\bibitem{jax2018github}
J.~Bradbury, R.~Frostig, P.~Hawkins, M.~J. Johnson, C.~Leary, D.~Maclaurin, G.~Necula, A.~Paszke, J.~Vander{P}las, S.~Wanderman-{M}ilne, and Q.~Zhang, ``{JAX}: composable transformations of {P}ython+{N}um{P}y programs,'' 2018.
\newblock \url{http://github.com/google/jax}.

\bibitem{2020SciPy-NMeth}
P.~Virtanen, R.~Gommers, T.~E. Oliphant, M.~Haberland, T.~Reddy, D.~Cournapeau, E.~Burovski, P.~Peterson, W.~Weckesser, J.~Bright, S.~J. {van der Walt}, M.~Brett, J.~Wilson, K.~J. Millman, N.~Mayorov, A.~R.~J. Nelson, E.~Jones, R.~Kern, E.~Larson, C.~J. Carey, {\.I}.~Polat, Y.~Feng, E.~W. Moore, J.~{VanderPlas}, D.~Laxalde, J.~Perktold, R.~Cimrman, I.~Henriksen, E.~A. Quintero, C.~R. Harris, A.~M. Archibald, A.~H. Ribeiro, F.~Pedregosa, P.~{van Mulbregt}, and {SciPy 1.0 Contributors}, ``{{SciPy} 1.0: Fundamental Algorithms for Scientific Computing in Python},'' \href{http://dx.doi.org/10.1038/s41592-019-0686-2}{{\em Nature Methods} {\bfseries 17} (2020) 261--272}.

\bibitem{Belitsky:2003nz}
A.~V. Belitsky, X.-d. Ji, and F.~Yuan, ``{Quark imaging in the proton via quantum phase space distributions},'' \href{http://dx.doi.org/10.1103/PhysRevD.69.074014}{{\em Phys. Rev. D} {\bfseries 69} (2004) 074014}, \href{http://arxiv.org/abs/hep-ph/0307383}{{\ttfamily arXiv:hep-ph/0307383}}.

\bibitem{Hagiwara:2016kam}
Y.~Hagiwara, Y.~Hatta, and T.~Ueda, ``{Wigner, Husimi, and generalized transverse momentum dependent distributions in the color glass condensate},'' \href{http://dx.doi.org/10.1103/PhysRevD.94.094036}{{\em Phys. Rev. D} {\bfseries 94} no.~9, (2016) 094036}, \href{http://arxiv.org/abs/1609.05773}{{\ttfamily arXiv:1609.05773 [hep-ph]}}.

\bibitem{negativity1}
A.~Kenfack and K.~Zyczkowski, ``Negativity of the wigner function as an indicator of non-classicality,'' \href{http://dx.doi.org/10.1088/1464-4266/6/10/003}{{\em Journal of Optics B: Quantum and Semiclassical Optics} {\bfseries 6} (08, 2004) 396}.

\bibitem{QSCOUT}
S.~M. Clark, D.~Lobser, M.~C. Revelle, C.~G. Yale, D.~Bossert, A.~D. Burch, M.~N. Chow, C.~W. Hogle, M.~Ivory, J.~Pehr, B.~Salzbrenner, D.~Stick, W.~Sweatt, J.~M. Wilson, E.~Winrow, and P.~Maunz, ``Engineering the quantum scientific computing open user testbed,'' \href{http://dx.doi.org/10.1109/TQE.2021.3096480}{{\em IEEE Transactions on Quantum Engineering} {\bfseries 2} (2021) 1--32}.

\end{thebibliography}\endgroup

\end{document}